\documentstyle[12pt,epsf]{article}

\newlength{\pubnumber} 

\catcode`\@=11
\@addtoreset{equation}{section}

\def\section{\@startsection{section}{1}{\z@}{3.5ex plus 1ex minus .2ex}
 {2.3ex plus .2ex}{\large\bf}}
\def\subsection{\@startsection{subsection}{2}{\z@}{2.3ex plus .2ex}
 {2.3ex plus .2ex}{\bf}}

\begin{document}

\begin{titlepage}
\samepage{
\setcounter{page}{1}
\rightline{ACT-2/99}
\rightline{CTP-TAMU-12/99}
\rightline{TPI-MINN-99/22}
\rightline{UMN-TH-1760-99}
\rightline{\tt hep-ph/9904301}
\rightline{May 2000}
\vfill
\begin{center}
 {\Large \bf A Minimal Superstring Standard Model \\I: Flat Directions\\ }
\vfill
\vskip .4truecm
\vfill {\large
        G.B. Cleaver,$^{1,2}$\footnote{gcleaver@rainbow.physics.tamu.edu}
        A.E. Faraggi,$^{3}$\footnote{faraggi@mnhepw.hep.umn.edu}
        and
        D.V. Nanopoulos$^{1,2,4}$\footnote{dimitri@soda.physics.tamu.edu}}
\\
\vspace{.12in}
{\it $^{1}$ Center for Theoretical Physics,
            Dept.\  of Physics, Texas A\&M University,\\
            College Station, TX 77843, USA\\}
\vspace{.06in}
{\it $^{2}$ Astro Particle Physics Group,
            Houston Advanced Research Center (HARC),\\
            The Mitchell Campus,
            Woodlands, TX 77381, USA\\}
\vspace{.06in}
{\it$^{3}$  Department of Physics, University of Minnesota, 
            Minneapolis, MN 55455, USA\\}
\vspace{.025in}
{\it$^{4}$  Academy of Athens, Chair of Theoretical Physics, 
            Division of Natural Sciences,\\
            28 Panepistimiou Avenue, Athens 10679, Greece\\}
\vspace{.025in}
\end{center}
\vfill
\begin{abstract}
Three family
$SU(3)_C\times SU(2)_L\times U(1)_Y$ string models in several constructions 
generically possess two features: (i) an extra local anomalous $U(1)_A$ and 
(ii) numerous (often fractionally charged) exotic particles beyond
those in the minimal supersymmetric model (MSSM). 
Recently, we demonstrated that
the observable sector effective field theory of such a free fermionic 
string model can reduce to that of
the MSSM, with the standard observable gauge group being just
$SU(3)_C\times SU(2)_L\times U(1)_Y$ and the
$SU(3)_C\times SU(2)_L\times U(1)_Y$--charged 
spectrum of the observable sector consisting solely of the MSSM spectrum.
An example of a model with this property was shown.
We continue our investigation of this model by presenting a large set
of different flat directions of the same model that all produce the MSSM
spectrum. Our results suggest that even after imposing the conditions for
the decoupling of exotic states, there may remain sufficient freedom
to satisfy the remaining phenomenological constraints imposed
by the observed data.  
\end{abstract}
\smallskip}
\end{titlepage}

\setcounter{footnote}{0}


\def\at{ }
\def\beq{\begin{equation}}
\def\eeq{\end{equation}}
\def\beqn{\begin{eqnarray}}
\def\eeqn{\end{eqnarray}}

\def\lt{<}
\def\eq#1{eq.\ (\ref{#1})}

\def\dag{\dagger}
\def\qandq{\quad {\rm and} \quad} 
\def\qand{\quad {\rm and} } 
\def\andq{ {\rm and} \quad } 
\def\qwithq{\quad {\rm with} \quad} 
\def\qwith{ \quad {\rm with} } 
\def\withq{ {\rm with} \quad} 

\def\no{\noindent }
\def\nolabel{\nonumber }
\def\ie{i.e., }
\def\eg{{\it e.g.}}
\def\fhalf{\frac{1}{2}}
\def\fsqrt{\frac{1}{\sqrt{2}}}
\def\half{{\textstyle{1\over 2}}}
\def\third{{\textstyle {1\over3}}}
\def\quarter{{\textstyle {1\over4}}}
\def\sixth{{\textstyle {1\over6}}}
\def\m{{\tt -}}
\def\p{{\tt +}}
\def\pps{\phantom{+}}

\def\zz{$Z_2\times Z_2$ }

\def\Tr{{\rm Tr}\, }
\def\tr{{\rm tr}\, }

\def\MP{M_{P}}
\def\hb{\bar{h}}
\def\slash#1{#1\hskip-6pt/\hskip6pt}
\def\slk{\slash{k}}
\def\GeV{\,{\rm GeV}}
\def\TeV{\,{\rm TeV}}
\def\y{\,{\rm y}}
\def\SM{Standard--Model }
\def\SUSY{supersymmetry }
\def\SSSM{supersymmetric standard model}
\def\MSSM{minimal supersymmetric standard model}
\def\smgg{ $SU(3)_C\times SU(2)_L\times U(1)_Y$ }
\def\vev#1{\left\langle #1\right\rangle}
\def\mvev#1{|\langle #1\rangle|^2}

\def\eps{\epsilon}
\def\UA{U(1)_{\rm A}}
\def\QA{Q^{(\rm A)}}
\def\mssm{SU(3)_C\times SU(2)_L\times U(1)_Y} 

\def\KM{Ka\v c--Moody }

\def\l{\langle}
\def\r{\rangle}
\def\o#1{\frac{1}{#1}}

\def\Htw{{\tilde H}}
\def\chibar{{\overline{\chi}}}
\def\qbar{{\overline{q}}}
\def\ibar{{\overline{\imath}}}
\def\jbar{{\overline{\jmath}}}
\def\Hbar{{\overline{H}}}
\def\Qbar{{\overline{Q}}}
\def\abar{{\overline{a}}}
\def\alphabar{{\overline{\alpha}}}
\def\betabar{{\overline{\beta}}}
\def\tautwo{{ \tau_2 }}
\def\thetatwo{{ \vartheta_2 }}
\def\thetathree{{ \vartheta_3 }}
\def\thetafour{{ \vartheta_4 }}
\def\ttwo{{\vartheta_2}}
\def\tthree{{\vartheta_3}}
\def\tfour{{\vartheta_4}}
\def\ti{{\vartheta_i}}
\def\tj{{\vartheta_j}}
\def\tk{{\vartheta_k}}
\def\calF{{\cal F}}
\def\smallmatrix#1#2#3#4{{ {{#1}~{#2}\choose{#3}~{#4}} }}
\def\ab{{\alpha\beta}}
\def\Minv{{ (M^{-1}_\ab)_{ij} }}
\def\ii{{(i)}}
\def\V{{\bf V}}
\def\N{{\bf N}}

\def\b{{\bf b}}
\def\S{{\bf S}}
\def\X{{\bf X}}
\def\I{{\bf I}}
\def\bone{{\mathbf 1}}
\def\bo{{\mathbf 0}}
\def\bs{{\mathbf S}}
\def\bb{{\mathbf b}}
\def\mb{{\mathbf b}}
\def\mS{{\mathbf S}}
\def\bS{{\mathbf S}}
\def\bs{{\mathbf S}}
\def\mX{{\mathbf X}}
\def\mI{{\mathbf I}}
\def\bI{{\mathbf I}}
\def\balpha{{\mathbf \alpha}}
\def\bbeta{{\mathbf \beta}}
\def\bgamma{{\mathbf \gamma}}
\def\bxi{{\mathbf \xi}}
\def\malpha{{\mathbf \alpha}}
\def\mbeta{{\mathbf \beta}}
\def\mgamma{{\mathbf \gamma}}
\def\mxi{{\mathbf \xi}}
\def\bphi{\overline{\Phi}}

\def\eps{\epsilon}

\def\t#1#2{{ \Theta\left\lbrack \matrix{ {#1}\cr {#2}\cr }\right\rbrack }}
\def\C#1#2{{ C\left\lbrack \matrix{ {#1}\cr {#2}\cr }\right\rbrack }}
\def\tp#1#2{{ \Theta'\left\lbrack \matrix{ {#1}\cr {#2}\cr }\right\rbrack }}
\def\tpp#1#2{{ \Theta''\left\lbrack \matrix{ {#1}\cr {#2}\cr }\right\rbrack }}
\def\l{\langle}
\def\r{\rangle}

\def\f#1{$\Phi_{#1}$}
\def\fb#1{$\overline{\Phi}_{#1}$}
\def\fp#1{$\Phi^{'}_{#1}$}
\def\fbp#1{$\overline{\Phi}^{'}_{#1}$}
\def\fpx#1{$\Phi^{(')}_{#1}$}
\def\fbpx#1{$\overline{\Phi}^{(')}_{#1}$}
\def\Hd#1{$H_{#1}$}
\def\Vd#1{$V_{#1}$}
\def\NC#1{$N^c_{#1}$}

\def\mf#1{\Phi_{#1}}
\def\mnfb#1{\overline{\Phi}_{#1}}
\def\mnfp#1{\Phi^{'}_{#1}}
\def\mfbp#1{\overline{\Phi}^{'}_{#1}}
\def\mfpx#1{\Phi^{(')}_{#1}}
\def\mfbpx#1{\overline{\Phi}^{(')}_{#1}}
\def\mH#1{H_{#1}}
\def\mV#1{V_{#1}}
\def\mNC#1{N^c_{#1}}

\def\Q#1{Q_{#1}}
\def\dc#1{d^{c}_{#1}}
\def\uc#1{u^{c}_{#1}}
\def\h#1{h_{#1}}
\def\hb#1{{\bar{h}}_{#1}}
\def\bh#1{{\bar{h}}_{#1}}
\def\L#1{L_{#1}}
\def\ec#1{e^{c}_{#1}}
\def\Nc#1{N^{c}_{#1}}
\def\H#1{H_{#1}}
\def\V#1{V_{#1}}
\def\Hs#1{H^{s}_{#1}}
\def\Vs#1{V^{s}_{#1}}
\def\sH#1{H^{s}_{#1}}
\def\sV#1{V^{s}_{#1}}
\def\P#1{\Phi_{#1}}
\def\p#1{\Phi_{#1}}
\def\pp#1{\Phi^{'}_{#1}}
\def\pb#1{{\overline{\Phi}}_{#1}}
\def\bp#1{{\overline{\Phi}}_{#1}}
\def\pbp#1{{\overline{\Phi}}^{'}_{#1}}
\def\ppb#1{{\overline{\Phi}}^{'}_{#1}}
\def\bpp#1{{\overline{\Phi}}^{'}_{#1}}
\def\bi#1{{\overline{\Phi}}^{'}_{#1}}

\def\obs{{\rm observable}}
\def\sig{{\rm singlets}}
\def\hid{{\rm hidden}}
\def\mix{{\rm mixed}}


\def\inbar{\,\vrule height1.5ex width.4pt depth0pt}

\def\IC{\relax\hbox{$\inbar\kern-.3em{\rm C}$}}
\def\IQ{\relax\hbox{$\inbar\kern-.3em{\rm Q}$}}
\def\IR{\relax{\rm I\kern-.18em R}}
 \font\cmss=cmss10 \font\cmsss=cmss10 at 7pt
\def\IZ{\relax\ifmmode\mathchoice
 {\hbox{\cmss Z\kern-.4em Z}}{\hbox{\cmss Z\kern-.4em Z}}
 {\lower.9pt\hbox{\cmsss Z\kern-.4em Z}}
 {\lower1.2pt\hbox{\cmsss Z\kern-.4em Z}}\else{\cmss Z\kern-.4em Z}\fi}

\def\AEF{A.E. Faraggi}
\def\AP#1#2#3{{\it Ann.\ Phys.}\/ {\bf#1} (19#2) #3}
\def\NPB#1#2#3{{\it Nucl.\ Phys.}\/ {\bf B#1} (19#2) #3}
\def\NPBPS#1#2#3{{\it Nucl.\ Phys.}\/ {{\bf B} (Proc. Suppl.) {\bf #1}} (19#2) 
 #3}
\def\PLB#1#2#3{{\it Phys.\ Lett.}\/ {\bf B#1} (19#2) #3}
\def\PRD#1#2#3{{\it Phys.\ Rev.}\/ {\bf D#1} (19#2) #3}
\def\PRL#1#2#3{{\it Phys.\ Rev.\ Lett.}\/ {\bf #1} (19#2) #3}
\def\PRT#1#2#3{{\it Phys.\ Rep.}\/ {\bf#1} (19#2) #3}
\def\PTP#1#2#3{{\it Prog.\ Theo.\ Phys.}\/ {\bf#1} (19#2) #3}
\def\MODA#1#2#3{{\it Mod.\ Phys.\ Lett.}\/ {\bf A#1} (#2) #3}
\def\IJMP#1#2#3{{\it Int.\ J.\ Mod.\ Phys.}\/ {\bf A#1} (19#2) #3}
\def\nuvc#1#2#3{{\it Nuovo Cimento}\/ {\bf #1A} (#2) #3}
\def\RPP#1#2#3{{\it Rept.\ Prog.\ Phys.}\/ {\bf #1} (19#2) #3}
\def\etal{{\it et al\/}}

\hyphenation{su-per-sym-met-ric non-su-per-sym-met-ric}
\hyphenation{space-time-super-sym-met-ric}
\hyphenation{mod-u-lar mod-u-lar--in-var-i-ant}


\setcounter{footnote}{0}

\section{Introduction}

Deriving the Standard Model
from heterotic string theory remains one of the strongly motivated
endevours in theoretical physics.
On the one hand, for over a quarter of a century now, the structure
of the Standard model itself suggests the realization of grand unified 
structures, most appealing in the context of $SO(10)$ 
unification \cite{granda,grandb}.
On the other hand supersymmetry, a key ingredient
in grand and string unification, continues to be the only extension of the 
Standard Model still consistent with the experimental data, 
whereas theories with
a low energy cutoff in general run into conflict
with experiment once detailed models are considered. 
Finally, string theory remains the only known theoretical
framework for the consistent unification of gravity
and the gauge interactions. 

There are two main approaches regarding how 
to derive the Standard Model from string theory.
The first approach asserts that we must first understand 
the nonperturbative formulation of string theory
and then the true string vacuum will be uniquely
revealed. The second approach argues that 
much can be learned about the string realization of the Standard Model
by studying the features and properties of 
phenomenologically promising perturbative string models.
Singling out such string models for study may be 
instrumental for learning about the nonperturbative
dynamics of the theory. 
Following the second line of thought,
a class of phenomenologically promising models has been
identified. These models, constructed in the free
fermionic formulation \cite{fff}, correspond to
$Z_2\times Z_2$ orbifold models with nontrivial
Wilson lines and background fields. Studying
F--theory compactification on the $Z_2\times Z_2$
orbifold, which is related to the free fermionic
models \cite{befnq}, indeed reveals new features that do
not arise in other similar F--theory studies \cite{ftheory}.
It is not unlikely that these new features will
eventually prove to be important for understanding
the issue of vacuum selection in string theory.

Connecting string theory to low energy experimental
data, parameterized by the Standard Model, remains vital.
It is important to point out that of the 
semi--realistic orbifold \cite{orbi} and free fermionic models
\cite{flipped,fny,eu,slm,so64,chl}, the NAHE--based $Z_2\times Z_2$
free fermionic models \cite{nahe,slm}
(or more generally $Z_2\times Z_2$ orbifold models  
at the free fermionic point) are the only ones that naturally give rise
to the $SO(10)$ unification structures. 
Thus, it seems of much value to continue improving    
our understanding of the more phenomenologically viable three generation 
$Z_2\times Z_2$ free fermionic models, as well as their
embeddings in compactifications of M-- and F--theory.   

It was stressed in the past for a variety of phenomenological reasons
that the canonical $SO(10)$ embedding of the weak--hypercharge in
string models is highly preferred \cite{ibanez}. 
In fact, it was even suggested that this must be the case
in the true string vacuum. 
We stress that this is precisely the embedding obtained in the 
$Z_2\times Z_2$ free fermionic models, in
contrast to other quasi--realistic orbifold \cite{orbi}
or free fermionic models \cite{chl}, 
which do not produce the standard $SO(10)$ embedding. 
The first phenomenological
criterion that a string model must satisfy are three chiral generations
with the standard $SO(10)$ embedding of the Standard Model gauge group.
While the $Z_2\times Z_2$ free fermionic models naturally give rise to the 
$SO(10)$ unification structure, the $SO(10)$ symmetry has to be broken at
the string level. Therefore, there is no explicit $SO(10)$ symmetry in the
effective field theory level. Nevertheless, the $SO(10)$ symmetry is still
reflected, for example, in some of the Yukawa coupling relations.

In realistic heterotic string models,
it is well known that modular invariance
constraints impose that the string spectrum contains
exotic fractionally charged states \cite{schellekens}. 
Such states indeed occur in all known string models built from
level--one \KM algebras,
and may appear in the massive or massless sectors. 
In many of the more realistic 
examples such states arise in vector--like representations
and therefore may obtain mass terms from cubic--level
or higher order terms in the superpotential.
It is clear that in the true string vacuum such
fractionally charged states must be either confined \cite{flipped}
or sufficiently massive \cite{fc,huet,nrt}.
Thus, the next non--trivial phenomenological criteria
on viable string models is that there
should be no free fractionally charged states
surviving to low energies. Furthermore, 
as the existence of fractionally
charged states and other states beyond
the minimal supersymmetric standard model
will in general affect the unification of the
the gauge couplings, an attractive
scenario is that all the states beyond the
MSSM decouple from the low energy spectrum at
the string scale. 

Recently, we have demonstrated the existence
of string models with the above properties \cite{cfn1}.
Studying the flat directions in the string model
of ref.\  \cite{fny} (referred to henceforth as the 
``FNY model''), we showed the existence
of one flat direction for which the massless
spectrum below the string scale consists
solely of the spectrum of the minimal supersymmetric
standard model. Furthermore, it was found that
in this particular model the $D$--flatness
constraints necessarily impose that either of the
$U(1)_{Z^\prime}$ or $U(1)_Y$ symmetries must
be broken by the choices of flat directions.
This is in fact an attractive situation
in which in the phenomenologically viable
case the surviving $SO(10)$ subgroup necessarily coincides 
with that of the Standard Model, and the $U(1)_{Z^\prime}$
is necessarily broken by the choices of flat directions.

We remark that
our model also contains, at the massless string level,
a number of electroweak Higgs doublets and 
a color triplet/anti--triplet pair beyond the MSSM. 
We show that by the same suitable choices
of flat directions that only one Higgs pair remains light below
the string scale. The additional color triplet/anti--triplet pair 
receives mass from a fifth order superpotential term. 
This results in the triplet pair receiving a mass that is 
slightly below the string scale and is perhaps smaller 
than the doublet and fractional exotic masses 
by a factor of around $(1/10-1/100)$.
We emphasize that the numerical estimate of the masses
arising from the singlet VEVs should be regarded only as
illustrative. The important result is the generation of
mass terms for all the states beyond the MSSM,
near the string scale. The actual masses of the extra
fields may be spread around the $M_U$ scale, thus
inducing small threshold corrections that are still
expected to be compatible with the low energy
experimental data.

The string solution found in ref.\  \cite{cfn1}
is the first known example of a minimal superstring
derived standard model, in which, of all the 
$SU(3)_C\times SU(2)_L \times U(1)_Y$--charged states,
only the MSSM spectrum remains light below the string scale. 
In this paper we expand the analysis of ref.\  \cite{cfn1}.
The first obvious question is naturally whether
the special solution found in ref.\  \cite{cfn1} is
an isolated example or whether there exists
an enlarged space of solutions giving rise
solely to the MSSM spectrum below the string scale. 
We recall that in the FNY model of ref.\  \cite{fny}
the condition for the decoupling of the exotic
fractionally charged states from the massless spectrum
is that a specific set of Standard Model singlet
fields \cite{fc}
acquire non--vanishing vacuum expectation values (VEVs) in the 
cancellation of the anomalous $U(1)_A$ Fayet--Iliopoulos (FI)
$D$--term. Additionally we showed that the 
same set of VEVs produce mass terms that lead
to the decoupling of the extra color triplets and
electroweak doublets, beyond the MSSM.
Thus, in any solution that incorporates 
those VEVs, the resulting spectrum below
the string scale consists solely of the MSSM
spectrum. Expanding the analysis of ref.\  \cite{cfn1}
we show that there exists, in fact, an extended space
of solutions which incorporate those VEVs. This
is a very promising situation, for it
demonstrates that there may still be sufficient freedom
to allow the possibility of accommodating the  
various phenomenological constraints in one of these
solutions. 

Our paper is organized as follows:
Section 2 is a brief review of the FNY model \cite{fny}, including a discussion
of the model's massless spectrum, prior to any states taking on VEVs. 
This is accompanied by Appendix A,
which lists all string quantum numbers, including the non--gauge charges,
of the massless states, 	
and by Appendix B, which contains the
complete renormalizable superpotential and all fourth through sixth
order non--renormalizable superpotential terms.
Section 3 reviews constraints on flat directions 
and presents a survey of viable flat directions that generate (near)
string--scale mass to all $SU(3)_C\times SU(2)_L\times U(1)_Y$--charged 
MSSM exotic states in the FNY model, 
producing an effective MSSM field theory below the string scale. 
Accompanying Section 3 are Appendices C and D.
Appendix C contains a SM--conserving basis set, 
of the ``maximally orthogonal'' class, 
for generating $D$--flat directions, while Appendix D contains the actual
tables of classes of $D$-- and $F$--flat directions. 
These tables indicate the order to which 
$F$--flatness is retained, the respective superpotential terms that break
$F$--flatness, the dimension of each flat direction, and the respective 
number of non--anomalous $U(1)_i$ that are broken by these directions.
Section 4 then discusses some distinguishing features of 
the various flat directions. Phenomenological implications of these
flat directions will be presented in \cite{cfn3}.
 
\section{The FNY Model (A Review)}

\subsection{Construction}
The more realistic free fermionic models, which utilize the
NAHE--set of boundary condition basis vectors,
admit the $SO(10)$ embedding of the Standard Model
gauge group. Aside from the phenomenological 
aspects, which motivate the need for the $SO(10)$ embedding,
another advantage of utilizing the standard $SO(10)$ embedding
and hence of the NAHE--set, is the decoupling \cite{decoup} of the
exotic fractionally charged states from the massless spectrum. 
This should be contrasted with models, like the free fermionic
models of \cite{chl}, which do not allow the $SO(10)$ embedding
and contain exotic fractionally charged states which cannot
decouple from the massless spectrum \cite{cceelw}. This distinction is an
important one, as it severely limits the number of phenomenologically
viable models, even among the three generation orbifold models that
are traditionally viewed as semi--realistic. The more realistic NAHE--based 
free fermionic models represent one class of string models that
still survives this requirement. 

For completeness we recall here the construction of the FNY model
and its distinctive properties. 
The boundary condition basis vectors which generate the more realistic 
free fermionic models are, in general, divided into two major subsets.
The first set consists of the NAHE set \cite{nahe,slm}, which is a set
of five boundary condition basis vectors denoted 
$\{{\bf 1},\bS,\bb_1,\bb_2,\bb_3\}$.
With `$\bo$' indicating Neveu--Schwarz boundary conditions
and `$\bone$' indicating Ramond boundary conditions, 
these vectors are as follows:
\beqn
 &&\begin{tabular}{c|c|ccc|c|ccc|c}
 ~ & $\psi^\mu$ & $\chi^{12}$ & $\chi^{34}$ & $\chi^{56}$ &
        $\bar{\psi}^{1,...,5} $ &
        $\bar{\eta}^1 $&
        $\bar{\eta}^2 $&
        $\bar{\eta}^3 $&
        $\bar{\phi}^{1,...,8} $ \\
\hline
\hline
      $\bone$ &  1 & 1&1&1 & 1,...,1 & 1 & 1 & 1 & 1,...,1 \\
         $\S$ &  1 & 1&1&1 & 0,...,0 & 0 & 0 & 0 & 0,...,0 \\
\hline
  ${\bb}_1$ &  1 & 1&0&0 & 1,...,1 & 1 & 0 & 0 & 0,...,0 \\
  ${\bb}_2$ &  1 & 0&1&0 & 1,...,1 & 0 & 1 & 0 & 0,...,0 \\
  ${\bb}_3$ &  1 & 0&0&1 & 1,...,1 & 0 & 0 & 1 & 0,...,0 \\
\end{tabular}
   \nonumber\\
   ~  &&  ~ \nonumber\\
   ~  &&  ~ \nonumber\\
     &&\begin{tabular}{c|cc|cc|cc}
 ~&      $y^{3,...,6}$  &
        $\bar{y}^{3,...,6}$  &
        $y^{1,2},\omega^{5,6}$  &
        $\bar{y}^{1,2},\bar{\omega}^{5,6}$  &
        $\omega^{1,...,4}$  &
        $\bar{\omega}^{1,...,4}$   \\
\hline
\hline
  {$\bone$} & 1,...,1 & 1,...,1 & 1,...,1 & 1,...,1 & 1,...,1 & 1,...,1 \\
   $\bS$    & 0,...,0 & 0,...,0 & 0,...,0 & 0,...,0 & 0,...,0 & 0,...,0 \\
\hline
${\bb}_1$ & 1,...,1 & 1,...,1 & 0,...,0 & 0,...,0 & 0,...,0 & 0,...,0 \\
${\bb}_2$ & 0,...,0 & 0,...,0 & 1,...,1 & 1,...,1 & 0,...,0 & 0,...,0 \\
${\bb}_3$ & 0,...,0 & 0,...,0 & 0,...,0 & 0,...,0 & 1,...,1 & 1,...,1 \\
\end{tabular}
\label{nahe}
\eeqn
with the following
choice of phases which define how the generalized GSO projections are to
be performed in each sector of the theory:
\beq
      C\left( \matrix{\bb_i\cr \bb_j\cr}\right)~=~
      C\left( \matrix{\bb_i\cr \bS\cr}\right) ~=~
      C\left( \matrix{\bone \cr \bone \cr}\right) ~= ~ -1~.
\label{nahephases}
\eeq
The remaining projection phases can be determined from those above through
the self--consistency constraints.
The precise rules governing the choices of such vectors and phases, as well
as the procedures for generating the corresponding space--time particle
spectrum, are given in refs.~\cite{fff}.

After imposing the NAHE set, the resulting model has gauge
group $SO(10)\times SO(6)^3\times E_8$ and $N=1$
space--time supersymmetry. The model contains 48
multiplets in the $16$ representation of $SO(10)$,
16 from each twisted sector $\bb_1$, $\bb_2$ and $\bb_3$.
In addition to the spin 2 multiplets and the space--time
vector bosons, the untwisted sector produces 
six multiplets in the vectorial 10 representation
of $SO(10)$ and a number of $SO(10)\times E_8$ singlets. 
As can be seen from Table (\ref{nahe}), the model
at this stage possesses a cyclic permutation symmetry
among the basis vectors $\bb_1$, $\bb_2$ and $\bb_3$,
which is also respected by the massless spectrum.

The second stage in the construction of these NAHE--based
free fermionic models consists
of adding three additional basis vectors to the above NAHE set.
These three additional basis vectors, which are often called
$\lbrace \balpha,\bbeta, \bgamma\rbrace$,
correspond to ``Wilson lines'' in the orbifold construction.
The allowed fermion boundary conditions in these additional basis vectors are
of course also constrained by the string consistency
constraints, and must preserve modular invariance and
world--sheet supersymmetry.
The choice of these additional basis vectors
$\lbrace \balpha,\bbeta,\bgamma\rbrace$ nevertheless distinguishes
between different models and determine their low--energy properties.
For example, three additional vectors are
needed to reduce the number of massless
generations to three, one from each sector $\bb_1$, $\bb_2$, and $\bb_3$,
and the choice of their boundary conditions for the internal fermions
${\{y,\omega\vert{\bar y},{\bar\omega}\}^{1,\cdots,6}}$
also determines the Higgs doublet--triplet splitting and
the Yukawa couplings. These
low--energy phenomenological requirements therefore impose strong
constraints \cite{slm} on the possible assignment of boundary conditions to the
set of internal world--sheet fermions
${\{y,\omega\vert{\bar y},{\bar\omega}\}^{1,\cdots,6}}$.

Consider the model in Table (\ref{fnymodel})
\beqn
 &\begin{tabular}{c|c|ccc|c|ccc|c}
 ~ & $\psi^\mu$ & $\chi^{12}$ & $\chi^{34}$ & $\chi^{56}$ &
        $\bar{\psi}^{1,...,5} $ &
        $\bar{\eta}^1 $&
        $\bar{\eta}^2 $&
        $\bar{\eta}^3 $&
        $\bar{\phi}^{1,...,8} $ \\
\hline
\hline
  ${\bb_4}$     &  1 & 1&0&0 & 1~1~1~1~1 & 1 & 0 & 0 & 0~0~0~0~0~0~0~0 \\
  ${\bbeta}$   &  1 & 0&0&1 & 1~1~1~0~0 & 1 & 0 & 1 & 1~1~1~1~0~0~0~0 \\
  ${\bgamma}$  &  1 & 0&1&0 &
		${1\over2}$~${1\over2}$~${1\over2}$~${1\over2}$~${1\over2}$
	      & ${1\over2}$ & ${1\over2}$ & ${1\over2}$ &
                ${1\over2}$~0~1~1~${1\over2}$~${1\over2}$~${1\over2}$~1 \\
\end{tabular}
   \nonumber\\
   ~  &  ~ \nonumber\\
   ~  &  ~ \nonumber\\
     &\begin{tabular}{c|c|c|c}
 ~&   $y^3{y}^6$
      $y^4{\bar y}^4$
      $y^5{\bar y}^5$
      ${\bar y}^3{\bar y}^6$
  &   $y^1{\omega}^6$
      $y^2{\bar y}^2$
      $\omega^5{\bar\omega}^5$
      ${\bar y}^1{\bar\omega}^6$
  &   $\omega^1{\omega}^3$
      $\omega^2{\bar\omega}^2$
      $\omega^4{\bar\omega}^4$
      ${\bar\omega}^1{\bar\omega}^3$ \\
\hline
\hline
$\bb_4$& 1 ~~~ 0 ~~~ 0 ~~~ 1  & 0 ~~~ 0 ~~~ 1 ~~~ 0  & 0 ~~~ 0 ~~~ 1 ~~~ 0 \\
$\bbeta$ & 0 ~~~ 0 ~~~ 0 ~~~ 1  & 0 ~~~ 1 ~~~ 0 ~~~ 1  & 1 ~~~ 0 ~~~ 1 ~~~ 0 \\
$\bgamma$& 0 ~~~ 0 ~~~ 1 ~~~ 1  & 1 ~~~ 0 ~~~ 0 ~~~ 1  & 0 ~~~ 1 ~~~ 0 ~~~ 0 \\
\end{tabular}
\label{fnymodel}
\eeqn
With the choice of generalized GSO coefficients:
\beqn
C\left(\matrix{\bb_4\cr
                                    \bb_j,\bbeta\cr}\right)&&=
-C\left(\matrix{\bb_4\cr
                                    {\bone}\cr}\right)=
-C\left(\matrix{\bbeta\cr
                                    {\bone}\cr}\right)=
C\left(\matrix{\bbeta\cr
                                    \bb_j\cr}\right)=\nonumber\\
-C\left(\matrix{\bbeta\cr
                                    \bgamma\cr}\right)&&=
C\left(\matrix{\bgamma\cr
                                    \bb_2\cr}\right)=
-C\left(\matrix{\bgamma\cr
                                    \bb_1,\bb_3,\bb_4,\bgamma\cr}\right)=
-1\nonumber
\eeqn
$(j=1,2,3),$
with the others specified by modular invariance and space--time
supersymmetry. Several properties of the boundary conditions, 
eq.\ (\ref{fnymodel}), that generate the FNY model distinguish it from
other standard--like models \cite{eu}. First the basis
vector $\balpha\equiv \bb_4$ does not break the $SO(10)$ symmetry.
In the models \cite{eu} both the basis vectors $\balpha$ and
$\bbeta$ break the $SO(10)$ symmetry to $SO(6)\times SO(4)$. 
This has the consequence that the combination $\balpha+\bbeta$
gives rise to states transforming only under the observable
($SO(10)\times SO(6)^3$) part of the gauge group, which
produce electroweak Higgs representations. Thus, we
may anticipate that the
structure of the Higgs mass matrix as well the fermion
mass matrices in the FNY model will differ from those
in the models of ref.\  \cite{eu}. Another distinction
between the models is in the pairing of left--moving and
right--moving fermions which produces Ising model operators
\cite{slm}.

\subsection{Gauge Group}

Before cancellation of the FI term by an $F$-- and $D$--flat
direction, the observable gauge group for the FNY model
consists of the universal $SO(10)$ sub--group,
$SU(3)_C\times SU(2)_L\times U(1)_C\times U(1)_L$,
generated by the five complex world--sheet fermions
${\bar\psi}^{1,\cdots,5}$, and six observable horizontal,
flavor--dependent, $U(1)$ symmetries $U(1)_{1,\cdots,6}$, generated by
$\{{\bar\eta}^1,{\bar\eta}^2,{\bar\eta}^3,{\bar y}^3{\bar y}^6,
{\bar y}^1{\bar\omega}^6,{\bar\omega}^1{\bar\omega}^3\}$,
respectively. The hidden sector gauge group is
the $E_8$ sub--group of 
$(SO(4)\sim SU(2)\times SU(2))\times SU(3)\times U(1)^4$, 
generated by ${\bar\phi}^{1,\cdots,8}$.

The weak hypercharge is given by 
\beq
U(1)_Y={1\over3}U(1)_C\pm{1\over2}U(1)_L,
\label{weakhyper}
\eeq
which has the standard effective level $k_1$ of $5/3$,
necessary for MSSM unification at $M_U$.
As we noted in \cite{cfn1}, the sign ambiguity in eq.\  (\ref{weakhyper})
can be understood in terms of the two alternative embeddings of $SU(5)$ 
within $SO(10)$, 
that produce either the standard or flipped $SU(5)$ \cite{flipeq}.   
Switching signs in (\ref{weakhyper}) flips the representations,
\beqn
     + &\leftrightarrow& -\\
 e_L^c &\leftrightarrow& N_L^c\nonumber\\
 u_L^c &\leftrightarrow& d_L^c\nonumber\\
     h &\leftrightarrow& {\bar h}\,\, .
\label{repflip}
\eeqn
In the case of $SU(5)$ string GUT models,
only the ``--'' (i.e., flipped version) is allowed, 
since there are no massless matter adjoint representations, which 
are needed to break the non--Abelian gauge symmetry of the
unflipped $SU(5)$, but are not needed for the flipped version.
For MSSM--like strings, either choice of sign is allowed
since the GUT non--Abelian symmetry is broken directly at the string level. 

The ``+'' sign was chosen for the hypercharge definition in \cite{fny}.
In \cite{cfn1}, we showed  that the choice of the sign
in eq.\  (\ref{weakhyper}) has interesting consequences
in terms of the decoupling of the exotic fractionally charged states.
The other combination of $U(1)_C$ and $U(1)_L$, which is
orthogonal to $U(1)_Y$, is given by
\beq
U(1)_{Z^\prime}=U(1)_C\mp U(1)_L\, .
\label{u1prime}
\eeq
In Section 3 we will show that
cancellation of the FI term by singlet fields along directions that are 
$D$--flat for all of the non--anomalous $U(1)$ requires that at least one
of the $U(1)_Y$ and $U(1)_{Z'}$ be broken. Proof of this is found in 
Table C.II in Appendix C.  
Therefore, in the phenomenologically viable case
we are forced to have only $SU(3)_C\times SU(2)_L\times U(1)_Y$
as the unbroken $SO(10)$ subgroup below the string scale, when only
singlets take on VEVs. Under the constraint that only singlets take
on VEVs, this is an interesting example of how string dynamics may
force the $SO(10)$ subgroup below the string scale
to coincide with the Standard Model gauge group.
(Complete proof that only one of $U(1)_Y$ and $U(1)_{Z'}$ can survive
would necessitate that non--Abelian state VEVs also be considered,
which is currently being investigated \cite{cfn4}.)

\subsection{Matter Spectrum and Superpotential}

The full massless spectrum of the model, together
with the quantum numbers under the right--moving
gauge group, were first presented in ref.\  \cite{fny}.
In our Tables A.I and A.II of Appendix A, we again
list these states and their gauge and global charges.
The gauge charges in Table A.I are expressed in the rotated $U(1)$ basis 
of eqs.\ (\ref{anomau1infny},\ref{nonau1}) 
(and of eqs.\ (18a-f) in \cite{fny}), rather than in the unrotated basis 
associated with the tables of \cite{fny}.  

In the FNY model, the three 
sectors $\bb_1$, $\bb_2$, and $\bb_3$ correspond to the three
twisted sectors of the $Z_2\times Z_2$ orbifold model, with
each sector producing one generation in the 16 representation,
($\Q{i}$, $\uc{i}$, $\dc{i}$, $\L{i}$, $\ec{i}$, $\Nc{i}$),
of $SO(10)$ decomposed under
$SU(3)_C\times SU(2)_L\times U(1)_C\times U(1)_L$,
with charges under the horizontal symmetries.
In addition to the gravity and gauge multiplets and several singlets (see below),  
the untwisted Neveu--Schwarz (NS) sector produces three pairs of 
electroweak 
scalar doublets $\{h_1, h_2, h_3, {\bar h}_1, {\bar h}_2, {\bar h}_3\}$.
Each NS electroweak doublet set $(h_i,\bar{h}_i)$ may be 
viewed as a pair of Higgs with the potential to give renormalizable 
(near EW scale) mass to the corresponding $\bb_i$--generation of MSSM matter. 
Thus, to reproduce the MSSM states and generate a viable three 
generation mass 
hierarchy,
two out of three of these Higgs pairs must become massive near the string/FI 
scale.
The twisted sector provides some additional $SU(3)_C\times SU(2)_L$
exotics:
one $SU(3)_C$ triplet/anti--triplet pair $\{\H{33},\,\, \H{40}\}$;  
one $SU(2)_L$ up--like doublet, $\H{34}$, and one down--like doublet, $\H{41}$;
and two pairs of vector--like $SU(2)_L$ doublets, 
$\{\V{45},\,\, \V{46}\}$ and $\{\V{51},\,\, \V{52}\}$,  
with fractional electric charges $Q_e= \pm\half$.

The FNY model contains a total of 63 non-Abelian singlets.
Three of these are the MSSM electron conjugate states 
($\ec{1}$, $\ec{2}$, $\ec{3}$) and 
another three are 
the neutrino $SU(2)_L$ singlets ($\Nc{1}$, $\Nc{2}$, $\Nc{3}$).
Of the remaining 57 singlets, 16 possess electric charge and
41 do not. The set of 16 are twisted sector states,\footnote{Vector--like 
representations of the hidden sector are denoted by a ``V'', 
while chiral representations are denoted by a ``H''. 
A superscript ``s'' indicates a non--Abelian singlet.}
eight of which carry $Q_e= \half$,
\beq
\{\Hs{3}, \Hs{5}, \Hs{7}, \Hs{9}, 
 \Vs{41}, \Vs{43}, \Vs{47}, \Vs{49}\},
\label{set8a}
\eeq
and another eight of which carry $Q_e= -\half$,
\beq
\{\Hs{4}, \Hs{6}, \Hs{8}, \Hs{10}, 
  \Vs{42}, \Vs{44}, \Vs{48}, \Vs{50}).
\label{set8b}
\eeq

Three of the 41 $Q_e= 0$ states, 
\beq
\{\Phi_1,\Phi_2,\Phi_3 \},
\label{set3a}
\eeq
are NS sector singlets of the entire four dimensional gauge group. 
Another fourteen of these singlets,
\beq
\{\p{12},\,\, \pb{12},\,\, \p{23},\,\, \pb{23},\,\, \p{13}\,\, ,\pb{13}\,\, ,
  \p{56},\,\, \pb{56},\,\, \pp{56},\,\, \ppb{56},\,\, 
   \p{4},\,\, \pb{4},\,\,  \pp{4},\,\, \ppb{4} \},
\label{set1a}
\eeq
form seven pairs, 
\beq
(\p{12},\pb{12}),\,\, (\p{23},\pb{23}),\,\, (\p{13},\pb{13})\,\, ,
(\p{56},\pb{56}),\,\, (\pp{56},\ppb{56}),\,\, 
(\p{4} ,\pb{4} ),\,\, (\pp{4},\ppb{4}),
\label{set1a2}
\eeq
that are vector--like, 
i.e., possessing charges of equal magnitude but opposite sign, 
for all local Abelian symmetries. 
(Note for later discussion that $\p{4}$ and $\pp{4}$ carry identical Abelian gauge charges,   
thereby resulting in six, rather than seven, 
{\it distinct} vector--like pairs.)  
The remaining 24 $Q_e= 0$ singlets,
\beqn
&&\{\Hs{15},\,\, \Hs{16},\,\, \Hs{17},\,\, \Hs{18},\,\, \Hs{19},\,\, 
     \Hs{20},\,\, \Hs{21},\,\, \Hs{22},\,\, \Hs{29},\,\,
     \Hs{30},\,\, \Hs{31},\,\, \Hs{32},\,\, \Hs{36},\,\,
     \Hs{37},\,\, 
\nolabel\\
&&\phantom{\{}
   \Hs{38},\,\, \Hs{39},\,\,  
   \Vs{1},\,\,  \Vs{2},\,\,  \Vs{11},\,\, \Vs{12},\,\, 
   \Vs{21},\,\, \Vs{22},\,\, \Vs{31},\,\, \Vs{32} \},
\label{set1b}
\eeqn
 are twisted sector states carrying
both observable and hidden sector Abelian charges.

The FNY model contains 34 hidden sector non--Abelian states, 
all of which also carry both observable and hidden $U(1)_i$ charges: 
Five of these are $SU(3)_H$ triplets, 
\beq
\{\H{42},\,\, \V{4},\,\, \V{14},\,\, \V{24},\,\, \V{34} \},
\label{set3b}
\eeq
while another five are anti--triplets,
\beq
\{\H{35},\,\, \V{3},\,\, \V{13},\,\, \V{23},\,\, \V{33} \}.
\label{set3c}
\eeq
The remaining 24 states include  
12 $SU(2)_H$ doublets, 
\beq
\{\H{1},\,\, \H{2},\,\,
\H{23},\,\, \H{26},\,\, \V{5},\,\, \V{7},\,\, \V{15},\,\, \V{17},\,\,  
              \V{25},\,\, \V{27},\,\, \V{39},\,\, \V{40} \}
\label{set2ha}
\eeq
and a corresponding 12 $SU(2)^{'}_H$ doublets,
\beq
\{\H{11},\,\, \H{13},\,\,
\H{25},\,\, \H{28},\,\, \V{9},\,\, \V{10},\,\, \V{19},\,\, \V{20},\,\,  
              \V{29},\,\, \V{30},\,\, \V{35},\,\, \V{37} \}.
\label{set2hb}
\eeq
The only hidden sector NA states with non--zero $Q_e$ are the four 
doublets $\H{1}$, $\H{2}$, $\H{11}$, and $\H{13}$, which  
carry $Q_e= \pm\half$. 
The sector origins of all exotics is discussed in \cite{fny,fc,cfn1},
and a general classification of exotic states in the more realistic
free fermionic models is discussed in ref.\  \cite{ssr}.

For a string derived MSSM to result from the FNY model, 
the several exotic MSSM--charged states must be eliminated from the 
low energy effective
field theory. Along with  
two linearly independent combinations of the $\h{i=1,2,3}$, and of the
$\hb{i=1,2,3}$, the entire set of states,
\beqn
&&\{ \H{33},\,\,  \H{40},\,\,  \H{34},\,\,  \H{41},\,\, \V{45},\,\, 
     \V{46},\,\,  \V{51},\,\,  \V{52},\,\,                \label{minvevsa}\\
&&\phantom{\{} 
     \H{1},\,\,   \H{2},\,\,   \H{11},\,\,  \H{12},\,\,   \label{minvevsb}\\
&&\phantom{\{}
     \Hs{3},\,\,  \Hs{5},\,\,  \Hs{7},\,\,  \Hs{9},\,\,   
     \Vs{41},\,\, \Vs{43},\,\, \Vs{47},\,\, \Vs{49},\,\,  \label{minvevsc}\\
&&\phantom{\{}
     \Hs{4},\,\,  \Hs{6},\,\,  \Hs{8},\,\,  \Hs{10},\,\,  
     \Vs{42},\,\, \Vs{44},\,\, \Vs{48},\,\, \Vs{50} \}.   
\label{minvevsd}
\eeqn
must be removed.

Examination of the MSSM--charged state superpotential\footnote{In
Appendix B, we present the FNY superpotential up to sixth order. 
Terms in the superpotential belong to one of four classes,
those containing, in addition to (possible) singlets:
(i) nothing else, 
(ii) only MSSM--charged states,
(iii) both MSSM and hidden sector charged--states, and
(iv) only hidden sector charged--states.}
shows that two out of three of each of the $h_i$ and $\bar{h}_i$ Higgs, 
and {\it all} of the (\ref{minvevsa}--\ref{minvevsd}) states, 
can, indeed, 
be decoupled from the low energy effective field theory via the terms,
\beqn
&& \p{12} \h{1} \hb{2} + \p{23} \h{3} \hb{2} 
+ \Hs{31} \h{2} \H{34} + \Hs{38} \hb{3} \H{41}+\label{minveva}\\ 
&&  \p{4} [ \V{45} \V{46} + \H{1} \H{2} ] +    \label{minvevb}\\
&& \pb{4} [ \Hs{3} \Hs{4} + \Hs{5} \Hs{6} + \Vs{41} \Vs{42} + \Vs{43} \Vs{44} ]
                                           +   \label{minvevc}\\
&& \pp{4}  [ \V{51} \V{52} + \Hs{7} \Hs{8} + \Hs{9} \Hs{10} ]+\label{minvevd}\\
&& \ppb{4} [ \Vs{47} \Vs{48} + \Vs{49} \Vs{50} +\H{11} \H{13}]+
\label{minveve}\\
&& \p{23} \Hs{31} \Hs{38} [ \H{33} \H{40} + \H{34} \H{41} ]\,\, .\label{minvevf}
\eeqn
This will occur if all states in the set 
\beq
\{ \p{4},\,\,  \pb{4},\,\, \pp{4},\,\, \ppb{4},\,\,
    \p{12},\,\, \p{23},\,\, \Hs{31},\,\, \Hs{38} \}
\label{minvev}
\eeq
take on near string scale VEVs through FI term anomaly cancellation. 
All but two of the terms in (\ref{minveva}-\ref{minvevf}) are of third order 
and will result in unsuppressed FI scale masses, while the remaining two terms 
are of fifth order. The $\H{33} \H{40}$ fifth order term  
may result in $\H{33}$ and $\H{40}$ receiving a slightly suppressed 
mass. On the other hand, the $\H{34} \H{41}$ term has only a
minor, perturbative effect on $\H{34}$ and $\H{41}$ masses,
since $\H{34}$ and $\H{41}$ both appear in third order mass terms as well. 

In Section 3, we present $D$-- and $F$--flat directions
that contain the required fields (\ref{minvev})
for decoupling all of the SM--charged exotics.
Some of these directions are flat to all orders in the 
superpotential, while others are flat only to finite order. 
Before discussing these directions though, 
we review the process by which they were found.

\subsubsection{Anomalous $U(1)_A$}

All known 
chiral three generation  
$SU(3)_C\times SU(2)_L \times U(1)_Y$
models, of lattice, orbifold, or free fermionic construction,
contain an anomalous local $U(1)_A$ \cite{anoma}. 
An anomalous $U(1)_A$ 
has non--zero trace of its charge over the massless 
states of the low energy effective field theory, 
\beq
         \Tr Q^{(A)} \neq 0\,\, .
\label{audef}
\eeq
String models often appear to have not just one, but several anomalous
Abelian symmetries $U(1)_{A,i}$ ($i= 1$ to $n$), 
each with $\Tr Q^{(A)}_i \neq 0$. 
However, there is always a rotation
that places the entire anomaly into a single $U(1)_{A}$,
uniquely defined by 
\beq
         U(1)_{\rm A} \equiv c_A\sum_{i=1}^n \{\Tr Q^{(A)}_{i}\}U(1)_{A,i},
\label{rotau}
\eeq
with $c_A$ a normalization coefficient. 
There are then $n-1$ traceless $U(1)'_j$ formed from linear combinations of
the $n$ $U(1)_{A,i}$ and orthogonal to $\UA$.

Prior to rotating the anomaly into a single $\UA$, 
six of the FNY model's twelve $U(1)$ symmetries are anomalous:
Tr${\, U_1=-24}$, Tr${\, U_2=-30}$, Tr${\, U_3=18}$,
Tr${\, U_5=6}$, Tr${\, U_6=6}$ and  Tr${\, U_8=12}$.
Thus, the total anomaly can be rotated into a single 
$U(1)_{\rm A}$ defined by 
\beq
U_A\equiv -4U_1-5U_2+3U_3+U_5+U_6+2U_8.
\label{anomau1infny}
\eeq
The five orthogonal linear combinations,
\beqn
U^{'}_1 &=& \hbox to 3.0truecm{$2 U_1  - U_2 + U_3$\,\, ;\hfill}\quad 
U^{'}_2= -U_1 + 5 U_2 + 7 U_3\,\, ;\nolabel\\
U^{'}_3 &=& \hbox to 3.0truecm{$U_5 - U_6$\,\, ;\hfill}\quad 
U^{'}_4= U_5 + U_6 - U_8\,\, \label{nonau1}\\
U^{'}_5 &=& 12 U_1 + 15 U_2 - 9 U_3 + 25 U_5 + 50 U_8\,\, .
\nolabel
\eeqn
are all traceless.

\section{Flat Directions}

\subsection{Constraints on Flat Directions}

\subsubsection{$D$--constraints}

A set of vacuum expectations values (VEVs) will automatically appear
in any string model with an anomalous $\UA$ as a result of the 
string theory anomaly cancellation mechanism \cite{u1a} .
Following the anomaly rotation of \eq{rotau}, the universal Green--Schwarz 
(GS) relations,
\beqn
\frac{1}{k_m k_A^{1/2}}\mathop{\Tr}_{G_m}\, 
T(R)Q_A &=& \frac{1}{3k_A^{3/2}}\Tr Q_A^3
                      = \frac{1}{k_i k_A^{1/2}}\Tr Q_i^2 Q_A 
                      =\frac{1}{24k_A^{1/2}}\Tr Q_A 
\nolabel\\
&\equiv& 8\pi^2 \delta_{\rm GS} \, ,
\label{gsa}\\
\frac{1}{k_m k_i^{1/2}}\mathop{\Tr}_{G_m}\, 
T(R)Q_i &=& \frac{1}{3k_i^{3/2}}\Tr Q_i^3
                     = \frac{1}{k_A k_i^{1/2}}\Tr Q_A^2 Q_i 
                     = \frac{1}{(k_i k_j k_A)^{1/2}}\Tr Q_i Q_{j\ne i} Q_A 
\nolabel\\ 
        &=&\frac{1}{24k_i^{1/2}}\Tr Q_i = 0 \, ,
\label{gsna}
\eeqn
where $k_m$ is the level of the non--Abelian gauge group $G_m$ and
$2 T(R)$ is the index of the representation $R$ of $G_m$, defined by
\beq
\Tr\, T^{(R)}_a T^{(R)}_b = T(R) \delta_{ab}\, ,
\label{tin}
\eeq
removes all Abelian triangle anomalies except those involving
either one or three $U_A$ gauge bosons. 
(The GS relations are a by--product of modular invariance constraints.)

The standard anomaly cancellation mechanism breaks $U_A$ and, 
in the process, generates a FI $D$--term, 
\beq
      \eps\equiv \frac{g^2_s M_P^2}{192\pi^2}\Tr Q^{(A)}\, ,
\label{fidt}
\eeq
where $g_{s}$ is the string coupling
and $M_P$ is the reduced Planck mass, 
$M_P\equiv M_{Planck}/\sqrt{8 \pi}\approx 2.4\times 10^{18}$ . 

The form of the FI--term was determined from string theory assumptions.
Therefore, a more encompassing $M$--theory \cite{mth} might
suggest modifications to this FI--term. 
However, recently it was argued that $M$--theory
does not appear to alter the form of the FI--term \cite{jmr}. 
Instead an $M$--theory FI--term should remain identical to the FI--term 
obtained for a weakly--coupled
$E_8\times E_8$ heterotic string, independent of the size of 
$M$--theory's 11$^{\rm th}$ dimension.

Spacetime supersymmetry is broken 
near the string scale by the FI $D_A$--term, unless a set of scalar
VEVs, $\{\vev{\varphi_m}\}$, 
carrying anomalous charges $Q^{(A)}_m$ can contribute a compensating
$\vev{D_{A}(\varphi_m)} \equiv \sum_\alpha Q^{(A)}_m |\vev{\varphi_{m}}|^2$ 
term to cancel the FI--term, i.e.,
\beq
\vev{D_{A}}= \sum_m Q^{(A)}_m |\vev{\varphi_{m}}|^2 
+ \eps  = 0\,\, ,
\label{daf}
\eeq
thereby restoring supersymmetry. 
The actual set of VEVs accomplishing this cancellation
will be dynamically determined by non--perturbative effects. Whichever VEV 
combination this may be, the phenomenology of the model will be drastically
altered from that which exists before the VEV is applied. 

Any set of scalar VEVs satisfying eq.\ (\ref{daf})
must also retain $D$--flatness for all of the non--anomalous Abelian
$U_i$ symmetries as well,\footnote{Here we consider flat directions 
involving only non--Abelian singlet fields. In cases where non--trivial
representations of the non--Abelian gauge groups are also allowed to take on 
VEVs, generalized non--Abelian $D$--flat constraints must also be imposed.
See, for example, \cite{cfn4}.}
\beq
\vev{D_i}= \sum_m Q^{(i)}_m |\vev{\varphi_{m}}|^2 = 0\,\, .
\label{dana}
\eeq

\subsubsection{$F$--constraints}

Each superfield $\Phi_{m}$ (containing a scalar field $\varphi_{m}$
and chiral spin--$\half$ superpartner $\psi_m$) that appears
in the superpotential imposes further constraints on the scalar VEVs. 
$F$--flatness will be broken (thereby destroying spacetime supersymmetry) at 
the scale of the VEVs unless,
\beq
\vev{F_{m}} \equiv \vev{\frac{\partial W}{\partial \Phi_{m}}} = 0; \,\, \vev{W}  =0.
\label{ff}
\eeq

$F$--flatness of a given set of VEVs
can be broken by two types of superpotential terms: 
(i)  those composed only of the VEV'd fields, and
(ii) those composed, in addition to the VEV'd fields, of 
a single field without a VEV. (These superpotential terms 
were called ``type A'' and ``type B'' respectively in \cite{cceel2}.)
Any power of a type A term appearing in a superpotential
will break $F$. For example if generic fields $\varphi_1$, $\varphi_2$ and
$\varphi_3$ take on VEVs in a $D$--flat direction, then any superpotential
term of the form
\beq      
\left( (\Phi_1)^{n_1} (\Phi_2)^{n_2} (\Phi_3)^{n_3} \right)^m 
\label{atypea}
\eeq
(with $\{n_{i=1,2,3}\}$ being a set of relative primes)   
eliminates $F$--flatness at respective order $m(n_1+n_2+n_3)$. Thus, all order 
$F$--flatness, requires that no terms of this form appear in the superpotential
for any combination of $m$, $n_{i=1,2,3}$ values.
On the other hand, when $\Phi_x$ lacks a VEV, 
we see from (\ref{ff}) that superpotential terms
$((\Phi_1)^{n_1} (\Phi_2)^{n_2} (\Phi_3)^{n_3}(\Phi_x)^{n_x})^m$,
(with $\{n_{i=1,2,3},n_x\}$ similarly a set of relative primes) 
$F$--flatness breaking type B terms only when $n_x=m=1$.      

Generically, there are many more $D$--flat directions 
that are simultaneously $F$--flat  
to a given order in the superpotential   
for the effective field theory of a string model
than for the field--theoretic counterpart.
In particular, there are usually several $D$--flat directions 
that are $F$--flat to all order in a string model,  
but only flat to some finite, often low, order in the 
corresponding field--theoretic model. 
This may be attributed to 
string world--sheet selection rules \cite{nahew5,wsc} which impose 
strong constraints on allowed superpotential terms beyond
gauge invariance \cite{cceelw}.
For example, for a given set of states, 
a comparison between the allowed terms in a stringy superpotential
and the corresponding gauge invariant field--theoretic superpotential 
is performed in \cite{lr}.

\subsection{$D$--flat Basis Sets}

We continue our flat direction discussion by
considering basis sets of $D$--flat directions.
Let $\{ \varphi_{m=1\,\, {\rm to}\,\, n} \}$ 
denote the $n$--dimension set of fields of that 
are allowed to possibly take on VEVs.
Further, let
\beqn
C_j &=& \{ |\vev{\varphi_{1}}_j|^2,\, |\vev{\varphi_{2}}_j|^2,
            \cdots , |\vev{\varphi_{n}}_j|^2 \} 
\label{cdf1}\\
  &\equiv& \{ a_{j,1}, a_{j,2}, \cdots , a_{j,n} \}\,\, , 
\label{cdf12}
\eeqn
with 
\beq
a_{j,x}\equiv |\vev{\varphi_{x}}_j|^2 \ge 0\,\, ,
\label{psd}
\eeq 
be a generic set of the norms of VEVs \footnote{Hereon, we will often, 
for sake of brevity, refer to the norms of the VEVs simply as the VEVs
when it is clear that the norms are implied.} 
that satisfy all non--anomalous $D$--flat constraints \eq{dana}.
A set $C_j$ of such VEVs corresponds to a polynomial of fields 
$\varphi_{1}^{a_{j,1}} \varphi_{2}^{a_{j,2}}
\cdots  \varphi_{n}^{a_{j,n}}$, 
invariant under all non--anomalous gauge symmetries \cite{dfset,dfset2,cceel2}.
A non--anomalous $D$--flat direction (\ref{cdf1}) possesses some    
number of overall scale degrees of freedom (DOFs), which is  
the dimension of the direction. For example, 
the norms of the VEVs of a one--dimensional $D$--flat direction can be 
expressed as a product of a single positive real overall scale factor $\alpha$ 
and positive semi--definite integral coefficients, $c^{\alpha}_{j,x} \ge 0$, 
which specify the ratios between the VEVs of the various $n$ fields, 
\beq
C^{1{\rm -dim}}_j = 
\{ c^{\alpha}_{j,1} \alpha,\, c^{\alpha}_{j,2}\alpha,\cdots , 
   c^{\alpha}_{j,n} \alpha \}\,\, . 
\label{cdf3}
\eeq
Similarly, a two-dimensional flat direction involves
independent scales $\alpha$ and $\beta$, i.e.,
\beq
C^{2{\rm -dim}}_j = \{ c^{\alpha}_{j,1} \alpha + c^{\beta}_{j,1} \beta ,\, 
c^{\alpha}_{j,2} \alpha + c^{\beta}_{j,2} \beta, \cdots , 
c^{\alpha}_{j,n} \ + c^{\beta}_{j,n} \beta \}\,\, . 
\label{cdf4}
\eeq

Any physical $D$--flat direction $C_j$ 
can be expressed
as a linear combination of a set of $D$--flat basis directions,
$\{B_1, B_2, B_3, \cdots , B_k \}$:
\beq
C_j = \sum_k w_{j,k} B_{k}\,\, ,
\label{cfb1}
\eeq
where $w_{j,k}$ are real weights.
A basis can always be formed in which each element 
\beqn
B_k& = &\{ |\vev{\varphi_{1}}_k|^2,\, |\vev{\varphi_{2}}_k|^2,
           \cdots , |\vev{\varphi_{n}}_k|^2 \}  
\label{bs2b}\\
  &\equiv& \{ b_{k,1}, b_{k,2}, \cdots , b_{k,n} \}\,\, , 
\label{bs2b2}
\eeqn
where $b_{k,x}\equiv |\vev{\varphi_{x}}_k|^2$ is an integer, 
is one--dimensional, i.e.,
\beq
B^{1{\rm -dim}}_k = \{ b^{'}_{k,1}  \gamma_k, b^{'}_{k,2} \gamma_k, \cdots , 
                 b^{'}_{k,n}  \gamma_k \}\,\, ,
\label{bs3b1}
\eeq
where $\gamma_k$ is the overall scale factor  
and $b^{'}_{k,x}$ are the relative ratios of the VEVs. 
Further, the scale factor of a basis element can always be normalized to 1, 
thereby leaving $B_k$ to be defined solely by the  
$b^{'}_{k,x}$,
\beq
B^{1{\rm -dim}}_k = \{ b^{'}_{k,1}, b^{'}_{k,2}, \cdots , b^{'}_{k,n} \}\,\, .
\label{bs3bd2}
\eeq
Neither the coefficients $b^{'}_{k,x}$ \cite{gcm} of a basis element $B_k$,
nor the weights $w_{j,k}$ \cite{cceel2} need all be non--negative, 
so long as the {\it total} contribution 
of all basis elements to an individual norm of a VEV,
$a_{j,x} \equiv  \sum_k w_{j,k} b^{'}_{k,x}$,
in a flat direction $C_j$ is non--negative \cite{cceel2}.
However, a basis vector $B_k$ that contains at least one negative coefficient 
$b^{'}_{k,x}< 0$ cannot be viewed as a physical one--dimensional $D$--flat 
direction. Instead it corresponds to a monomial of fields containing 
at least one field with a negative power, 
$ \varphi_{1}^{b^{'}_{k,1}} \varphi_{2}^{b^{'}_{k,2}}
\cdots \varphi_{n}^{b^{'}_{k,n}}/\varphi_{x}^{|b^{'}_{k,x}|}$ 

Two types of basis sets were primarily discussed in \cite{cceel2}:
(i) a basis composed of a maximal set of linearly independent 
{\it physical} (i.e., all $b^{'}_{k,x}\ge 0$) 
one--dimensional $D$--flat directions\footnote{See Table IV of \cite{cceel2}
for an FNY non-anomalous flat direction basis of this type.}
and 
(ii) a ``superbasis'' composed of the set of 
all (and therefore not all linearly independent) 
one--dimensional flat directions.
The dimension $d_B$ of a linearly independent physical basis set is less than
or equal to $N_{VEV} - N_{d}$, where $N_{VEV}=n$ is the number of fields 
allowed VEVs and $N_{d}$ is the number of independent 
non--anomalous $D$--constraints from the set of $N_U$ non--anomalous 
$U(1)_i$.\footnote{For a proper subset of states in a model, all 
$U(1)_i$ charges may not be independent.}
A necessity for saturation of the upper bound for $d_B$, is that 
there is at least one pair of states with charges of opposite sign
for each non--anomalous $U(1)_i$.
When there is some $U(1)_s$ under which  
$N_s$ states all have charges of the same sign and the remaining 
$N_{VEV} - N_s$ states are uncharged,
none of the $N_s$ states may take on a VEV and the number of relevant
independent $U(1)_i$ is reduced by one,
$d_B \le N^{'}_{VEV} - N^{'}_{d} = (N_{VEV}- N_s) - (N_{d} - 1) 
\le N_{VEV} - N_{d}$. 
In general, determining the effective $N^{'}_{VEV}$ and $N^{'}_{d}$ 
is an iterative process:
After the $N_s$ states are removed and $N_{d}$ decreased by one for 
$U(1)_s$, then we must check again that there are 
 no other $U(1)_i$ without a pair of states of oppositely sign charges,
etc $\cdots$. Let $N^{'}_{VEV}$ and $N^{'}_d$ be the number of states 
allowed VEVs and number of independent $D$--constraints 
after ``eliminating'' all $U(1)_i$ that lack pairs of states with charges 
of opposite sign.
As we discuss in greater detail in the next subsection,
$d_B$ will generally be less than $N^{'}_{VEV} - N^{'}_{d}$ 
because of the very non--trivial positive semi--definite requirement 
(\ref{psd}) on the $\mvev{\phi_k}$ in (\ref{cdf1},\ref{cdf12}).

There is one possible disadvantage to choosing a basis set of 
linearly independent physical VEVs to generate higher dimensional physical
flat directions $C^{n{\rm -dim}}_j$.    
That is, there are usually physical flat directions $C_j$ 
that require some negative $w_{j,k}$ coefficients in (\ref{cfb1}) 
when $C_j$ is expressed in terms of physical basis directions.
This can complicate the systematic generation of $D$--flat directions.
On the other hand, using a superbasis of not all linearly independent
physical VEVs overcomes this possible difficulty, 
since all higher dimensional physical $D$--flat directions, 
$C^{n\ge 2{\rm -dim}}_i$, 
can be constructed from superbasis elements $B_k$ 
using only non--negative coefficients $|w_{j,k}|$ \cite{cceel2},
\beq
C_j= \sum_k |w_{j,k}| B_k\,\, .
\label{gdf}
\eeq
In practice, however, application of the superbasis approach can, at times,
have its own complication: 
the number of one--dimensional flat directions composing a given superbasis 
can be extremely large (on the order of several hundred or more) 
in some string models such as the FNY model.
Extremely large dimensions of a superbasis
can make systematic generation of $D$--flat directions unwieldy.

An alternative to these two types of $D$--flat basis sets 
is what we term the ``maximally orthogonal'' basis set.
This type of basis may, in fact, coincide with a linearly independent
physical basis in some models, but more commonly is slightly larger in
dimension.  
In a maximally orthogonal basis, each $D$--flat basis element  
$B^{mo}_k$ has one non--zero {\it positive} coefficient 
$b^{mo'}_{k,x}$ for which the corresponding $b^{mo'}_{k',x}$ are zero 
for all other basis elements $B^{mo}_{k'\neq k}$. 
Each $B^{mo}_k$ becomes associated with a particular field $\varphi_{x}$.
Hence, all $w_{j,k}$ defining a physical non--anomalous 
$D$--flat direction $C_j$ must be non--negative since
\beq
a_{j,x}\equiv |\vev{\varphi_{x}}_j|^2 = w_{j,k} b^{mo'}_{k,x} \ge 0\,\, .
\label{mopf}
\eeq
Otherwise a flat direction involving a negative $w_{j,k}$
weight would have at least one VEV with a negative norm and would, therefore,
be unphysical.

Associating one component $b^{mo'}_{k,x}$  
of a maximally orthogonal basis element $B^{mo}_k$ with the 
field $\varphi_{x}$ often results in
a few of the other components $b^{mo'}_{k,x'\ne x}$ being negative.
Hence, several $B^{mo}_k$ may correspond to unphysical directions. 
This does not present any real difficulty though. Rather, 
production of only physical directions $C_j$ then
simply places some constraints on linear combinations of the $B_k$.
A ``maximally orthogonal'' basis set of flat directions has 
essentially the same advantage as a superbasis, 
yet can keep the dimension of the basis set reasonable 
when the dimension of the superbasis may be unfeasibly large.  

There is one case in which the positivity constraint (\ref{psd}) discussed 
above is effectively relaxed. The exception to (\ref{psd})
occurs when two states can be combined into vector--like pairs.
Let us denote a generic pair as $\varphi_{m}$ and $\varphi_{-m}$,
For this pair, $a_{j,m}$ in (\ref{cdf12}) 
may take on negative values in a physical flat direction, because 
a vector--like pair of states acts effectively like a single state
with regard to all $D$--constraints.
For a flat direction, $C_j$, the contribution of this pair to each $D$--term is
\beqn
\vev{D_i}(\varphi_{m},\varphi_{-m}) &=&
Q^{(i)}_{m_{n}} \mvev{\varphi_{m}}_j + Q^{(i)}_{-m} \mvev{\phi_{-m}}_j 
\label{dppb1}\\
&\equiv&   
Q^{(i)}_{m} a_{j,m} + Q^{(i)}_{-m} a_{j,-m}\,\, .    
\label{dppb2}
\eeqn
Since by definition $Q^{(i)}_{-m} = - Q^{(i)}_{m}$, eq.\ (\ref{dppb2})
can be rewritten as
\beqn
\vev{D_i}(\varphi_{m},\varphi_{-m}) 
 &=& Q^{(i)}_{m}  (a_{j,m} - a_{j,-m}) 
\label{vdppb4}\\
 &\equiv& Q^{(i)}_{m}  a^{'}_{j,m}\,\, ,
\label{vdppb5}
\eeqn
where $a^{'}_{j,m}\equiv (a_{j,m} - a_{j,-m})$ 
may be positive, negative, or zero.

We can consider $a^{'}_{j,m}$ 
as originating from a single field 
$\vev{\varphi^{'}_{m}}$. 
This allows us to 
reduce the effective number of nontrivial states, $N_{VEV}$,
by one for each vector pair \cite{slm,gcm}. 
This reduction is compensated by an additional {\it trivial} 
$D$--flat direction
basis element, $\vev{\varphi_{m}} = \vev{\varphi_{-m}}$, 
(corresponding to the binomial $\varphi_{m} \varphi_{-m}$) 
formed from both components of a vector pair.
The FNY model contains several vector--like pairs and 
provides an excellent example of reduction of effective nontrivial states. 

Having all states in vector--like pairs
is equivalent to totally relaxing the ``positivity'' constraint.
As a general rule, the more vector pairs of non--Abelian
singlets there are, the more likely a $D$--flat FI--term cancelling
direction can be formed. 

\subsubsection{Maximally Orthogonal Basis Sets Via Singular Value 
Decomposition}

One method for generating a maximally orthogonal basis set of 
$D$--flat directions for the non--anomalous $U(1)_i$ involves
singular value decomposition (SVD) of a matrix \cite{pftv}.
While the matrix \cite{gcm} method 
and the more standard monomial approach \cite{dfset,dfset2,cceel2} 
are essentially different languages for the same process,
a strength of the matrix decomposition method is that it generates 
a complete basis of $D$--flat directions for non--Abelian singlet states
{\it en masse}. We briefly discuss the SVD approach 
here, since it provides a somwhat new interpretation to flat directions. 

SVD is based on the mathematical fact that
any $(M\times N)$--dimensional matrix $\bf D$ whose number of rows $M$ is
greater than or equal to its number of columns $N$, can be written as 
the product of an $M\times N$ column--orthogonal matrix $\bf U$, an $N\times N$
diagonal matrix $\bf W$ containing only semi--positive--definite elements, 
and the transpose of an $N\times N$ orthogonal matrix $\bf V$ \cite{pftv},
\beq
{\bf D}_{M\times N} = 
{\bf U}_{M\times N}\cdot  {\bf W}^{\rm diag}_{N\times N} \cdot
{\bf V}^T_{N\times N}, \quad {\rm for~} M\ge N\, .  
\label{decomp}
\eeq
This decomposition is always possible, no matter how singular 
the matrix is. The decomposition is also nearly unique, up to
(i) making the same permutation of the columns of $U$, 
diagonal elements of $W$, and columns of $V$, or (ii) forming linear
combinations of any columns of $U$ and $V$ whose corresponding elements
of $W$ are degenerate. If initially $M < N$, then a 
$(N-M)\times N$ zero--matrix
can always be appended onto $\bf D$ so this decomposition can be performed:
${\bf D}_{(M<N),N}\rightarrow {\bf D}^{'}_{(M=N),N}$.

SVD is extremely useful when the matrix $\bf D$\footnote{We 
assume from hereon that $\bf D$ has been enhanced by a zero submatrix 
if necessary so that $M\ge N$.}
is associated with a set of $M$ simultaneous linear equations expressed by,
\beq
{\bf D}\cdot \vec{x} = \vec{b},
\label{sle}
\eeq  
where $x$ and $b$ are vectors.
Eq.\ (\ref{sle}) defines a linear mapping from $N$--dimensional vector space $x$ to 
$M$--dimensional vector--space $b$.

For $M=N$, $\bf D$ is singular when
at least one of the $M$ constraints is not linearly independent.
Associated with a singular $\bf D$ is a subspace of $\vec x$ termed the 
{\it nullspace}, ${\cal M}_{null}$,
that is mapped to $\vec 0$ in $b$--space by $\bf D$. 
The dimension of this nullspace is referred to as the {\it nullity}. 
The subspace of $\vec b$ that {\it can} be reached by the matrix $\bf D$ acting
on $\vec x$ is called the {\it range} of $\bf D$. The dimension of the 
range is denoted as the {\it rank} of $\bf D$ and is equal to the
number of independent constraint equations $\equiv  M'\le M$.  
Clearly 
\beq
{\rm rank}\,{\bf D} + {\rm nullity}\,{\bf D} = N\,\, .
\label{svd1}
\eeq
 
In the decomposition of $\bf D$ in (\ref{decomp}), 
the columns of $\bf U$ corresponding to the non--zero diagonal components
of $\bf W$ form an orthonormal set of basis vectors that span the
range of $D$. Alternately, the columns of $\bf V$ corresponding 
to the zero diagonal components of $\bf W$ 
form an orthonormal basis for the nullspace.

As is perhaps obvious,
this method is directly applicable to constructing $D$--flat directions,
especially when only non--Abelian singlet states are allowed VEVs.
Let $M= N_U$ ($M_I= N_d$) denote the number of (independent) $D$--flat constraints and
$N= N_{VEV}$, the number of fields allowed to take on VEVs. Then
the $D_{i,j}$ component of the matrix $\bf D$ is the $Q^{(i)}_j$
charge of the state $\varphi_j$.
($i$ takes on the value $A$ for the anomalous $U(1)_A$ and values 
$\{a=1 {\rm ~to~} M-1\}$ for the set of non--anomalous $U(1)_i$.) 
The components of the vector $x$ are the values of $\mvev{\phi_j}$, and $b$ 
has all zero--components except in its $U(1)_A$ position. 
The value of $b$ in the anomalous position is $-\epsilon$ from
eq.\ (\ref{daf}). 

Let $\bf D'$ be the matrix that excludes the row of anomalous charges 
in $\bf D$. 
In this language, the dimension $d_B$ of the moduli space  
of flat directions (not necessarily all physical) 
for the $M'\equiv M-1 = N_U - 1 $ non--anomalous $U(1)_{i= 1 {\rm ~to~} M'}$ 
is the nullity of matrix ${\bf D}'$, denoted as dim ${\cal M}^{'}_{null}$,
formed from the $M'$ non--anomalous $D$--flat constraints. 
In other words, the nullity of ${\bf D}'$ is all 
VEVs formed from combinations of states that have zero net charge in
each non--anomalous direction. 
dim ${\cal M}^{'}_{null}$ is in the range 
\beq
N-M'\le {\rm dim}\,{{\cal M}^{'}_{null}}= N-M'_I\le N,  
\label{mrange}
\eeq
where $N$ is the number of states allowed VEVs and $M'_I$ is the number of
independent non--anomalous constraints.

For the matrix $\bf D$, which contains the anomalous charges,
the elements of the range corresponding to anomaly cancelling 
$\bf D$--flat directions are those formed solely from linear combinations of
elements of the nullity of ${\bf D}'$ that generate an anomalous
component for $\vec{b}$ of opposite sign to the FI term, $\epsilon$.
The nullspace of $\bf D$ will likewise 
be formed from linear combinations of ${\bf D}'$s nullity elements that 
generate a zero anomalous component for $\vec{b}$. 
The dimension of the ${\bf D}'$ nullity subset that projects into the range 
of $\bf D$, denoted by ${\rm dim}\,{\cal M}^{'}_R$, is 1 (since the anomalous
constraint must necessarily be independent of the non--anomalous constraints).
An $(N-M'_I-1)$--dimensional subset of 
${\cal M}^{'}_{null}$ forms the nullity of $\bf D$.

Our maximally orthogonal basis set for FNY was obtained using the SVD
routine in \cite{pftv}. After eliminating the row of the ${\bf D}_{(10+23)\times 33}^{fny}$ charge matrix
corresponding to the anomalous $U(1)_A$ $D$--constraint, singular value decomposition was performed
on the reduced matrix ${\bf D}_{9\times 33}^{fny'}$,
\beq
{\bf D}_{(9+24)\times 33}^{fny'} = {\bf U}_{(9+24)\times 33}^{fny'}
               \cdot {\bf W}_{33\times 33}^{fny'}\cdot {{\bf V}^{fny'}}^{T}_{33\times 33}\,\, .
\label{fnysvg}
\eeq
As discussed in the following subsections,
33 is the number of FNY nontrivial singlets allowed VEVs and 
 9 is the number of effective $D$--constraints. 
An initial basis set of $D$--flat directions
of dimension $d_b = 24 = {\rm dim}\,\, {\cal M}^{'}_{null} =  33 - 9$, 
was obtained. 
The components of these 24 basis directions are the components of the
24 columns of $\bf V$ for which the diagonal components of $\bf W$ is zero.  
While these 24 basis directions were not initially in maximally orthogonal form, a simple rotation
transformed them into this.

\subsection{FNY Flat Directions}

\subsubsection{$D$--flat Basis}

The FNY model contains a total of 63 non--Abelian singlets. 
These fields, along with their local $U(1)$ and global world--sheet charges,
are listed in Appendix A. 
Of the 63 singlets, 
14 can be used to form seven pairs of vector--like singlets:
$(\p{12},\pb{12})$, $(\p{13},\pb{13})$, $(\p{23},\pb{23})$,
$(\p{56},\pb{56})$, $(\pp{56},\ppb{56})$, $(\p{4},\pb{4})$,
$(\pp{4},\ppb{4})$.
The two $\p{4}$--related pairs possess identical gauge charges, 
so we will refer to the model as having six {\it distinct} vector--like pairs
of singlets and 49 non--vector--like singlets. 

Since we wish the SM gauge group to survive after cancellation of the
FI term, we investigate flat directions involving only 
singlets not carrying hypercharge.  
The set of hypercharge--free singlets is composed of 
the six distinct vector--like pairs and 27 non--vectors,  
$\Hs{15\,\, {\rm to }\,\, 22}$, 
$\Hs{29\,\, {\rm to }\,\, 32}$,
$\Hs{36\,\, {\rm to }\,\, 39}$,
$\Vs{ 1\,\, {\rm to }\,\,  2}$,
$\Vs{11\,\, {\rm to }\,\, 12}$,
$\Vs{21\,\, {\rm to }\,\, 22}$,
$\Vs{31\,\, {\rm to }\,\, 32}$, and
$\Nc{ 1\,\, {\rm to }\,\, 3}$.   
These 33 vector--like and non--vector--like 
singlets carry varying combinations of $\UA$ 
and nine other $U(1)_i$ charges.\footnote{While there are ten 
non--anomalous $U(1)_i$ besides hypercharge, 
all singlets charged under $U(1)_9$ of the hidden sector also carry 
non--zero hypercharge. 
Thus, none of the singlets we allow to take on a VEV carry $U(1)_9$ charge.}
Hence, from these singlets we can form 
$33 - 9 = 24$ non--trivial $D$--flat basis directions (some physical, others 
not physical). 

A non--trivial basis set may be generated by several methods, which can
give differing properties to the set. 
As discussed in the preceding subsection, 
we chose the singular value decomposition method
to generate a maximally orthogonal basis set,  
for which there is a one--to--one correspondence between the 
24 flat directions in the basis set and 24 of the 33 distinct states allowed VEVs. 
Eight more 
{\it trivial} basis elements are formed from vector--like pairs.
There are eight rather than six vector--like pairs because of the 
gauge charge redundancy between the two vector--like pairs,
$(\Phi_4,\bphi_4)$ and $(\Phi^{'}_4,\bphi^{'}_4)$. 

In Table C.I, the first six components of each basis
element correspond to the norms of the VEVs of the six distinct vector--like pairs. 
The norms of the VEVs of these fields may be either positive or negative
in a physical flat direction. A positive norm implies a field $\phi_m$
takes on the VEV, while a negative norm implies the vector partner field
$\phi_{-m}$ of opposite charge takes on the VEV.
The remaining 27 components of each basis element give the norms of the
the non--vector--like singlet fields. 

A true physical flat direction formed from a combination of basis elements
of $D$--flat directions must have positive semi--definite norms for all of its 
non--vector--like fields. However, as discussed earlier, corresponding components of 
the basis elements need not. A basis element with a negative norm of a 
non--vector field simply is not a physical flat direction. 
Our maximally orthogonal method associates as many basis
directions with non--vector fields as possible. In the FNY model, this
is possible to do for 23 of the 24 basis directions.
Thus, for the FNY model, a given basis direction could have up to
$33-23-6=4$ VEV non--vector--like negative component norms, 
any number of which would imply a flat direction  
is ``non--physical.'' The four non--vector--like fields 
with potentially negative norms are
$\Nc{1}$, $\Hs{39}$, $\Hs{16}$, and $\Nc{3}$.\footnote{The VEVs of these four 
fields appear as the last components of the basis directions in Table C.I of 
Appendix C.} 
Thus, 
the physical constraints,
\beq
\sum_{k= 1}^{24} w_k b^{mo'}_{k,{\Nc{ 1}}}\ge 0\,\,;\quad
\sum_{k= 1}^{24} w_k b^{mo'}_{k,{\Hs{39}}}\ge 0\,\,;\quad
\sum_{k= 1}^{24} w_k b^{mo'}_{k,{\Hs{16}}}\ge 0\,\,;\quad
\sum_{k= 1}^{24} w_k b^{mo'}_{k,{\Nc{ 3}}}\ge 0\,\, ,
\label{wkcon2}
\eeq
must be imposed 
upon the weight factors $w_k$ in (\ref{cfb1})
through which physical flat directions $C_j$ 
are formed from the basis elements $B^{mo}_k$.

An additional constraint on physical flat directions is, of course, that
the net anomalous charge $Q^{(A)}$ must be negative,
\beq
Q^{(A)}(C_j) =\sum_{k= 1}^{24} w_k Q^{(A)}(B_k) < 0 \,\, ,
\label{wkcon3}
\eeq
since $\eps > 0$ in eq.\ (\ref{daf}).
Of the 24 non--trivial basis directions, nine carry negative
anomalous charge, while eight carry positive anomalous charge and
seven do not carry this charge.

The SVD approach was also used to generate a set of basis directions
that simultaneously conserve both $U(1)_Y$ and $U(1)_{Z'}$. 
The elements of this five--dimensional set are presented in Table C.II. 
Interestingly, none of the these 
basis directions have negative anomalous charge:
all are, in fact, chargeless under $U(1)_A$! Thus, FI--term cancelling 
$D$--flat directions can never be formed by non--Abelian singlets
if both $U(1)_Y$ and $U(1)_{Z'}$ are to survive.  In other
words, singlet flat directions imply the reduction of observable
$SO(10)$ to exactly the SM, $SU(3)_C\times SU(2)_L \times U(1)_Y$. 

\subsubsection{$F$--flat Directions}

In Table D.I of Appendix D, we present several classes of directions in the 
parameter space of VEVs that are flat up to at least sixth order in the FNY 
superpotential. These classes are defined solely by the states
that take on VEVs, rather than the specific ratios of the VEVs. 
The first three classes of directions are flat to {\it all} order in the 
superpotential.
The fourth class is broken by two twelfth order type A terms, 
denoted as ``12--1'' and ``12--2'' in Table D.II,
while the fifth through eighth classes are all broken at seventh order by a single 
term, denoted as ``7--1'' in Table D.II.
Numerous (specifically 22) classes are broken at sixth order, again
all by a single term, ``6--1.''
Three of the sixth order classes
are simultaneously broken by an additional sixth order term,
``6--2.''  
Thus, it appears likely that FNY $D$--flat directions producing the MSSM spectrum 
are either $F$--flat 
to all finite orders or experience $F$--breaking at twelfth order or lower. 

Our MSSM flat directions were found via a computer search that generated
combinations of the maximally orthogonal $D$--flat basis directions $B^{mo}_k$
given in Table C.I of Appendix C. 
Linear combinations of up to nine $L_j$--class basis directions  
were surveyed, with the range of the non--zero integer weights, $w_k$, 
in (\ref{cfb1}) being from one to ten. 
To eliminate redundancy of flat directions, 
the set of non--zero $w_k$
in a given linear combination was required to be relatively prime, i.e., 
the greatest common factor among any permitted set of $w_k$ was 1. 
The combinations of basis directions producing our 30 classes of
flat directions is given in Table D.IV.
The examples in each of the 30 flat direction classes were formed from five, six, or 
seven $L_j$--class basis directions. 
We found that all $D$--flat directions involving eight or more 
maximally orthogonal basis directions  
experienced breaking of $F$--flatness at fifth order or lower.   

In our computer search, we required flat directions 
to minimally contain VEVs for the set of states,
\beq
\{ \p{4},\,\,  \pb{4},\,\, \pp{4},\,\, \ppb{4},\,\,
    \p{12},\,\, \p{23},\,\, \Hs{31},\,\, \Hs{38} \}\,\, ,
\nolabel
\eeq
necessary for decoupling of all 32 SM--charged MSSM exotics, comprised of the  
four extra Higgs doublets and the 28 exotics identified in 
(\ref{minvevsa}--\ref{minvevsd}). Imposing this, we found 
VEVs of $\Hs{15}$ and $\Hs{30}$ were also always present.
We refer to the set of VEVs of these ten fields in Appendix $D$ as ``$\{ VEV_1\}$.''
All of the eight directions broken at seventh order or higher additionally 
contained the VEV of $\pb{56}$. 
The three classes of directions $F$--flat to all finite order also 
involved (i) no other VEVs, (ii) the VEV of $\pbp{56}$, and
(iii) the VEV of $\Hs{19}$, respectively.  
The class broken at twelfth order additionally included the VEV of $\Hs{20}$.
Note that the trilinear superpotential term $\pbp{56}\Hs{19}\Hs{20}$ 
allows only one of $\pbp{56}$, $\Hs{19}$, and $\Hs{20}$
to receive a VEV in any flat direction.

In all directions with seventh order flatness, the sneutrino $SU(2)_L$ singlet $\Nc{1}$
takes on a VEV, as do one of $\pbp{56}$, $\Hs{19}$, or $\Hs{20}$ and/or
$\Vs{31}$. 
For the 22 sixth order classes,    
subsets of $\{\Nc{3}$, $\Hs{17}$, $\Hs{18}$, $\Hs{21}$, $\Hs{39}$, $\Vs{12}\}$
obtain VEVs,
along with various combinations from 
$\{\pbp{56}$, $\pp{56}$, $\Hs{19}$, $\Hs{20}$, $\Vs{31}$, $\Nc{1}\}$.

Systematic generation of flat directions of the FNY model
is efficiently performed using a maximally orthogonal basis. 
However, the dimension (i.e., the number of VEV scale degrees of freedom) 
$Dim$ of a given direction is not always apparent from this approach.
To determine $Dim$, we also  
express each flat direction class in terms of 
its embedded {\it physical} one--dimensional $D$--flat directions.
(See Tables D.IV and D.V.)
Prior to FI term cancellation, the dimension of a given MSSM flat 
direction equals the number of embedded 
physical dimension--one $D$--flat directions. 
Cancellation of the FI term removes one degree of freedom, 
so the dimension after FI term cancellation, $Dim_{FI}$,
is one less than $Dim$. 

The number $N_B$ of non--anomalous $U(1)_i$ broken along a given direction 
is the difference between 
the number of independent VEVs, $N_{VEV}$, and the dimension $Dim$, 
\beq
N_{VEV} - Dim = N_{B}.
\label{nbr1}
\eeq
Or, equivalently,
\beq
N_{VEV} - Dim_{FI} = N_{B} + 1.
\label{nbr2}
\eeq
As Table D.VI shows, the all--order  
and twelfth order $F$--flat directions  
break seven non--anomalous $U(1)_i$, 
while the seventh order flat directions break eight. 
The sixth order directions remove anywhere from eight to ten non--anomalous
$U(1)_i$. 

\setcounter{footnote}{0}

A pair of physical one--dimensional flat directions, 
denoted ``$X$'' and ``$Y$'',  
are at the heart of 27 (out of 30) of our MSSM directions.  
$X$ is identified with the class 1 all--order flat direction\footnote{The 
class 1 flat direction was first formed in \cite{cceel2}
from the combination of VEVs denoted $M_6$, $M_7$ and 
$R_{10}$. Table V of \cite{cceel2} also contains 
several FNY $Dim_{FI}=0$ non--MSSM flat directions.}
and is the root of the three all--order, the one twelfth order,
the four seventh order, and five of the sixth order flat directions.
The seventh and higher order directions contain anywhere from zero to three 
additional physical dimension--one directions.
At the root of 14 sixth order directions is the direction $Y$.
$X$ and $Y$ are simultaneously embedded in five of the sixth order 
directions, while three of the sixth order directions contain neither.

Our MSSM flat direction classes have a property not common to 
generic stringy flat directions. Specifically, $F$--flatness of the MSSM directions 
is broken by the stringy superpotential at exactly the same level as it would be in a   
field--theoretic gauge invariant superpotential. That is, 
stringy world sheet constraints do not remove all of the 
lowest order dangerous gauge--invariant terms.
The equivalent string and gauge--invariant  
seventh and twelfth order $F$--breaking 
essentially results from all of the
$\p{4}$, $\pp{4}$, $\pb{4}$, and $\ppb{4}$ states necessarily taking on VEVs.

\section{Discussion}

We have investigated $D$-- and $F$--flat directions in the FNY model
of \cite{fny,fc}. The FNY model possesses two aspects generic to many 
classes of three family $SU(3)_C\times SU(2)_L\times U(1)_Y$ string 
models: both an extra local anomalous $U(1)_A$ and 
numerous (often fractionally charged) exotic particles beyond
the MSSM.   
We found several flat directions involving only non--Abelian singlet
fields that near the string scale can
simultaneously break the anomalous $U(1)_A$ and 
give mass to {\it all} exotic SM--charged 
observable particles,  
decoupling them from the low energy spectrum.
We were thus able to produce the first known examples of
Minimal Superstring Standard Models. 
Some of our flat directions were shown to be flat to {\it all}
finite orders in the superpotential.

The models produced by our flat directions are consistent with, and may 
in fact offer the first potential realizations of, 
the recent conjecture by Witten of possible equivalence 
between the string scale and the minimal supersymmetric standard model
unification scale $M_U\approx 2.5\times 10^{16}$ GeV. 
This conjecture indeed suggests that the observable gauge group 
just below the string scale should be 
$SU(3)_C\times SU(2)_L\times U(1)_Y$ and that the 
$SU(3)_C\times SU(2)_L\times U(1)_Y$--charged spectrum of the 
observable sector should consist solely of the MSSM spectrum.

We have also discovered that
the FNY model provides an interesting example of how string dynamics
may force the $SO(10)$ subgroup below the string scale to coincide with the
SM gauge group. When only non--Abelian singlets take on VEVs, we have shown that
$U(1)_Y$ or $U(1)_{Z'}$ of
$SU(3)_C\times SU(2)_L\times U(1)_Y\times U(1)_{Z'}\in SO(10)$ is necessarily
broken. Reversing the roles of $U(1)_Y$ and $U(1)_{Z'}$ 
corresponds to flipping the components in each $SU(2)_L$ doublet. 

The phenomenology obtained from our various flat directions
will be studied in \cite{cfn3}.
In particular, we will examine the mass hierarchies of the three
generations of SM quarks and leptons, the hidden sector dynamics,
and issues such as proton decay. For each flat direction, we will
present the resulting superpotential after 
decoupling of the states turned massive via FI--term cancelling VEVs.
The rich space of flat directions that we found in the present paper
suggests the exciting prospect that one of these flat directions
may accommodate all of the phenomenological constraints imposed
by the Supersymmetric Standard Model phenomenology. Furthermore,
the rich space of solutions may be even further enlarged
by adding VEVs of the non--Abelian fields. It ought to
be emphasized that it is the promising structure,
afforded by the NAHE set, which enables this promising
scenario. To highlight this important fact, NAHE--based models
should be contrasted with the non--NAHE based models,
which although having three generations with the Standard Model
gauge group, do not allow the standard $SO(10)$ embedding of the
Standard Model spectrum and contain massless exotic
states that cannot be decoupled. Thus, we emphasize once again,
that although suggesting a specific three generation model
as the true string vacuum, seems still premature,
the concrete results, obtained in the analysis of specific
models, highlight the underlying, phenomenologically successful,
structure generated by the NAHE set. Therefore, it suggests
that the true string vacuum could be in the vicinity of these
models. That is, it is a $Z_2\times Z_2$ model in the vicinity
of the free fermionic point in the Narain moduli space,
also containing several, perhaps still unknown, additional Wilson lines.
Such Wilson lines correspond in the fermionic language to the
boundary condition basis vectors beyond the NAHE set.

\section{Acknowledgments}
This work is supported in part
by DOE Grants No. DE--FG--0294ER40823 (AF)
and DE--FG--0395ER40917 (GC,DVN). GC wishes to thank 
Alexander Maslikov for helpful discussions.
\newpage

\appendix
\section{String Quantum Number of All FNY Massless Fields}

{\def\half{\frac{1}{2}}
\def\mhalf{-\frac{1}{2}}
\def\hfw{$\frac{1}{2}$}
\def\malf{-\frac{1}{2}}
\def\mfw{$-\frac{1}{2}$}
\def\sixth{\frac{1}{6}}
\def\third{\frac{1}{3}}
\def\mthird{-\frac{1}{3}}
\def\mbd{$\frac{2,-1}{3}$}
\def\mtd{$-\frac{1}{3}$}
\def\td{$\frac{1}{3}$}
\def\ttd{$\frac{2}{3}$}
\def\mttd{$-\frac{2}{3}$}
\def\mtwothird{\frac{2}{3}}
\def\mtwothird{-\frac{2}{3}}
\def\pmh{$\pm\half$}
\def\sutc{$SU(3)_C$}
\def\sutl{$SU(2)_L$}
\def\suth{$SU(3)_H$}
\def\sutl{$SU(2)_H$}
\def\sutn{$SU(2)^{'}_H$}
\def\UP#1{$U^{'}_{#1}$}
\def\U#1{$U_{#1}$}
\def\UC{$U_C$}
\def\UL{$U_L$}
\def\Ua{$U_A$}

\def\tb{$\bar{3}$}
\def\tbn{\bar{3}}

\def\T#1{$T_{#1}$}
\def\S#1{$S_{#1}$}
\def\H#1{$H_{#1}$}
\def\UR#1{$U_{#1}$}
\def\R#1{$R_{#1}$}
\def\b#1{$b_{#1}$}
\def\hv#1{$h_{#1}$}

\def\p#1{$\Phi_{#1}$}
\def\pb#1{$\bar{\Phi}_{#1}$}
\def\pp#1{$\Phi^{'}_{#1}$}
\def\pbp#1{$\bar{\Phi}^{'}_{#1}$}
\def\h#1{$h_{#1}$}
\def\hb#1{$\bar{h}_{#1}$}
\def\L#1{$L_{#1}$}
\def\ec#1{$e^c_{#1}$}
\def\Nc#1{$N^c_{#1}$}
\def\Q#1{$Q_{#1}$}
\def\dc#1{$d^c_{#1}$}
\def\uc#1{$u^c_{#1}$}
\def\Hs#1{$H^s_{#1}$}
\def\V#1{$V_{#1}$}
\def\Vs#1{$V^s_{#1}$}

\def\K#1{$K_{#1}$}

\begin{flushleft}
\begin{tabular}{|l||r|l|rrrrrrrrr|lccc|}
\hline 
\hline
State         &\U{E}&$(C,L)_Y$ &\Ua &\UC&\UL  &\UP{1}&\UP{2}&\UP{3}&\UP{4}&\UP{5}&\U{4}&$(3,2,2^{'})_H$&\U{7}&\U{H}&\U{9}\\
\hline
\hline
\Q{1}&  \mbd&  $(3,2)_{\sixth}$&       8&     2&     0&    -4&     2&     0&     0&   -24&     2&     (1,1,1)&   0&     0&     0\\ 
\Q{2}&  \mbd&  $(3,2)_{\sixth}$&      12&     2&     0&     2&   -10&     2&     2&    20&     0&     (1,1,1)&   0&     0&     0\\ 
\Q{3}&  \mbd&  $(3,2)_{\sixth}$&       8&     2&     0&     2&    14&    -2&     2&    32&     0&     (1,1,1)&   0&     0&     0\\ 
\hline
\dc{1}& \td&  $(\tbn,1)_{\third}$&    8&    -2&     4&    -4&     2&     0&     0&   -24&    -2&     (1,1,1)&   0&     0&     0\\ 
\dc{2}& \td&  $(\tbn,1)_{\third}$&    8&    -2&     4&     2&   -10&    -2&    -2&   -80&     0&     (1,1,1)&   0&     0&     0\\ 
\dc{3}& \td&  $(\tbn,1)_{\third}$&    4&    -2&     4&     2&    14&     2&    -2&   -68&     0&     (1,1,1)&   0&     0&     0\\

\hline
\uc{1}&\mttd&$(\tbn,1)_{\mtwothird}$&  8&    -2&    -4&    -4&     2&     0&     0&   -24&    -2&     (1,1,1)&   0&     0&     0\\ 
\uc{2}&\mttd&$(\tbn,1)_{\mtwothird}$& 12&    -2&    -4&     2&   -10&     2&     2&    20&     0&     (1,1,1)&   0&     0&     0\\ 
\uc{3}&\mttd&$(\tbn,1)_{\mtwothird}$&  8&    -2&    -4&     2&    14&    -2&     2&    32&     0&     (1,1,1)&   0&     0&     0\\ 
\hline
\H{33}&\mtd&   $   (3,1)_{\mthird}$&   8&    -1&    -2&    -2&   -11&     2&    -4&    32&     0&     (1,1,1)&  -1&     3&     0\\ 
\H{40}&\td&    $(\tbn,1)_{\third} $&   0&     1&     2&     2&   -13&    -2&    -4&    56&     0&     (1,1,1)&   1&    -3&     0\\ 
\hline
\L{1}& 0,-1& $(1,2)_{\mhalf}$&       8&    -6&     0&    -4&     2&     0&     0&   -24&     2&     (1,1,1)&   0&     0&     0\\
\L{2}& 0,-1& $(1,2)_{\mhalf}$&       8&    -6&     0&     2&   -10&    -2&    -2&   -80&     0&     (1,1,1)&   0&     0&     0\\ 
\L{3}& 0,-1& $(1,2)_{\mhalf}$&       4&    -6&     0&     2&    14&     2&    -2&   -68&     0&     (1,1,1)&   0&     0&     0\\ 
\hline
\h{1}& 0,-1& $(1,2)_{\mhalf}$&      16&     0&    -4&    -8&     4&     0&     0&   -48&     0&     (1,1,1)&   0&     0&     0\\ 
\h{2}& 0,-1& $(1,2)_{\mhalf}$&     -20&     0&    -4&    -4&    20&     0&     0&    60&     0&     (1,1,1)&   0&     0&     0\\ 
\h{3}& 0,-1& $(1,2)_{\mhalf}$&     -12&     0&    -4&    -4&   -28&     0&     0&    36&     0&     (1,1,1)&   0&     0&     0\\ 
\hb{1}&1, 0& $(1,2)_{ \half}$&     -16&     0&     4&     8&    -4&     0&     0&    48&     0&     (1,1,1)&   0&     0&     0\\
\hb{2}&1, 0& $(1,2)_{ \half}$&      20&     0&     4&     4&   -20&     0&     0&   -60&     0&     (1,1,1)&   0&     0&     0\\
\hb{3}&1, 0& $(1,2)_{ \half}$&      12&     0&     4&     4&    28&     0&     0&   -36&     0&     (1,1,1)&   0&     0&     0\\
\hline
\H{34}&1,0& $(1,2)_{ \half}$&       8&     3&     2&    -2&   -11&     2&    -4&    32&     0&     (1,1,1)&  -1&     3&     0\\ 
\H{41}&0,-1&$(1,2)_{\mhalf}$&       0&    -3&    -2&     2&   -13&    -2&    -4&    56&     0&     (1,1,1)&   1&    -3&     0\\ 
\hline
\V{45}&\pmh &$(1,2)_{0}$&          12&     0&     0&     2&   -10&    -2&     2&    20&    -2&     (1,1,1)&   2&     0&    -2\\ 
\V{46}&\pmh &$(1,2)_{0}$&         -12&     0&     0&    -2&    10&     2&    -2&   -20&    -2&     (1,1,1)&  -2&     0&     2\\ 
\V{51}&\pmh &$(1,2)_{0}$&          -4&     0&     0&    -2&   -14&     2&     2&    68&    -2&     (1,1,1)&  -2&     0&     2\\
\V{52}&\pmh &$(1,2)_{0}$&           4&     0&     0&     2&    14&    -2&    -2&   -68&    -2&     (1,1,1)&   2&     0&    -2\\ 
\hline
$\tr Q_o$&&&                      392&   -36&     0&     0&   -36&     0&   -24&  -168&   -12&&               0&     0&     0\\ 
\hline
\hline
\end{tabular}
\end{flushleft}

\no Table A.I.a: Gauge Charges of FNY Observable Sector $SU(3)_C\times SU(2)_L\times U(1)_Y$ 
Non-Abelian (NA) States. (Charges of both electron conjugates $e^c$ 
and neutrino singlets $N^c$ appear in Table A.I.b with non-Abelian singlets.) 
The names of the states appear in the first column, with the states' 
various charges appearing in the other columns.
The entries under $(C,L)_Y$ denote Standard Model charges, while
the entries under $(3,2,2')$ denote hidden sector $SU(3)_H\times SU(2)_H\times SU(2)^{'}_H$ charges.
The entries in the last row give the traces of the $U(1)_i$ over these states.
(Note that all Table A.I $U_A$ through $U_9$ charges have been multiplied by 
a factor of 4 compared to those charges given in ref.\  \cite{fny}.)

\begin{flushleft}
\begin{tabular}{|l||r|l|rrrrrrrrr|lccc|}
\hline 
\hline
State       &\U{E}&$(C,L)_Y$      &  \Ua    &\UC   &\UL   &\UP{1}&\UP{2}&\UP{3}&\UP{4}&\UP{5}&\U{4}&$(3,2,2^{'})_{H}$&\U{7}&\U{H}&\U{9}\\
\hline
\hline
\ec{1}&   1& $(1,1)_{1}$&            8&     6&     4&    -4&     2&     0&     0&   -24&    -2&     (1,1,1)&   0&     0&     0\\ 
\ec{2}&   1& $(1,1)_{1}$&           12&     6&     4&     2&   -10&     2&     2&    20&     0&     (1,1,1)&   0&     0&     0\\ 
\ec{3}&   1& $(1,1)_{1}$&            8&     6&     4&     2&    14&    -2&     2&    32&     0&     (1,1,1)&   0&     0&     0\\ 
\hline
\Hs{3}&\hfw & $(1,1)_{\half}$&       -8&     3&     2&     2&    11&     2&     4&   -32&     2&    (1,1,1)&   -1&    -3&     2\\ 
\Hs{4}&\mfw & $(1,1)_{\mhalf}$&       8&    -3&    -2&    -2&   -11&    -2&    -4&    32&     2&    (1,1,1)&    1&     3&    -2\\ 
\Hs{5}&\hfw & $(1,1)_{\half}$&       -4&     3&     2&     2&    11&    -2&    -4&    68&     2&    (1,1,1)&   -1&    -3&    -2\\ 
\Hs{6}&\mfw & $(1,1)_{\mhalf}$&       4&    -3&    -2&    -2&   -11&     2&     4&   -68&     2&    (1,1,1)&    1&     3&     2\\ 
\Hs{7}&\hfw & $(1,1)_{\half}$&        4&     3&     2&     2&   -13&    -2&     0&   156&    -2&    (1,1,1)&   -1&    -3&     2\\ 
\Hs{8}&\mfw & $(1,1)_{\mhalf}$&      -4&    -3&    -2&    -2&    13&     2&     0&  -156&    -2&    (1,1,1)&    1&     3&    -2\\ 
\Hs{9}&\hfw & $(1,1)_{\half}$&       -8&     3&     2&     2&   -13&     2&     0&  -144&    -2&    (1,1,1)&   -1&    -3&    -2\\ 
\Hs{10}&\mfw &$(1,1)_{\mhalf}$&       8&    -3&    -2&    -2&    13&    -2&     0&   144&    -2&    (1,1,1)&    1&     3&     2\\ 
\Vs{41}&\hfw &$(1,1)_{\half}$&      -12&     0&     4&    -2&    10&     2&    -2&   -20&     2&    (1,1,1)&    2&     0&    -2\\ 
\Vs{42}&\mfw &$(1,1)_{\mhalf}$&      12&     0&    -4&     2&   -10&    -2&     2&    20&     2&    (1,1,1)&   -2&     0&     2\\ 
\Vs{43}&\hfw &$(1,1)_{\half}$&        8&     0&     4&     2&   -10&     2&    -2&   -80&     2&    (1,1,1)&   -2&     0&     2\\ 
\Vs{44}&\mfw &$(1,1)_{\mhalf}$&      -8&     0&    -4&    -2&    10&    -2&     2&    80&     2&    (1,1,1)&    2&     0&    -2\\ 
\Vs{47}&\hfw &$(1,1)_{\half}$&        8&     0&     4&     2&    14&     2&     2&    32&     2&    (1,1,1)&   -2&     0&     2\\ 
\Vs{48}&\mfw &$(1,1)_{\mhalf}$&      -8&     0&    -4&    -2&   -14&    -2&    -2&   -32&     2&    (1,1,1)&    2&     0&    -2\\ 
\Vs{49}&\hfw &$(1,1)_{\half}$&       -4&     0&     4&    -2&   -14&     2&     2&    68&     2&    (1,1,1)&    2&     0&    -2\\ 
\Vs{50}&\mfw &$(1,1)_{\mhalf}$&       4&     0&    -4&     2&    14&    -2&    -2&   -68&     2&    (1,1,1)&   -2&     0&     2\\ 
\hline
\Nc{1}&   0& $(1,1)_{0}$&            8&     6&    -4&    -4&     2&     0&     0&   -24&    -2&     (1,1,1)&   0&     0&     0\\ 
\Nc{2}&   0& $(1,1)_{0}$&            8&     6&    -4&     2&   -10&    -2&    -2&   -80&     0&     (1,1,1)&   0&     0&     0\\ 
\Nc{3}&   0& $(1,1)_{0}$&            4&     6&    -4&     2&    14&     2&    -2&   -68&     0&     (1,1,1)&   0&     0&     0\\ 
\hline 
\p{1}&     0& $(1,1)_{0}$&            0&     0&     0&     0&     0&     0&     0&     0&     0&    (1,1,1)&    0&     0&     0\\ 
\p{2}&     0& $(1,1)_{0}$&            0&     0&     0&     0&     0&     0&     0&     0&     0&    (1,1,1)&    0&     0&     0\\ 
\p{3}&     0& $(1,1)_{0}$&            0&     0&     0&     0&     0&     0&     0&     0&     0&    (1,1,1)&    0&     0&     0\\ 
\hline
\p{23}&   0& $(1,1)_{0}$&           -8&     0&     0&     0&    48&     0&     0&    24&     0&    (1,1,1)&    0&     0&     0\\ 
\pb{23}&  0& $(1,1)_{0}$&            8&     0&     0&     0&   -48&     0&     0&   -24&     0&    (1,1,1)&    0&     0&     0\\ 
\p{13}&   0& $(1,1)_{0}$&          -28&     0&     0&     4&   -32&     0&     0&    84&     0&    (1,1,1)&    0&     0&     0\\ 
\pb{13}&  0& $(1,1)_{0}$&           28&     0&     0&    -4&    32&     0&     0&   -84&     0&    (1,1,1)&    0&     0&     0\\ 
\p{12}&   0& $(1,1)_{0}$&          -36&     0&     0&     4&    16&     0&     0&   108&     0&    (1,1,1)&    0&     0&     0\\ 
\pb{12}&  0& $(1,1)_{0}$&           36&     0&     0&    -4&   -16&     0&     0&  -108&     0&    (1,1,1)&    0&     0&     0\\ 
\p{4}&    0& $(1,1)_{0}$&            0&     0&     0&     0&     0&     0&     0&     0&     4&    (1,1,1)&    0&     0&     0\\ 
\pp{4}&   0& $(1,1)_{0}$&            0&     0&     0&     0&     0&     0&     0&     0&     4&    (1,1,1)&    0&     0&     0\\ 
\pb{4}&    0& $(1,1)_{0}$&           0&     0&     0&     0&     0&     0&     0&     0&    -4&    (1,1,1)&    0&     0&     0\\
\pbp{4}&   0& $(1,1)_{0}$&           0&     0&     0&     0&     0&     0&     0&     0&    -4&    (1,1,1)&    0&     0&     0\\ 
\p{56}&   0& $(1,1)_{0}$&            8&     0&     0&     0&     0&     0&     8&   200&     0&    (1,1,1)&    0&     0&     0\\ 
\pb{56}&  0& $(1,1)_{0}$&           -8&     0&     0&     0&     0&     0&    -8&  -200&     0&    (1,1,1)&    0&     0&     0\\ 
\pp{56}&  0& $(1,1)_{0}$&            0&     0&     0&     0&     0&     8&     0&     0&     0&    (1,1,1)&    0&     0&     0\\ 
\pbp{56}& 0& $(1,1)_{0}$&            0&     0&     0&     0&     0&    -8&     0&     0&     0&    (1,1,1)&    0&     0&     0\\ 
\hline
\end{tabular}
\end{flushleft}

\begin{flushleft}
\begin{tabular}{|l||r|l|rrrrrrrrr|lccc|}
\hline 
\hline
State       &\U{E}&$(C,L)_Y$      &  \Ua    &\UC   &\UL   &\UP{1}&\UP{2}&\UP{3}&\UP{4}&\UP{5}&\U{4}&$(3,2,2^{'})_{H}$&\U{7}&\U{H}&\U{9}\\
\hline
\hline
\Hs{15}&   0& $(1,1)_{0}$&           -8&     3&    -2&     0&    -3&    -4&    -2&   136&    -2&    (1,1,1)&   -1&     3&     0\\ 
\Hs{16}&   0& $(1,1)_{0}$&            8&    -3&     2&     0&     3&    -4&     2&  -136&     2&    (1,1,1)&    1&    -3&     0\\ 
\Hs{17}&   0& $(1,1)_{0}$&           -4&     3&    -2&     0&    -3&     0&     2&   236&     2&    (1,1,1)&   -1&     3&     0\\ 
\Hs{18}&   0& $(1,1)_{0}$&           12&    -3&     2&     0&     3&     0&     6&   -36&    -2&    (1,1,1)&    1&    -3&     0\\ 
\Hs{19}&   0& $(1,1)_{0}$&          -16&     3&    -2&     0&    -3&     4&     2&   -64&     2&    (1,1,1)&   -1&     3&     0\\ 
\Hs{20}&   0& $(1,1)_{0}$&           16&    -3&     2&     0&     3&     4&    -2&    64&    -2&    (1,1,1)&    1&    -3&     0\\ 
\Hs{21}&   0& $(1,1)_{0}$&          -12&     3&    -2&     0&    -3&     0&     6&    36&    -2&    (1,1,1)&   -1&     3&     0\\ 
\Hs{22}&   0& $(1,1)_{0}$&           20&    -3&     2&     0&     3&     0&     2&   164&     2&    (1,1,1)&    1&    -3&     0\\ 
\Hs{29}&   0& $(1,1)_{0}$&           -4&     3&    -2&     6&   -15&    -2&     0&   180&     0&    (1,1,1)&   -1&     3&     0\\ 
\Hs{30}&   0& $(1,1)_{0}$&          -24&    -3&     2&    -2&   -17&     2&     0&   -96&     0&    (1,1,1)&    1&    -3&     0\\ 
\Hs{31}&   0& $(1,1)_{0}$&           12&    -3&     2&     6&    -9&    -2&     4&   -92&     0&    (1,1,1)&    1&    -3&     0\\ 
\Hs{32}&   0& $(1,1)_{0}$&            0&    -3&     2&     2&    11&    -2&     0&   168&     0&    (1,1,1)&   -3&    -3&     0\\ 
\Hs{36}&   0& $(1,1)_{0}$&           20&    -3&     2&    -6&    -9&     2&     0&   108&     0&    (1,1,1)&    1&    -3&     0\\ 
\Hs{37}&   0& $(1,1)_{0}$&           16&     3&    -2&     2&    -7&    -2&     0&  -216&     0&    (1,1,1)&   -1&     3&     0\\ 
\Hs{38}&   0& $(1,1)_{0}$&          -12&     3&    -2&    -6&   -15&     2&     4&   -20&     0&    (1,1,1)&   -1&     3&     0\\ 
\Hs{39}&   0& $(1,1)_{0}$&            8&     3&    -2&    -2&    13&     2&     0&   144&     0&    (1,1,1)&    3&     3&     0\\ 
\Vs{1}&    0& $(1,1)_{0}$&           16&     0&     0&     4&     4&     0&     0&   -48&     2&    (1,1,1)&    2&     6&     0\\ 
\Vs{2}&    0& $(1,1)_{0}$&           16&     0&     0&     4&     4&     0&     0&   -48&    -2&    (1,1,1)&   -2&    -6&     0\\ 
\Vs{11}&   0& $(1,1)_{0}$&           16&     0&     0&    -2&    16&     2&     2&     8&     0&    (1,1,1)&    2&     6&     0\\ 
\Vs{12}&   0& $(1,1)_{0}$&           12&     0&     0&    -2&    16&    -2&    -2&   -92&     0&    (1,1,1)&   -2&    -6&     0\\ 
\Vs{21}&   0& $(1,1)_{0}$&           20&     0&     0&    -2&    -8&    -2&     2&    -4&     0&    (1,1,1)&    2&     6&     0\\ 
\Vs{22}&   0& $(1,1)_{0}$&           16&     0&     0&    -2&    -8&     2&    -2&  -104&     0&    (1,1,1)&   -2&    -6&     0\\ 
\Vs{31}&   0& $(1,1)_{0}$&           -4&     0&     0&     0&    24&     0&     0&    12&     2&    (1,1,1)&   -2&     6&     0\\ 
\Vs{32}&   0& $(1,1)_{0}$&           -4&     0&     0&     0&    24&     0&     0&    12&    -2&    (1,1,1)&    2&    -6&     0\\ 
\hline
$\tr Q_s$&&&                       168&    36&     0&     0&    36&     0&    24&   168&    12&&               0&     0&     0\\ 
\hline
\hline
\end{tabular}
\end{flushleft}

\no Table A.I.b: Same as in Table A.I.a except for Non-Abelian Singlet States.
An ``$s$'' superscript indicates that the various $H$ and $V$ states are singlets. 

\begin{flushleft}
\begin{tabular}{|l||r|l|rrrrrrrrr|lccc|}
\hline 
\hline
State       &\U{E}&$(C,L)_Y$      &  \Ua    &\UC   &\UL   &\UP{1}&\UP{2}&\UP{3}&\UP{4}&\UP{5}&\U{4}&$(3,2,2^{'})_{H}$&\U{7}&\U{H}&\U{9}\\
\hline 
\hline 
\H{1}&\hfw &  $(1,1)_{\half}$&      -4&     3&     2&     2&    11&     2&     2&    68&    -2&     (1,2,1)&  1&     3&     0\\
\H{2} &\mfw & $(1,1)_{\mhalf}$&      4&    -3&    -2&    -2&   -11&    -2&    -2&   -68&    -2&     (1,2,1)& -1&    -3&     0\\
\H{11}&\hfw & $(1,1)_{\half}$&       0&     3&     2&     2&   -13&    -2&     2&    56&     2&     (1,1,2)&  1&     3&     0\\
\H{13}&\mfw &$(1,1)_{\mhalf}$&       0&    -3&    -2&    -2&    13&     2&    -2&   -56&     2&     (1,1,2)& -1&    -3&     0\\
\hline
\H{42}&  0&   $(1,1)_{0}$&           8&     3&    -2&    -2&    13&     2&     0&   144&     0&     (3,1,1)& -1&    -1&     0\\
\V{4} &  0&   $(1,1)_{0}$&          16&     0&     0&     4&     4&     0&     0&   -48&    -2&     (3,1,1)& -2&     2&     0\\
\V{14}&  0&   $(1,1)_{0}$&          16&     0&     0&    -2&    16&     2&     2&     8&     0&     (3,1,1)& -2&     2&     0\\
\V{24}&  0&   $(1,1)_{0}$&          20&     0&     0&    -2&    -8&    -2&     2&    -4&     0&     (3,1,1)& -2&     2&     0\\
\V{34}&  0&   $(1,1)_{0}$&           4&     0&     0&     0&   -24&     0&     0&   -12&    -2&     (3,1,1)&  2&     2&     0\\
\hline
\H{35}&  0&   $(1,1)_{0}$&           0&    -3&     2&     2&    11&    -2&     0&   168&     0&   (\tb,1,1)&  1&     1&     0\\
\V{3} &  0&   $(1,1)_{0}$&          16&     0&     0&     4&     4&     0&     0&   -48&     2&   (\tb,1,1)&  2&    -2&     0\\
\V{13}&  0&   $(1,1)_{0}$&          12&     0&     0&    -2&    16&    -2&    -2&   -92&     0&   (\tb,1,1)&  2&    -2&     0\\
\V{23}&  0&   $(1,1)_{0}$&          16&     0&     0&    -2&    -8&     2&    -2&  -104&     0&    (\tb,1,1)& 2&    -2&     0\\
\V{33}&  0&   $(1,1)_{0}$&           4&     0&     0&     0&   -24&     0&     0&   -12&     2&   (\tb,1,1)& -2&    -2&     0\\
\hline
\H{25}&  0&   $(1,1)_{0}$&           8&     3&    -2&    -2&   -11&    -2&     2&    32&     0&     (1,2,1)&  1&    -3&    -2\\ 
\H{28}&  0&   $(1,1)_{0}$&           0&    -3&     2&     2&   -13&     2&     2&    56&     0&     (1,2,1)& -1&     3&    -2\\ 
\V{9 }&  0&   $(1,1)_{0}$&          12&     0&     0&     4&     4&     0&     2&  -148&    -2&     (1,2,1)&  0&     0&     2\\ 
\V{10}&  0&   $(1,1)_{0}$&          20&     0&     0&     4&     4&     0&    -2&    52&     2&     (1,2,1)&  0&     0&    -2\\ 
\V{19}&  0&   $(1,1)_{0}$&          16&     0&     0&    -2&    16&    -2&    -4&     8&     0&     (1,2,1)&  0&     0&    -2\\ 
\V{20}&  0&   $(1,1)_{0}$&          12&     0&     0&    -2&    16&     2&     4&   -92&     0&     (1,2,1)&  0&     0&     2\\ 
\V{29}&  0&   $(1,1)_{0}$&          24&     0&     0&    -2&    -8&    -2&     0&    96&     0&     (1,2,1)&  0&     0&    -2\\ 
\V{30}&  0&   $(1,1)_{0}$&          12&     0&     0&    -2&    -8&     2&     0&  -204&     0&     (1,2,1)&  0&     0&     2\\ 
\V{39}&  0&   $(1,1)_{0}$&          20&     0&     0&     4&     4&     0&    -2&    52&     2&     (1,2,1)&  0&     0&     2\\ 
\V{40}&  0&   $(1,1)_{0}$&         -12&     0&     0&    -4&    -4&     0&    -2&   148&     2&     (1,2,1)&  0&     0&     2\\ 
\hline
\H{23}&  0&   $(1,1)_{0}$&           8&     3&    -2&    -2&   -11&    -2&     2&    32&     0&     (1,1,2)&  1&    -3&     2\\ 
\H{26}&  0&   $(1,1)_{0}$&           0&    -3&     2&     2&   -13&     2&     2&    56&     0&     (1,1,2)& -1&     3&     2\\ 
\V{5 }&  0&   $(1,1)_{0}$&          20&     0&     0&     4&     4&     0&    -2&    52&    -2&     (1,1,2)&  0&     0&     2\\ 
\V{7 }&  0&   $(1,1)_{0}$&          12&     0&     0&     4&     4&     0&     2&  -148&     2&     (1,1,2)&  0&     0&    -2\\ 
\V{15}&  0&   $(1,1)_{0}$&          16&     0&     0&    -2&    16&    -2&    -4&     8&     0&     (1,1,2)&  0&     0&     2\\ 
\V{17}&  0&   $(1,1)_{0}$&          12&     0&     0&    -2&    16&     2&     4&   -92&     0&     (1,1,2)&  0&     0&    -2\\ 
\V{25}&  0&   $(1,1)_{0}$&          24&     0&     0&    -2&    -8&    -2&     0&    96&     0&     (1,1,2)&  0&     0&     2\\ 
\V{27}&  0&   $(1,1)_{0}$&          12&     0&     0&    -2&    -8&     2&     0&  -204&     0&     (1,1,2)&  0&     0&    -2\\ 
\V{35}&  0&   $(1,1)_{0}$&          20&     0&     0&     4&     4&     0&    -2&    52&    -2&     (1,1,2)&  0&     0&    -2\\ 
\V{37}&  0&   $(1,1)_{0}$&         -12&     0&     0&    -4&    -4&     0&    -2&   148&    -2&     (1,1,2)&  0&     0&    -2\\ 
\hline
$\tr Q_h$&&&                      784&     0&     0&     0&     0&     0&     0&     0&     0&&              0&     0&     0\\
\hline
\hline
\end{tabular}
\end{flushleft}

\no Table A.I.c: Same as in Table A.I.a except for  
Hidden Sector $SU(3)_H\times SU(2)_H\times SU(2)^{'}_H$ States (that are $SU(3)_C\times SU(2)_L$ singlets). 

\no Note: In \cite{fny} and \cite{fc} both components of several hidden sector $SU(2)_H$ doublets appeared in the tables
of states and in the superpotential terms. The doublets formed by these pairs of components were 
$(H_{11},H_{12})$, $(H_{13},H_{14})$, $(H_{23},H_{24})$, $(H_{26},H_{27})$,  
$(V_{5},V_{6})$, $(V_{7},V_{8})$, $(V_{15},V_{16})$, $(V_{17},V_{18})$, $(V_{25},V_{26})$, $(V_{27},V_{28})$, 
$(V_{35},V_{36})$, $(V_{37},V_{38})$. Throughout this paper, each doublet is denoted by its component with lower subscript.   

{
\def\sp{\sigma^{+}}  
\def\sm{\sigma^{-}}  
\def\T#1{T_{#1}}
\def\S#1{S_{#1}}

\def\H#1{H_{#1}}

\def\UR#1{Q_{#1}}
\def\R#1{R_{#1}}
\def\b#1{b_{#1}}
\def\hv#1{h_{#1}}

\def\p#1{\Phi_{#1}}
\def\pb#1{\bar{\Phi}_{#1}}
\def\pp#1{\Phi^{'}_{#1}}
\def\pbp#1{\bar{\Phi}^{'}_{#1}}
\def\h#1{h_{#1}}
\def\hb#1{\bar{h}_{#1}}
\def\ec#1{e^c_{#1}}
\def\Nc#1{N^c_{#1}}
\def\Q#1{Q_{#1}}
\def\dc#1{d^c_{#1}}
\def\uc#1{u^c_{#1}}

\def\L#1{L_{#1}}
\def\V#1{V_{#1}}
\def\Vs#1{V^s_{#1}}
\def\Hs#1{H^s_{#1}}

\def\fp{$.5$}
\def\fm{$-.5$}
\def\msp{$\sigma^{+}$}
\def\msm{$\sigma^{-}$}

\def\ibf{$\bar{f}$}
\def\BS{\bar{S}}
\def\xre{${r_1}:{r_2}= $} 

\def\by{{\bar{y}}}
\def\w{\omega}
\def\bw{{\bar{\omega}}}

\begin{flushleft}
\begin{tabular}{|l|rrrrrr|cccccc|}
\hline\hline
    &        Q  &       &       &       &       &       &     I &       &       &       &       &      \\
                &$x_1$&$y_1$ &$\w_1$&$x_3$&$y_3$&$x_5$&$  y_2$&$ \w_2$&$  y_4$&$ \w_4$&$  y_5$&$ \w_5$\\ 
State           &$x_2$&$\w_6$&$\w_3$&$x_4$&$y_6$&$x_6$&$\by_2$&$\bw_2$&$\by_4$&$\bw_4$&$\by_5$&$\bw_5$\\ 
\hline\hline
$\p{1}$   &    \fm    &       &       &\fp    &       &\fp    & \ibf  & \ibf  &       &       &       &      \\  
$\p{2}$   &    \fp    &       &       &\fm    &       &\fp    &       &       & \ibf  & \ibf  &       &      \\  
$\p{3}$   &    \fp    &       &       &\fp    &       &\fm    &       &       &       &       & \ibf  & \ibf \\  
\hline
$\p{4}$   &    \fp    &       &       &\fm    &       &\fp    &       & \ibf  &       &       &       &      \\  
$\pb{4}$  &    \fp    &       &       &\fm    &       &\fp    &       & \ibf  &       &       &       &      \\  
$\pp{4}$  &    \fp    &       &       &\fp    &       &\fm    & \ibf  &       &       &       &       &      \\  
$\pbp{4}$ &    \fp    &       &       &\fp    &       &\fm    & \ibf  &       &       &       &       &      \\  
$\p{56}$  &    \fm    &       &       &\fp    &       &\fp    &       &       &       &       &       &      \\  
$\pb{56}$ &    \fm    &       &       &\fp    &       &\fp    &       &       &       &       &       &      \\  
$\pp{56}$ &    \fm    &       &       &\fp    &       &\fp    &       &       &       &       &       &      \\  
$\pbp{56}$&    \fm    &       &       &\fp    &       &\fp    &       &       &       &       &       &      \\  
$\p{23}$  &    \fm    &       &       &\fp    &       &\fp    &       &       &       &       &       &      \\  
$\pb{23}$ &    \fm    &       &       &\fp    &       &\fp    &       &       &       &       &       &      \\  
$\p{12}$  &    \fp    &       &       &\fp    &       &\fm    &       &       &       &       &       &      \\  
$\pb{12}$ &    \fp    &       &       &\fp    &       &\fm    &       &       &       &       &       &      \\  
$\p{13}$  &    \fp    &       &       &\fm    &       &\fp    &       &       &       &       &       &      \\  
$\pb{13}$ &    \fp    &       &       &\fm    &       &\fp    &       &       &       &       &       &      \\  
\hline 
$\Hs{3}$  &           &\fp    &       &\fp    &       &       &       & \msm  &       &       & \msm  &      \\  
$\Hs{4} $ &           &\fm    &       &\fp    &       &       &       & \msp  &       &       & \msm  &      \\  
$\Hs{5}$  &           &\fm    &       &\fp    &       &       &       & \msm  &       &       & \msm  &      \\  
$\Hs{6} $ &           &\fp    &       &\fp    &       &       &       & \msp  &       &       & \msm  &      \\  
$\Hs{7}$  &           &       &\fp    &       &       &\fp    & \msp  &       & \msp  &       &       &      \\  
$\Hs{8} $ &           &       &\fm    &       &       &\fp    & \msm  &       & \msp  &       &       &      \\  
$\Hs{9}$  &           &       &\fm    &       &       &\fp    & \msp  &       & \msp  &       &       &      \\  
$\Hs{10}$ &           &       &\fp    &       &       &\fp    & \msm  &       & \msp  &       &       &      \\  
$\Hs{15}$ &    \fp    &       &       &       &\fm    &       &       &       & \msm  &       &       & \msm \\  
$\Hs{16}$ &    \fp    &       &       &       &\fp    &       &       &       & \msm  &       &       & \msm \\  
$\Hs{17}$ &    \fp    &       &       &       &\fm    &       &       &       & \msp  &       &       & \msp \\  
$\Hs{18}$ &    \fp    &       &       &       &\fp    &       &       &       & \msp  &       &       & \msp \\  
$\Hs{19}$ &    \fp    &       &       &       &\fm    &       &       &       &       & \msm  & \msp  &      \\  
$\Hs{20}$ &    \fp    &       &       &       &\fp    &       &       &       &       & \msm  & \msp  &      \\  
$\Hs{21}$ &    \fp    &       &       &       &\fm    &       &       &       &       & \msp  & \msm  &      \\  
$\Hs{22}$ &    \fp    &       &       &       &\fp    &       &       &       &       & \msp  & \msm  &      \\  
$\Hs{29}$ &           &\fm    &       &\fp    &       &       & \msm  &       &       &       & \msp  &      \\  
$\Hs{30}$ &           &\fp    &       &\fp    &       &       & \msm  &       &       &       & \msp  &      \\  
$\Hs{31}$ &           &\fp    &       &\fp    &       &       & \msp  &       &       &       & \msp  &      \\  
$\Hs{32}$ &           &\fp    &       &\fp    &       &       & \msm  &       &       &       & \msp  &      \\  
\hline
\end{tabular}
\end{flushleft}

\begin{flushleft}
\begin{tabular}{|l|rrrrrr|cccccc|}
\hline\hline
   &      $Q$  &       &       &       &       &       &    $I$ &       &       &       &       &      \\
                &$x_1$&$y_1$ &$\w_1$&$x_3$&$y_3$&$x_5$&$  y_2$&$ \w_2$&$  y_4$&$ \w_4$&$  y_5$&$ \w_5$\\ 
State           &$x_2$&$\w_6$&$\w_3$&$x_4$&$y_6$&$x_6$&$\by_2$&$\bw_2$&$\by_4$&$\bw_4$&$\by_5$&$\bw_5$\\ 
\hline\hline
$\Hs{36}$ &           &       &\fp    &       &       &\fp    &       & \msp  & \msm  &       &       &      \\  
$\Hs{37}$ &           &       &\fm    &       &       &\fp    &       & \msp  & \msm  &       &       &      \\  
$\Hs{38}$ &           &       &\fm    &       &       &\fp    &       & \msm  & \msm  &       &       &      \\  
$\Hs{39}$ &           &       &\fm    &       &       &\fp    &       & \msp  & \msm  &       &       &      \\  
\hline
$\Vs{1}$  &    \fp    &       &       &       &\fp    &       &       &       & \msp  &       & \msp  &      \\  
$\Vs{2}$  &    \fp    &       &       &       &\fm    &       &       &       & \msm  &       & \msm  &      \\  
$\Vs{11}$ &           &\fp    &       &\fp    &       &       & \msp  &       &       &       &       & \msp \\  
$\Vs{12}$ &           &\fm    &       &\fp    &       &       & \msp  &       &       &       &       & \msp \\  
$\Vs{21}$ &           &       &\fp    &       &       &\fp    &       & \msm  &       & \msp  &       &      \\  
$\Vs{22}$ &           &       &\fm    &       &       &\fp    &       & \msm  &       & \msp  &       &      \\  
$\Vs{31}$ &    \fp    &       &       &       &\fp    &       &       &       &       & \msp  &       & \msm \\  
$\Vs{32}$ &    \fp    &       &       &       &\fm    &       &       &       &       & \msp  &       & \msm \\  
$\Vs{41}$ &           &\fp    &       &\fp    &       &       &       & \msp  &       &       &       & \msm \\  
$\Vs{42}$ &           &\fm    &       &\fp    &       &       &       & \msm  &       &       &       & \msm \\  
$\Vs{43}$ &           &\fm    &       &\fp    &       &       &       & \msp  &       &       &       & \msm \\  
$\Vs{44}$ &           &\fp    &       &\fp    &       &       &       & \msm  &       &       &       & \msm \\  
$\Vs{47}$ &           &       &\fm    &       &       &\fp    & \msp  &       &       & \msm  &       &      \\ 
$\Vs{48}$ &           &       &\fp    &       &       &\fp    & \msm  &       &       & \msm  &       &      \\ 
$\Vs{49}$ &           &       &\fp    &       &       &\fp    & \msp  &       &       & \msm  &       &      \\  
$\Vs{50}$ &           &       &\fm    &       &       &\fp    & \msm  &       &       & \msm  &       &      \\  
\hline\hline
\end{tabular}
\end{flushleft}
\noindent Table A.II.a: 
The non-gauge world--sheet charges of the 63 non-Abelian singlets
in model FNY, except for the three generations of $e^c$ and $N^c$, 
whose charges appear in Table A.II.b.
The superscript ``$s$'' for the 
$H$ and $V$ states denotes non-Abelian singlets.
For the heading, in the ``$Q$'' section the two real fermions
specifying a column are the two components of a complex left-moving fermion, 
and in the ``$I$'' section they denote
the components of a non-chiral Ising fermion. 
A global $U(1)$ charge $Q$ carried 
by a singlet state is listed in the column of the
complex world--sheet fermion associated with the charge. 
Likewise, a conformal field $I\in \{ f, \bar{f}, \sp, \sm \}$ of a
non-chiral Ising fermion 
carried by a singlet is listed in the column of
the appropriate Ising fermion.

\begin{flushleft}
\begin{tabular}{|l|rrrrrr|cccccc|}
\hline\hline
 &       $Q$  &       &       &       &       &       &    $I$ &       &       &       &       &      \\
                &$x_1$&$y_1$ &$\w_1$&$x_3$&$y_3$&$x_5$&$  y_2$&$ \w_2$&$  y_4$&$ \w_4$&$  y_5$&$ \w_5$\\ 
State           &$x_2$&$\w_6$&$\w_3$&$x_4$&$y_6$&$x_6$&$\by_2$&$\bw_2$&$\by_4$&$\bw_4$&$\by_5$&$\bw_5$\\ 
\hline\hline
$\Q{1}$  &    \fp    &       &       &       &\fm    &       &       &       & \msm  &       & \msm  &      \\  
$\dc{1}$ &    \fp    &       &       &       &\fp    &       &       &       & \msp  &       & \msp  &      \\  
$\uc{1}$ &    \fp    &       &       &       &\fp    &       &       &       & \msm  &       & \msm  &      \\  
$\L{1}$  &    \fp    &       &       &       &\fm    &       &       &       & \msp  &       & \msp  &      \\  
$\ec{1}$ &    \fp    &       &       &       &\fp    &       &       &       & \msm  &       & \msm  &      \\  
$\Nc{1}$ &    \fp    &       &       &       &\fp    &       &       &       & \msp  &       & \msp  &      \\  
\hline
$\Q{2}$  &           &\fm    &       &\fp    &       &       & \msp  &       &       &       &       & \msp \\  
$\dc{2}$ &           &\fp    &       &\fp    &       &       & \msp  &       &       &       &       & \msp \\  
$\uc{2}$ &           &\fp    &       &\fp    &       &       & \msm  &       &       &       &       & \msp \\  
$\L{2}$  &           &\fm    &       &\fp    &       &       & \msm  &       &       &       &       & \msp \\  
$\ec{2}$ &           &\fp    &       &\fp    &       &       & \msm  &       &       &       &       & \msp \\  
$\Nc{2}$ &           &\fp    &       &\fp    &       &       & \msp  &       &       &       &       & \msp \\  
\hline
$\Q{3}$  &           &       &\fm    &       &       &\fp    &       & \msm  &       & \msp  &       &      \\  
$\dc{3}$ &           &       &\fp    &       &       &\fp    &       & \msm  &       & \msp  &       &      \\  
$\uc{3}$ &           &       &\fp    &       &       &\fp    &       & \msm  &       & \msp  &       &      \\  
$\L{3}$  &           &       &\fp    &       &       &\fp    &       & \msm  &       & \msp  &       &      \\  
$\ec{3}$ &           &       &\fp    &       &       &\fp    &       & \msp  &       & \msp  &       &      \\  
$\Nc{3}$ &           &       &\fp    &       &       &\fp    &       & \msp  &       & \msp  &       &      \\  
\hline 
$\H{40}$  &           &       &\fm    &       &       &\fp    &       & \msp  & \msm  &       &       &      \\  
$\H{33}$  &           &\fp    &       &\fp    &       &       & \msm  &       &       &       & \msp  &      \\  
\hline 
$\h{1}$   &    \fm    &       &       &\fp    &       &\fp    &       &       &       &       &       &      \\  
$\h{2}$   &    \fp    &       &       &\fm    &       &\fp    &       &       &       &       &       &      \\  
$\h{3}$   &    \fp    &       &       &\fp    &       &\fm    &       &       &       &       &       &      \\ 
$\hb{1}$  &    \fm    &       &       &\fp    &       &\fp    &       &       &       &       &       &      \\  
$\hb{2}$  &    \fp    &       &       &\fm    &       &\fp    &       &       &       &       &       &      \\  
$\hb{3}$  &    \fp    &       &       &\fp    &       &\fm    &       &       &       &       &       &      \\  
\hline 
$\H{34}$&           &\fm    &       &\fp    &       &       & \msp  &       &       &       & \msp  &      \\  
$\H{41}$&           &       &\fp    &       &       &\fp    &       & \msm  & \msm  &       &       &      \\  
\hline 
$\V{45}$&           &\fp    &       &\fp    &       &       &       & \msp  &       &       &       & \msm \\ 
$\V{46}$&           &\fm    &       &\fp    &       &       &       & \msm  &       &       &       & \msm \\  
$\V{51}$&           &       &\fm    &       &       &\fp    & \msm  &       &       & \msm  &       &      \\ 
$\V{52}$&           &       &\fp    &       &       &\fp    & \msp  &       &       & \msm  &       &      \\  
\hline\hline
\end{tabular}
\end{flushleft}
\noindent Table A.II.b:
Same as Table A.II.a except for the 
non-trivial $SU(3)_C\times SU(2)_L$ states, along with those for the $e^c$ and
$N^c$ states.

\begin{flushleft}
\begin{tabular}{|l|rrrrrr|cccccc|}
\hline\hline
      &      $Q$  &       &       &       &       &       &   $I$ &       &       &       &       &      \\
                &$x_1$&$y_1$ &$\w_1$&$x_3$&$y_3$&$x_5$&$  y_2$&$ \w_2$&$  y_4$&$ \w_4$&$  y_5$&$ \w_5$\\ 
State           &$x_2$&$\w_6$&$\w_3$&$x_4$&$y_6$&$x_6$&$\by_2$&$\bw_2$&$\by_4$&$\bw_4$&$\by_5$&$\bw_5$\\ 
\hline\hline
$\H{1}$   &           &\fp    &       &\fp    &       &       &       & \msm  &       &       & \msp  &      \\  
$\H{2} $  &           &\fm    &       &\fp    &       &        &       & \msp  &       &       & \msp  &      \\  
$\H{11}$  &           &       & \fp   &       &       &       &  \msp &       &  \msm &       &       &      \\  
$\H{13}$  &           &       &\fm    &       &       &\fp     & \msm  &       & \msm  &       &       &      \\  
$\H{23}$  &           &\fm    &       &\fp    &       &       &  \msm &       &       &       & \msm  &      \\  
$\H{25}$  &           &\fm    &       &\fp    &       &       &  \msm &       &       &       & \msm  &      \\  
$\H{26}$  &           &       &\fm    &       &       &\fp     &       & \msm  & \msp  &       &       &      \\
$\H{28}$  &           &       &\fp    &       &       &\fp     &       & \msp  & \msp  &       &       &      \\
$\H{35}$  &           &\fm    &       &\fp    &       &        & \msp  &       &       &       & \msp  &      \\  
$\H{42}$  &           &       &\fp    &       &       &\fp    &       & \msm  & \msm  &       &       &      \\  
$\V{3}$   &    \fp    &       &       &       &\fp    &       &       &       & \msm  &       & \msm  &      \\  
$\V{4}$   &    \fp    &       &       &       &\fm    &       &       &       & \msp  &       & \msp  &      \\  
$\V{5} $  &    \fp    &       &       &       &\fp    &        &       &       & \msm  &       & \msp  &      \\  
$\V{7} $  &    \fp    &       &       &       &\fm    &        &       &       & \msp  &       & \msm  &      \\  
$\V{9} $  &    \fp    &       &       &       &\fp    &        &       &       & \msp  &       & \msm  &      \\  
$\V{10}$  &    \fp    &       &       &       &\fm    &        &       &       & \msm  &       & \msp  &      \\  
$\V{13}$  &           &\fp    &       &\fp    &       &       & \msm  &       &       &       &       & \msp \\  
$\V{14}$  &           &\fm    &       &\fp    &       &       & \msm  &       &       &       &       & \msp \\  
$\V{15}$  &           &\fp    &       &\fp    &       &        & \msp  &       &       &       &       & \msm \\  
$\V{17}$  &           &\fm    &       &\fp    &       &        & \msp  &       &       &       &       & \msm \\  
$\V{19}$  &           &\fm    &       &\fp    &       &        & \msm  &       &       &       &       & \msm \\  
$\V{20}$  &           &\fp    &       &\fp    &       &        & \msm  &       &       &       &       & \msm \\  
$\V{23}$  &           &       &\fp    &       &       &\fp     &       & \msp  &       & \msp  &       &      \\  
$\V{24}$  &           &       &\fm    &       &       &\fp     &       & \msp  &       & \msp  &       &      \\  
$\V{25}$  &           &       &\fp    &       &       &\fp     &       & \msp  &       & \msm  &       &      \\  
$\V{27}$  &           &       &\fm    &       &       &\fp     &       & \msp  &       & \msm  &       &      \\  
$\V{29}$  &           &       &\fm    &       &       &\fp     &       & \msm  &       & \msm  &       &      \\  
$\V{30}$  &           &       &\fp    &       &       &\fp     &       & \msm  &       & \msm  &       &      \\  
$\V{33}$  &    \fp    &       &       &       &\fp    &        &       &       &       & \msm  &       & \msp \\  
$\V{34}$  &    \fp    &       &       &       &\fp    &        &       &       &       & \msp  &       & \msm \\  
$\V{35}$  &    \fp    &       &       &       &\fm    &        &       &       &       & \msp  &       & \msp \\  
$\V{37}$  &    \fp    &       &       &       &\fm    &        &       &       &       & \msm  &       & \msm \\  
$\V{39}$  &    \fp    &       &       &       &\fp    &        &       &       &       & \msp  &       & \msp \\  
$\V{40}$  &    \fp    &       &       &       &\fp    &        &       &       &       & \msm  &       & \msm \\  
\hline\hline
\end{tabular}
\end{flushleft}
\noindent Table A.II.c: Same as Table A.II.a except for the hidden sector non-trivial NA states. 
}
  
\section{Renormalizable and Fourth through Sixth Order
Nonrenormalizable Terms in FNY Superpotential}

 
{ 
\def\qp{+}
\def\T#1{T_{#1}}
\def\S#1{S_{#1}}

\def\H#1{H_{#1}}
\def\UR#1{U_{#1}}
\def\R#1{R_{#1}}
\def\b#1{b_{#1}}
\def\hv#1{h_{#1}}

\def\Q#1{Q_{#1}}
\def\dc#1{d^{c}_{#1}}
\def\uc#1{u^{c}_{#1}}
\def\h#1{h_{#1}}
\def\hb#1{{\bar{h}}_{#1}}
\def\L#1{L_{#1}}
\def\ec#1{e^{c}_{#1}}
\def\Nc#1{N^{c}_{#1}}
\def\H#1{H_{#1}}
\def\V#1{V_{#1}}
\def\Hs#1{H^{s}_{#1}}
\def\Vs#1{V^{s}_{#1}}
\def\p#1{\Phi_{#1}}
\def\pp#1{\Phi^{'}_{#1}}
\def\pb#1{{\overline{\Phi}}_{#1}}
\def\pbp#1{{\overline{\Phi}}^{'}_{#1}}

\def\obs{{\rm observable}}
\def\sig{{\rm singlets}}
\def\hid{{\rm hidden}}

\def\ba#1{\hbox to 2.8truecm{#1\hfill}}
\def\bb#1{\hbox to 3.4truecm{#1\hfill}}
\def\bc#1{\hbox to 4.3truecm{#1\hfill}}


\no Superpotential terms containing only non-Abelian singlet fields
(including neutrino $SU(2)_L$ singlets, $\Nc{i=1,2,3}$). 
Coupling constants are implicit for non-renormalizable terms.

\def\pl{\phantom{\left[\right.}}
\def\ppsl{\phantom{+\left[\right.}}
\def\ppst{\phantom{+ .}}

\vskip .5truecm
\no $W_3(\sig)$: (with $g^{'}_{s}\equiv g_{s} \sqrt{2}$)  
\beqn 
\begin{tabular}{rl}
$\pps g_{s}$& 
$\left[\right.\ppst \pp{4} \{ \Hs{7} \Hs{8} + \Hs{9} \Hs{10}\} 
   +\pb{4} \{ \Hs{3} \Hs{4} + \Hs{5} \Hs{6}+\Vs{41}\Vs{42}+ \Vs{43} \Vs{44}\}$
\\
&$\pl+\pbp{4}\{  \Vs{47}\Vs{48} + \Vs{49} \Vs{50}\}\left.\right]$\\
$+g^{'}_{s}$ 
&$\left[\right.\ppst\p{1} \{ \pb{4}  \pp{4}   + \pbp{4}  \p{4}\}    
  +\p{12} \{\pb{13}\pb{23} + \Hs{37} \Hs{36}    + \Vs{21} \Vs{22}\}
  +\pb{12}  \p{13} \p{23}$ \\
&$\pl+\p{13} \Vs{11} \Vs{12} +\pb{13} \Hs{29} \Hs{30}     
  +\pb{23} \Vs{31} \Vs{32}$\\
&$\pl+\pb{56} \{ \Hs{17} \Hs{18} + \Hs{21} \Hs{22}\}   
  +\pp{56} \Hs{15} \Hs{16}  
  +\pbp{56}\Hs{19} \Hs{20}\left.\right]$\\     
\end{tabular}
\nolabel\\
\label{w3sig2}
\eeqn

\no $W_4(\sig)$:
\beqn 
\begin{tabular}{lllll}
$\pps\Nc{1}\Hs{30}\Hs{32} \Vs{1} $   
&$+\Nc{3} \Hs{30} \Hs{32} \Vs{21}$   
&$+\Hs{3} \Hs{4}  \Hs{7}  \Hs{8} $   
&$+\Hs{3} \Hs{4}  \Hs{9}  \Hs{10}$   
&$+\Hs{4} \Hs{21} \Vs{32} \Vs{43}$\\
 $+\Hs{5} \Hs{6}  \Hs{7}  \Hs{8} $  
&$+\Hs{5} \Hs{6}  \Hs{9}  \Hs{10}$   
&$+\Hs{6} \Hs{15} \Vs{2}  \Vs{41}$   
&$+\Hs{7} \Hs{8}  \Vs{41} \Vs{42}$   
&$+\Hs{7} \Hs{8}  \Vs{43} \Vs{44}$\\   
 $+\Hs{9} \Hs{10} \Vs{41} \Vs{42}$   
&$+\Hs{9} \Hs{10} \Vs{43} \Vs{44}$  
&$+\Hs{30}\Hs{32} \Hs{37} \Hs{39}$     
\end{tabular}
\nolabel\\
\label{w4sig}
\eeqn

\no $W_5(\sig)$:
\beqn 
\begin{tabular}{llll}
   $\pps\Nc{1}  \p{12} \sH{4 } \sH{18} \sV{43}$  
  &$+ \Nc{2} \bpp{4} \sH{22} \sH{30} \sV{31}  $  
  &$+ \P{1 } \sH{3 } \sH{4 } \sH{7 } \sH{8 }  $  
  &$+ \P{1 } \sH{3 } \sH{4 } \sH{9 } \sH{10}  $\\ 
   $+ \P{1 } \sH{5 } \sH{6 } \sH{7 } \sH{8 }  $  
  &$+ \P{1 } \sH{5 } \sH{6 } \sH{9 } \sH{10}  $  
  &$+ \P{1 } \sH{7 } \sH{8 } \sV{41} \sV{42}  $  
  &$+ \P{1 } \sH{7 } \sH{8 } \sV{43} \sV{44}  $\\ 
   $+ \P{1 } \sH{9 } \sH{10} \sV{41} \sV{42}  $  
  &$+ \P{1 } \sH{9 } \sH{10} \sV{43} \sV{44}  $  
  &$+\bp{12} \sH{29} \sH{30} \sV{31} \sV{32}  $  
  &$+ \p{13} \sH{36} \sH{37} \sV{31} \sV{32}  $\\  
   $+ \p{13} \sV{21} \sV{22} \sV{31} \sV{32}  $  
  &$+ \p{23} \sH{4 } \sH{6 } \sH{7 } \sH{9 }  $ 
  &$+ \p{23} \sH{29} \sH{30} \sH{36} \sH{37}  $  
  &$+ \p{23} \sH{29} \sH{30} \sV{21} \sV{22}  $\\  
   $+\bp{23} \sH{3 } \sH{5 } \sH{8 } \sH{10}  $  
  &$+ \pp{4} \sH{15} \sH{30} \sV{2 } \sV{11}  $ 
  &$+\bpp{4} \sH{17} \sH{30} \sV{1 } \sV{12}  $  
  &$+\bp{56} \sH{31} \sH{32} \sH{38} \sH{39}  $\\ 
\end{tabular}
\nolabel\\
\label{w5sig}
\eeqn
\newpage

\no $W_6(\sig)$:
\beqn
\begin{tabular}{lll}
  $\pps\Nc{1}\P{3 }  \P{3 } \sH{30} \sH{32} \sV{1 } $               
 &$+ \Nc{1}  \P{3 }  \p{12} \sH{4 } \sH{18} \sV{43} $               
 &$+ \Nc{1}  \p{12} \bp{12} \sH{30} \sH{32} \sV{1 } $\\               
  $+ \Nc{1}  \pp{4} \bpp{4} \sH{30} \sH{32} \sV{1 } $               
 &$+ \Nc{1} \sH{16} \sH{20} \sH{29} \sH{30} \sV{31} $               
 &$+ \Nc{2}  \P{3 } \bpp{4} \sH{22} \sH{30} \sV{31} $\\               
  $+ \Nc{2} \sH{6 } \sH{7 } \sH{8 } \sH{32} \sV{41} $               
 &$+ \Nc{2} \sH{6 } \sH{9 } \sH{10} \sH{32} \sV{41} $               
 &$+ \Nc{2} \sH{7 } \sH{8 } \sH{22} \sH{30} \sV{31} $\\               
  $+ \Nc{2} \sH{8 } \sH{15} \sH{22} \sH{30} \sV{47} $               
 &$+ \Nc{2} \sH{9 } \sH{10} \sH{22} \sH{30} \sV{31} $               
 &$+ \Nc{2} \sH{10} \sH{30} \sV{2 } \sV{31} \sV{49} $\\               
  $+ \Nc{3}  \P{1 }  \P{1 } \sH{30} \sH{32} \sV{21} $               
 &$+ \Nc{3}  \p{13} \bp{4 } \sH{16} \sH{36} \sV{31} $               
 &$+ \Nc{3}  \p{23} \bp{23} \sH{30} \sH{32} \sV{21} $\\               
  $+ \Nc{3}  \p{56} \bp{56} \sH{30} \sH{32} \sV{21} $               
 &$+ \Nc{3} \pp{56} \bi{56} \sH{30} \sH{32} \sV{21} $               
 &$+ \Nc{3} \sH{4 } \sH{7 } \sH{18} \sH{30} \sV{31} $\\               
  $+ \Nc{3} \sH{4 } \sH{15} \sH{18} \sH{30} \sV{47} $               
 &$+ \Nc{3} \sH{10} \sH{30} \sH{32} \sH{37} \sV{49} $               
 &$+ \Nc{3} \sH{15} \sH{30} \sH{31} \sH{36} \sV{31} $\\               
  $+ \P{1 }  \P{1 } \sH{3 } \sH{4 } \sH{7 } \sH{8 } $               
 &$+ \P{1 }  \P{1 } \sH{3 } \sH{4 } \sH{9 } \sH{10} $               
 &$+ \P{1 }  \P{1 } \sH{5 } \sH{6 } \sH{7 } \sH{8 } $\\
\end{tabular}
\label{w6sig}
\eeqn
\newpage

\no Superpotential terms containing observable sector 
$SU(3)_C\times SU(2)_L$--charged fields (along with the electron conjugates,
$\ec{i=1,2,3}$).
Coupling constants are implicit for non-renormalizable terms.


\vskip .5truecm
\no $W_3(\obs)$:
\beqn
\begin{tabular}{rl}
$\pps g^{'}_{s}$  
& $\left[ \right.\ppst  
   \hb{1} \{ \Q{1}  \uc{1}+ \Nc{1} \L{1} \}$\\    
&$\pl+\h{2}\{ \Q{2}  \dc{2} + \L{2} \ec{2} + \H{34} \Hs{31} + \V{45} \Vs{43} \}
  +\h{3}  \{ \Q{3}  \dc{3} + \L{3} \ec{3} + \V{52}\Vs{47}\}$\\   
&$\pl+\h{1} \{  \hb{2} \p{12}  + \hb{3} \p{13}  \}  
  +\h{2} \{  \hb{1} \pb{12} + \hb{3} \pb{23} \}   
  +\h{3} \{  \hb{1} \pb{13} + \hb{2} \p{23}  \}$\\   
&$\pl+\hb{2} \V{46} \Vs{44} 
  +\hb{3} \{ \H{41} \Hs{38} + \V{51} \Vs{48} \}\left.\right]$\\    
$+ g_{s}$ 
&$\left[\right.\ppst \V{45} \V{46} \p{4} + \V{51} \V{52} \pp{4} \left.\right]$ 
\end{tabular}
\nolabel\\
\label{w3ssm}
\eeqn

\no $W_4(\obs)$:
\beqn
\begin{tabular}{lllll}
  $\pps\Q{1}\dc{3} \H{41} \Hs{21}$   
 &$+\Q{1}  \L{3}  \H{40}  \Hs{21}$   
 &$+\Q{3}  \Q{3}  \H{33}  \Hs{30}$   
 &$+\Q{3}  \uc{3} \H{34}  \Hs{30}$   
 &$+\Q{3}  \dc{1} \H{41}  \Hs{19}$\\    
  $+\uc{3} \dc{1} \H{40}  \Hs{19}$   
 &$+\uc{3} \ec{3} \H{33}  \Hs{30}$   
 &$+\L{1}  \V{52} \Hs{7}  \Hs{19}$
 &$+\V{45} \V{46} \Vs{47} \Vs{48}$   
 &$+\V{45} \V{46} \Vs{49} \Vs{50}$\\   
  $+\V{45} \V{51} \Vs{41} \Vs{50}$   
 &$+\V{46} \V{52} \Vs{42} \Vs{49}$ 
 &$+\V{51} \V{52} \Hs{3}  \Hs{4} $   
 &$+\V{51} \V{52} \Hs{5}  \Hs{6} $   
 &$+\V{51} \V{52} \Vs{41} \Vs{42}$\\   
  $+\V{51} \V{52} \Vs{43} \Vs{44}$\\  
\end{tabular}
\nolabel\\
\label{w4ssm}
\eeqn

\no $W_5(\obs)$:
\beqn
\begin{tabular}{llll}
   $\pps\Q{1 }  \Q{2 }  \uc{1}  \dc{2}  \p{12}  $  
  &$+ \Q{1 }  \Q{3 }  \uc{1}  \dc{3}  \p{13}  $  
  &$+ \Q{1 }  \uc{1}  \L{2 }  \ec{2}  \p{12}  $  
  &$+ \Q{1 }  \uc{1}  \L{3 }  \ec{3}  \p{13}  $\\ 
   $+ \Q{1 }  \uc{1} \bh{3 } \sH{29} \sH{30}  $  
  &$+ \Q{1 }  \uc{1}  \H{34}  \p{12} \sH{31}  $  
  &$+ \Q{1 }  \uc{1}  \V{45}  \p{12} \sV{43}  $  
  &$+ \Q{1 }  \uc{1}  \V{52}  \p{13} \sV{47}  $\\ 
   $+ \Q{1 }  \uc{2}  \L{2 }  \ec{1}  \p{12}  $  
  &$+ \Q{1 }  \uc{2} \bh{3 } \sH{15} \sH{30}  $  
  &$+ \Q{1 }  \uc{3}  \L{3 }  \ec{1}  \p{13}  $  
  &$+ \Q{2 }  \uc{1}  \H{34}  \p{12} \sH{16}  $\\ 
   $+ \Q{2 }  \dc{1}  \L{1 }  \Nc{2}  \p{12}  $  
  &$+ \Q{2 }  \dc{2}  \L{1 }  \Nc{1}  \p{12}  $  
  &$+ \Q{2 }  \dc{2}  \h{3 } \sV{31} \sV{32}  $  
  &$+ \Q{2 }  \dc{2}  \H{41}  \p{23} \sH{38}  $\\ 
   $+ \Q{2 }  \dc{2}  \V{51}  \p{23} \sV{48}  $  
  &$+ \Q{2 }  \L{2 }  \H{40}  \p{23} \sH{38}  $  
  &$+ \Q{3 }  \uc{3}  \H{34}  \P{1 } \sH{30}  $  
  &$+ \Q{3 }  \uc{3}  \V{46} \bp{23} \sV{41}  $\\ 
   $+ \Q{3 }  \dc{1}  \L{1 }  \Nc{3}  \p{13}  $  
  &$+ \Q{3 }  \dc{1}  \H{41}  \P{2 } \sH{19}  $  
  &$+ \Q{3 }  \dc{3}  \L{1 }  \Nc{1}  \p{13}  $  
  &$+ \Q{3 }  \dc{3}  \h{1 } \sH{29} \sH{30}  $\\ 
   $+ \Q{3 }  \dc{3}  \V{46} \bp{23} \sV{44}  $  
  &$+ \Q{3 }  \L{1 }  \H{40} \bp{4 } \sH{19}  $  
  &$+ \uc{1}  \dc{3}  \H{40}  \p{4 } \sH{21}  $  
  &$+ \uc{1}  \ec{2}  \H{33}  \p{12} \sH{16}  $\\ 
   $+ \uc{2}  \dc{2}  \H{40}  \p{23} \sH{38}  $  
  &$+ \uc{2}  \ec{1}  \H{33}  \p{12} \sH{16}  $  
  &$+ \uc{3}  \dc{1}  \H{40}  \P{2 } \sH{19}  $  
  &$+ \dc{2}  \H{33}  \pp{4} \sH{21} \sV{32}  $\\
\end{tabular}
\nolabel\\
\label{w5ssm}
\eeqn
\newpage

\no $W_6(\obs)$:
\beqn
\begin{tabular}{lll}
  $\pps\Q{1 }\Q{1 }  \Q{2 }  \L{2 }  \p{12} \bpp{4} $               
 &$+ \Q{1 }  \Q{1 }  \Q{3 }  \L{3 }  \p{13} \bp{4 } $               
 &$+ \Q{1 }  \Q{1 }  \uc{2}  \dc{2}  \p{12} \bpp{4} $\\               
  $+ \Q{1 }  \Q{1 }  \uc{3}  \dc{3}  \p{13} \bp{4 } $               
 &$+ \Q{1 }  \Q{1 }  \H{33}  \p{12} \bpp{4} \sH{31} $               
 &$+ \Q{1 }  \Q{2 }  \Q{3 }  \L{3 } \sH{15} \sH{30} $\\               
  $+ \Q{1 }  \Q{2 }  \uc{3}  \dc{3} \sH{15} \sH{30} $               
 &$+ \Q{1 }  \Q{2 }  \H{33}  \p{12} \bpp{4} \sH{16} $               
 &$+ \Q{1 }  \Q{3 }  \uc{1}  \dc{3} \sH{29} \sH{30} $\\               
  $+ \Q{1 }  \Q{3 }  \uc{2}  \dc{3} \sH{15} \sH{30} $               
 &$+ \Q{1 }  \uc{1}  \L{3 }  \ec{3} \sH{29} \sH{30} $               
 &$+ \Q{1 }  \uc{1}  \V{46} \sH{29} \sH{31} \sV{41} $\\               
  $+ \Q{1 }  \uc{1}  \V{52} \sH{29} \sH{30} \sV{47} $               
 &$+ \Q{1 }  \uc{2}  \L{3 }  \ec{3} \sH{15} \sH{30} $               
 &$+ \Q{1 }  \uc{2} \bh{3 }  \P{3 } \sH{15} \sH{30} $\\               
  $+ \Q{1 }  \uc{2}  \H{34}  \p{12} \bpp{4} \sH{16} $               
 &$+ \Q{1 }  \uc{2}  \V{46} \sH{15} \sH{31} \sV{41} $               
 &$+ \Q{1 }  \uc{2}  \V{52} \sH{7 } \sH{30} \sV{31} $\\               
  $+ \Q{1 }  \uc{2}  \V{52} \sH{15} \sH{30} \sV{47} $               
 &$+ \Q{1 }  \uc{3}  \L{3 }  \ec{1} \sH{29} \sH{30} $               
 &$+ \Q{1 }  \uc{3}  \L{3 }  \ec{2} \sH{15} \sH{30} $\\               
  $+ \Q{1 }  \dc{1}  \L{2 } \sH{19} \sH{29} \sV{32} $               
 &$+ \Q{1 }  \dc{1}  \H{41} \sH{3 } \sH{8 } \sH{29} $               
 &$+ \Q{1 }  \dc{2}  \L{3 } \sH{29} \sH{38} \sV{32} $\\
\end{tabular}
\nolabel
\eeqn
\newpage

\no $W_6(\obs)$ continued:
\beqn
\begin{tabular}{lll}
  $+ \uc{1}  \uc{3}  \H{40} \sH{21} \sV{41} \sV{43} $               
 &$+ \uc{1}  \dc{1}  \dc{3} \sH{3 } \sH{29} \sV{48} $               
 &$+ \uc{1}  \dc{2}  \dc{2} \sH{17} \sH{19} \sV{32} $\\               
  $+ \uc{1}  \dc{2}  \dc{3} \sH{7 } \sH{19} \sV{44} $               
 &$+ \uc{1}  \dc{2}  \H{40} \sH{3 } \sH{8 } \sH{17} $               
 &$+ \uc{1}  \dc{3}  \H{40} \sH{3 } \sH{4 } \sH{21} $\\               
  $+ \uc{1}  \dc{3}  \H{40} \sH{5 } \sH{6 } \sH{21} $               
 &$+ \uc{1}  \dc{3}  \H{40} \sH{21} \sV{41} \sV{42} $               
 &$+ \uc{1}  \dc{3}  \H{40} \sH{21} \sV{43} \sV{44} $\\               
  $+ \uc{1}  \dc{3}  \H{40} \sH{21} \sV{47} \sV{48} $               
 &$+ \uc{1}  \dc{3}  \H{40} \sH{21} \sV{49} \sV{50} $               
 &$+ \uc{1}  \ec{1}  \H{33}  \p{12}  \pp{4} \sH{31} $\\               
  $+ \uc{1}  \ec{2}  \H{33}  \P{3 }  \p{12} \sH{16} $               
 &$+ \uc{1}  \H{33} \sH{3 } \sH{15} \sH{31} \sV{41} $               
 &$+ \uc{1}  \H{40}  \H{40} \sH{18} \sH{19} \sV{31} $\\               
  $+ \uc{2}  \uc{3}  \ec{1}  \H{40} \sH{30} \sV{31} $               
 &$+ \uc{2}  \dc{1}  \dc{1}  \Nc{2}  \p{12}  \pp{4} $               
 &$+ \uc{2}  \dc{1}  \dc{2}  \Nc{1}  \p{12}  \pp{4} $\\               
  $+ \uc{2}  \dc{1}  \dc{3} \sH{3 } \sH{15} \sV{48} $               
 &$+ \uc{2}  \dc{1}  \H{40} \sH{21} \sV{41} \sV{50} $               
 &$+ \uc{2}  \dc{2}  \H{40}  \P{1 }  \p{23} \sH{38} $\\               
  $+ \uc{2}  \dc{2}  \H{40} \sH{38} \sV{31} \sV{32} $               
 &$+ \uc{2}  \ec{1}  \H{33}  \P{3 }  \p{12} \sH{16} $               
 &$+ \uc{3}  \dc{1}  \dc{1}  \Nc{3}  \p{13}  \p{4 } $\\
\end{tabular}
\nolabel
\eeqn
\newpage

\no $W_6(\obs)$ continued:
\beqn
\begin{tabular}{lll}
  $+ \L{2 }  \V{46} \sH{5 } \sH{31} \sH{38} \sH{39} $               
 &$+ \L{2 }  \V{46} \sH{9 } \sH{16} \sH{17} \sH{39} $               
 &$+ \L{2 }  \V{46} \sH{17} \sH{37} \sV{32} \sV{49} $\\               
  $+ \L{2 }  \V{51} \sH{5 } \sH{19} \sV{21} \sV{32} $               
 &$+ \L{2 }  \V{51} \sH{5 } \sH{38} \sV{1 } \sV{32} $               
 &$+ \L{2 }  \V{51} \sH{17} \sH{37} \sV{32} \sV{41} $\\               
  $+ \L{2 }  \V{52} \sH{3 } \sH{15} \sH{19} \sH{36} $               
 &$+ \L{2 }  \V{52} \sH{17} \sH{38} \sV{32} \sV{43} $               
 &$+ \L{3 }  \ec{1}  \h{2 }  \p{13} \sH{22} \sH{37} $\\               
  $+ \L{3 }  \ec{1}  \H{41} \sH{3 } \sH{4 } \sH{21} $               
 &$+ \L{3 }  \ec{1}  \H{41} \sH{5 } \sH{6 } \sH{21} $               
 &$+ \L{3 }  \ec{1}  \H{41} \sH{21} \sV{41} \sV{42} $\\               
  $+ \L{3 }  \ec{1}  \H{41} \sH{21} \sV{43} \sV{44} $               
 &$+ \L{3 }  \ec{1}  \H{41} \sH{21} \sV{47} \sV{48} $               
 &$+ \L{3 }  \ec{1}  \H{41} \sH{21} \sV{49} \sV{50} $\\               
  $+ \L{3 }  \ec{1}  \V{45} \sH{17} \sH{30} \sV{50} $               
 &$+ \L{3 }  \ec{1}  \V{46} \sH{17} \sH{31} \sV{48} $               
 &$+ \L{3 }  \ec{1}  \V{52} \sH{17} \sH{30} \sV{42} $\\               
  $+ \L{3 }  \Nc{1}  \V{46} \sH{16} \sH{29} \sV{49} $               
 &$+ \L{3 }  \Nc{1}  \V{51} \sH{16} \sH{29} \sV{41} $               
 &$+ \L{3 }  \Nc{3}  \H{34}  \p{56} \bi{56} \sH{30} $\\               
  $+ \L{3 }  \Nc{3}  \V{51} \sH{15} \sH{16} \sV{49} $               
 &$+ \L{3 } \bh{2 }  \p{4 } \sH{15} \sH{38} \sV{32} $               
 &$+ \L{3 }  \H{34} \sH{3 } \sH{15} \sH{18} \sV{48} $\\
\end{tabular}
\nolabel
\eeqn
\newpage

\no $W_6(\obs)$ continued:
\beqn
\begin{tabular}{lll}
  $+ \V{45}  \V{51}  \p{23} \bp{56} \sV{42} \sV{49} $               
 &$+ \V{45}  \V{51}  \p{56} \bp{56} \sV{41} \sV{50} $               
 &$+ \V{45}  \V{51} \pp{56} \bi{56} \sV{41} \sV{50} $\\               
  $+ \V{45}  \V{51} \sH{5 } \sH{30} \sV{11} \sV{50} $               
 &$+ \V{45}  \V{51} \sV{31} \sV{32} \sV{43} \sV{48} $               
 &$+ \V{46}  \V{51} \sH{5 } \sH{18} \sV{1 } \sV{48} $\\               
  $+ \V{46}  \V{51} \sH{5 } \sH{31} \sV{11} \sV{48} $               
 &$+ \V{46}  \V{52}  \P{1 }  \P{1 } \sV{42} \sV{49} $               
 &$+ \V{46}  \V{52}  \P{1 }  \p{56} \sV{43} \sV{48} $\\               
  $+ \V{46}  \V{52}  \p{23} \bp{23} \sV{42} \sV{49} $               
 &$+ \V{46}  \V{52} \bp{23}  \p{56} \sV{41} \sV{50} $               
 &$+ \V{46}  \V{52}  \p{56} \bp{56} \sV{42} \sV{49} $\\               
  $+ \V{46}  \V{52} \pp{56} \bi{56} \sV{42} \sV{49} $               
 &$+ \V{46}  \V{52} \sH{6 } \sH{29} \sV{12} \sV{49} $               
 &$+ \V{46}  \V{52} \sH{16} \sH{19} \sH{29} \sH{36} $\\               
  $+ \V{46}  \V{52} \sH{17} \sH{18} \sV{43} \sV{48} $               
 &$+ \V{46}  \V{52} \sH{17} \sH{22} \sH{30} \sH{37} $               
 &$+ \V{46}  \V{52} \sH{21} \sH{22} \sV{43} \sV{48} $\\               
  $+ \V{51}  \V{52}  \P{1 }  \P{1 } \sH{3 } \sH{4 } $               
 &$+ \V{51}  \V{52}  \P{1 }  \P{1 } \sH{5 } \sH{6 } $               
 &$+ \V{51}  \V{52}  \P{1 }  \P{1 } \sV{41} \sV{42} $\\               
  $+ \V{51}  \V{52}  \P{1 }  \P{1 } \sV{43} \sV{44} $               
 &$+ \V{51}  \V{52}  \p{23} \bp{23} \sH{3 } \sH{4 } $               
 &$+ \V{51}  \V{52}  \p{23} \bp{23} \sH{5 } \sH{6 } $\\
\end{tabular}
\label{w6ssm}
\eeqn
\newpage


\no Superpotential terms containing both 
observable sector $SU(3)_C\times SU(2)_L$--charged fields
and non-Abelian hidden sector fields.  
Coupling constants are implicit for non-renormalizable terms.

\vskip .5truecm
\no $W_4(\mix)$:
\beqn
\begin{tabular}{lllll}
  $ \H{41} \V{46} \H{11} \V{17}$   
 &$+\V{45} \V{46 }\H{11} \H{13}$
 &\phantom{$+\V{45} \V{46 }\H{11} \H{13}$}
 &\phantom{$+\V{45} \V{46 }\H{11} \H{13}$}
 &\phantom{$+\V{45} \V{46 }\H{11} \H{13}$}\\ 
\end{tabular}
\nolabel\\
\label{w4mix}
\eeqn

\no $W_5(\mix)$:
\beqn
\begin{tabular}{llll}
   $\pps\Q{1 }\h{2 }  \H{40}  \H{28}  \V{9 }  $  
  &$+ \Q{3 }  \dc{3}  \h{2 }  \V{34}  \V{33}  $  
  &$+ \dc{1}  \H{33}  \p{12}  \H{23}  \V{7 }  $  
  &$+ \L{1 }  \h{2 } \sH{9 }  \H{11}  \V{5 }  $\\ 
   $+ \L{1 }  \h{3 } \sH{5 }  \H{1 }  \V{9 }  $  
  &$+ \L{1 }  \H{34}  \p{12}  \H{25}  \V{9 }  $  
  &$+ \L{2 } \bh{1 } \sH{38}  \H{42}  \V{13}  $ 
  &$+ \L{2 } \bh{3 } \sH{19}  \H{23}  \V{37}  $\\  
   $+ \L{3 }  \ec{3}  \h{2 }  \V{34}  \V{33}  $  
  &$+ \ec{1}  \h{2 }  \H{41}  \H{26}  \V{7 }  $  
  &$+ \ec{1}  \p{12} \sH{6 }  \H{2 }  \V{10}  $ 
  &$+ \ec{1}  \p{13} \sH{10}  \H{13}  \V{7 }  $\\  
   $+ \h{2 }  \V{52} \sV{47}  \V{34}  \V{33}  $  
  &$+ \h{3 }  \V{46} \sH{16}  \H{1 }  \V{10}  $ 
  &$+\bh{1 }  \H{41} \sH{38}  \V{14}  \V{13}  $  
  &$+\bh{1 }  \H{41} \sH{38}  \V{19}  \V{20}  $\\  
   $+\bh{1 }  \H{41} \sH{38}  \V{15}  \V{17}  $  
  &$+\bh{1 }  \V{45} \sH{38}  \H{13}  \V{15}  $ 
  &$+\bh{1 }  \V{46} \sV{44}  \V{24}  \V{23}  $  
  &$+\bh{1 }  \V{46} \sV{44}  \V{29}  \V{30}  $\\  
   $+\bh{1 }  \V{46} \sV{44}  \V{25}  \V{27}  $  
  &$+\bh{1 }  \V{46} \sV{48}  \V{20}  \V{29}  $ 
  &$+\bh{1 }  \V{46} \sV{48}  \V{17}  \V{25}  $  
  &$+\bh{1 }  \V{51} \sV{44}  \V{19}  \V{30}  $\\  
   $+\bh{1 }  \V{51} \sV{44}  \V{15}  \V{27}  $  
  &$+\bh{1 }  \V{51} \sV{48}  \V{14}  \V{13}  $ 
  &$+\bh{1 }  \V{51} \sV{48}  \V{19}  \V{20}  $  
  &$+\bh{1 }  \V{51} \sV{48}  \V{15}  \V{17}  $\\  
   $+\bh{3 }  \V{45} \sH{19}  \H{2 }  \V{40}  $  
  &$+\bh{3 }  \V{46} \sV{44}  \V{34}  \V{33}  $ 
  &$+ \V{45}  \V{46}  \P{1 }  \H{11}  \H{13}  $\\  
\end{tabular}
\nolabel\\
\label{w5mix}
\eeqn

\no $W_6(\mix)$: 
\beqn
\begin{tabular}{lll}
  $\pps\Q{1} \uc{3}  \L{2 } \sH{9 }  \H{1 }  \V{40} $               
 &$+ \Q{1 }  \uc{3}  \V{46} \sH{30}  \H{11}  \V{35} $               
 &$+ \Q{1 }  \dc{2}  \h{3 }  \p{12}  \V{5 }  \V{17} $\\               
  $+ \Q{1 }  \dc{3}  \h{2 }  \p{13}  \V{9 }  \V{29} $               
 &$+ \Q{1 }  \h{2 }  \H{40} \bp{4 }  \H{26}  \V{7 } $               
 &$+ \Q{1 }  \H{40}  \V{46} \sV{44}  \H{28}  \V{9 } $\\               
  $+ \Q{1 }  \H{40}  \V{52} \sH{21}  \H{13}  \H{26} $               
 &$+ \Q{2 }  \dc{1}  \h{3 }  \p{12}  \V{7 }  \V{15} $               
 &$+ \Q{3 }  \uc{3}  \V{46} \sV{41}  \V{34}  \V{33} $\\               
  $+ \Q{3 }  \dc{1}  \h{2 }  \p{13}  \V{10}  \V{30} $               
 &$+ \Q{3 }  \dc{2}  \L{3 } \sH{19}  \H{23}  \V{37} $               
 &$+ \Q{3 }  \dc{3}  \L{2 } \sH{19}  \H{23}  \V{37} $\\               
  $+ \Q{3 }  \dc{3}  \V{45} \sH{19}  \H{2 }  \V{40} $               
 &$+ \Q{3 }  \dc{3}  \V{46} \sV{44}  \V{34}  \V{33} $               
 &$+ \Q{3 }  \L{1 }  \H{40} \sH{19}  \H{1 }  \H{2 } $\\               
  $+ \Q{3 }  \H{40}  \V{46} \sH{6 }  \V{7 }  \V{37} $               
 &$+ \Q{3 }  \H{40}  \V{46} \sV{44}  \H{28}  \V{30} $               
 &$+ \Q{3 }  \H{40}  \V{46} \sV{44}  \H{26}  \V{27} $\\               
  $+ \Q{3 }  \H{40}  \V{46} \sV{48}  \H{28}  \V{20} $               
 &$+ \Q{3 }  \H{40}  \V{46} \sV{48}  \H{26}  \V{17} $               
 &$+ \uc{1}  \dc{3}  \H{40} \sH{21}  \H{11}  \H{13} $\\               
  $+ \uc{3}  \dc{2}  \dc{3} \sH{19}  \H{23}  \V{37} $               
 &$+ \uc{3}  \H{33} \sH{9 } \sH{16}  \H{1 }  \V{40} $               
 &$+ \dc{1}  \H{33}  \p{12}  \pp{4}  \H{25}  \V{9 } $\\
\end{tabular}
\nolabel
\eeqn
\newpage

\no $W_6(\mix)$ continued: 
\beqn
\begin{tabular}{lll}
  $+ \Nc{1} \sH{30}  \V{4 }  \H{35}  \H{11}  \H{13} $               
 &$+ \Nc{1} \sH{31} \sV{41} \sV{50}  \H{26}  \V{37} $               
 &$+ \Nc{2} \sH{3 } \sH{10} \sH{30}  \H{28}  \V{19} $\\               
  $+ \Nc{2} \sH{3 } \sH{10} \sV{41}  \H{2 }  \H{28} $               
 &$+ \Nc{2} \sH{5 } \sH{10} \sH{30}  \H{28}  \V{20} $               
 &$+ \Nc{2} \sH{5 } \sH{10} \sH{30}  \H{26}  \V{17} $\\               
  $+ \Nc{2} \sH{8 } \sH{32} \sV{41}  \H{23}  \H{26} $               
 &$+ \Nc{2} \sH{8 } \sH{32} \sV{49}  \V{9 }  \V{40} $               
 &$+ \Nc{2} \sH{10} \sH{30} \sV{47}  \V{7 }  \V{37} $\\               
  $+ \Nc{2} \sH{15} \sH{18} \sV{41}  \H{13}  \H{26} $               
 &$+ \Nc{2} \sH{18} \sV{41} \sV{50}  \H{26}  \V{37} $               
 &$+ \Nc{2} \sH{22} \sH{30} \sV{31}  \H{1 }  \H{2 } $\\               
  $+ \Nc{3}  \h{1 } \bh{1 } \sH{30}  \V{24}  \H{35} $               
 &$+ \Nc{3}  \h{2 } \bh{2 } \sH{16}  \H{26}  \V{37} $               
 &$+ \Nc{3}  \V{45}  \V{46} \sH{16}  \H{28}  \V{40} $\\               
  $+ \Nc{3}  \V{51}  \V{52} \sH{16}  \H{28}  \V{40} $               
 &$+ \Nc{3}  \P{1 }  \P{1 } \sH{30}  \V{24}  \H{35} $               
 &$+ \Nc{3}  \P{2 }  \P{2 } \sH{16}  \H{26}  \V{37} $\\               
  $+ \Nc{3}  \P{2 } \bp{4 } \sH{16}  \H{28}  \V{40} $               
 &$+ \Nc{3}  \p{13} \bp{13} \sH{16}  \H{26}  \V{37} $               
 &$+ \Nc{3}  \p{23} \bp{23} \sH{30}  \V{24}  \H{35} $\\               
  $+ \Nc{3}  \p{4 } \bp{4 } \sH{16}  \H{26}  \V{37} $               
 &$+ \Nc{3}  \p{56} \bp{56} \sH{30}  \V{24}  \H{35} $               
 &$+ \Nc{3} \pp{56} \bi{56} \sH{30}  \V{24}  \H{35} $\\
\end{tabular}
\label{w6mix}
\eeqn
\newpage

\no Superpotential terms containing non-Abelian hidden sector fields. 
Coupling constants are implicit for non-renormalizable terms.


\vskip .5truecm
\no $W_3(\hid)$:
\beqn
\begin{tabular}{rl}
$\pps  g^{'}_{s}$  
&$\left[\right.   
 \p{12} \{ \V{23} \V{24} + \V{29} \V{30} + \V{25} \V{27} \}
+\p{13} \{ \V{13} \V{14} + \V{19} \V{20} + \V{15} \V{17} \}      
+\p{23} \V{33} \V{34} \left.\right]$\\
$+ g_{s}$  
&$\left[\right.\p{4} \H{1} \H{2} + \pbp{4} \H{11} \H{13}\left.\right]$\\   
\end{tabular}
\nolabel\\
\label{w3hid}
\eeqn

\no $W_4(\hid)$:
\beqn
\begin{tabular}{lllll}
$\pps\Nc{3} \V{24}  \H{35 }  \sH{30} $
&$+ \Nc{3}  \H{26}  \V{37 }  \sH{16} $
&$+ \H{42}  \H{35}  \sH{30}  \sH{37} $
&$+ \V{4 }  \V{13}  \sH{17}  \sH{30} $
&$+ \V{14}  \V{3 }  \sH{15}  \sH{30} $\\
 $+ \V{34}  \H{35}  \sH{19}  \sV{12} $
&$+ \V{34}  \V{13}  \sH{19}  \sH{32} $
&$+ \H{1 }  \H{2 }  \H{11 }  \H{13}  $
&$+ \H{1 }  \H{2 }  \sV{47}  \sV{48} $
&$+ \H{1 }  \H{2 }  \sV{49}  \sV{50} $\\
 $+ \H{1 }  \V{29}  \sH{30}  \sV{50} $
&$+ \H{1 }  \V{30}  \sH{32}  \sV{48} $
&$+ \H{2 }  \V{9 }  \sH{17}  \sV{41} $
&$+ \H{2 }  \V{39}  \sH{21}  \sV{41} $
&$+ \H{28}  \V{40}  \sH{37}  \sV{32} $\\
 $+ \V{9 }  \V{19}  \sH{17}  \sH{30} $
&$+ \V{10}  \V{20}  \sH{15}  \sH{30} $
&$+ \V{19}  \V{39}  \sH{21}  \sH{30} $
&$+ \H{23}  \H{26}  \sH{5 }  \sH{8}  $
&$+ \H{23}  \V{35}  \sH{30}  \sV{31} $\\
 $+ \H{26}  \V{15}  \sH{9 }  \sV{44} $
&$+ \V{15}  \V{37}  \sH{19}  \sH{31} $\\
\end{tabular}
\nolabel\\
\label{w4hid}
\eeqn

\no $W_5(\hid)$:
\beqn
\begin{tabular}{llll}
   $\pps\Nc{1}  \V{4 }  \H{35}  \pp{4} \sH{30}$  
  &$+ \Nc{3}  \V{24}  \H{35}  \P{1 } \sH{30}  $  
  &$+ \Nc{3}  \H{28}  \V{40} \bp{4 } \sH{16}  $  
  &$+ \Nc{3}  \H{26}  \V{37}  \P{2 } \sH{16}  $\\ 
   $+ \H{42}  \H{35}  \P{1 } \sH{30} \sH{37}  $  
  &$+ \H{42}  \H{35} \bp{56} \sH{31} \sH{38}  $  
  &$+ \V{14}  \V{34}  \V{13}  \V{33}  \p{12}  $  
  &$+ \V{14}  \V{3 }  \P{3 } \sH{15} \sH{30}  $\\ 
   $+ \V{24}  \V{23}  \p{13} \sV{31} \sV{32}  $  
  &$+ \V{24}  \V{23}  \p{23} \sH{29} \sH{30}  $  
  &$+ \V{34}  \V{33}  \V{19}  \V{20}  \p{12}  $  
  &$+ \V{34}  \V{33}  \V{15}  \V{17}  \p{12}  $\\ 
   $+ \V{34}  \V{33}  \p{12} \sV{11} \sV{12}  $  
  &$+ \H{35}  \V{13}  \V{33} \bpp{4} \sH{19}  $  
  &$+ \H{1 }  \H{2 }  \H{11}  \H{13}  \P{1 }  $  
  &$+ \H{1 }  \H{2 }  \P{1 } \sV{47} \sV{48}  $\\ 
   $+ \H{1 }  \H{2 }  \P{1 } \sV{49} \sV{50}  $  
  &$+ \H{2 }  \V{9 }  \P{3 } \sH{17} \sV{41}  $ 
  &$+ \H{2 }  \V{30}  \p{23} \sH{29} \sV{49}  $  
  &$+ \H{2 }  \V{39}  \P{3 } \sH{21} \sV{41}  $\\ 
   $+ \H{25}  \H{28} \bp{56} \sH{3 } \sH{10}  $  
  &$+ \H{25}  \V{39} \bpp{4} \sH{30} \sV{31}  $  
  &$+ \H{28}  \V{20} \bp{56} \sH{7 } \sV{44}  $  
  &$+ \H{28}  \V{30}  \p{4 } \sH{15} \sV{32}  $\\ 
   $+ \H{28}  \V{40}  \P{2 } \sH{37} \sV{32}  $  
  &$+ \V{9 }  \V{19}  \P{3 } \sH{17} \sH{30}  $  
  &$+ \V{19}  \V{39}  \P{3 } \sH{21} \sH{30}  $ 
  &$+ \V{19}  \V{40} \bpp{4} \sH{19} \sH{31}  $\\  
   $+ \V{29}  \V{30}  \p{13} \sV{31} \sV{32}  $  
  &$+ \V{29}  \V{30}  \p{23} \sH{29} \sH{30}  $  
  &$+ \H{23}  \H{26}  \p{23} \sH{4 } \sH{9 }  $ 
  &$+ \H{23}  \V{35}  \P{3 } \sH{30} \sV{31}  $\\
\end{tabular}
\nolabel\\
\label{w5hid}
\eeqn

\no $W_6(\hid)$:
\beqn
\begin{tabular}{lll}
  $\pps\H{42}\H{35}  \P{1 }  \P{1 } \sH{30} \sH{37} $         
 &$+ \H{42}  \H{35}  \p{23} \bp{23} \sH{30} \sH{37} $         
 &$+ \H{42}  \H{35}  \p{56} \bp{56} \sH{30} \sH{37} $\\         
  $+ \H{42}  \H{35} \pp{56} \bi{56} \sH{30} \sH{37} $         
 &$+ \H{42}  \H{35} \sH{9 } \sH{30} \sV{21} \sV{50} $         
 &$+ \H{42}  \V{13}  \H{26}  \V{7 } \sH{15} \sH{30} $\\         
  $+ \H{42}  \V{23}  \p{13} \bp{4 } \sH{16} \sV{31} $         
 &$+ \H{42}  \V{23} \sH{15} \sH{30} \sH{31} \sV{31} $         
 &$+ \V{4 }  \H{35}  \V{20}  \V{40} \sH{9 } \sV{48} $\\         
  $+ \V{4 }  \H{35}  \V{30}  \V{40} \sH{9 } \sV{44} $         
 &$+ \V{4 }  \H{35} \sH{38} \sV{32} \sV{43} \sV{48} $         
 &$+ \V{4 }  \V{3 }  \H{2 }  \V{40} \sH{21} \sV{41} $\\         
  $+ \V{4 }  \V{3 }  \V{19}  \V{40} \sH{21} \sH{30} $         
 &$+ \V{4 }  \V{3 }  \H{23}  \V{37} \sH{30} \sV{31} $         
 &$+ \V{4 }  \V{3 } \sH{5 } \sH{10} \sH{30} \sH{38} $\\         
  $+ \V{4 }  \V{13}  \V{10}  \V{40} \sH{21} \sH{30} $         
 &$+ \V{4 }  \V{13}  \H{26}  \V{37} \sH{3 } \sV{48} $         
 &$+ \V{4 }  \V{13}  \P{3 }  \P{3 } \sH{17} \sH{30} $\\         
  $+ \V{4 }  \V{13}  \p{12} \bp{12} \sH{17} \sH{30} $         
 &$+ \V{4 }  \V{13}  \pp{4} \bpp{4} \sH{17} \sH{30} $         
 &$+ \V{4 }  \V{13} \sH{15} \sH{19} \sH{22} \sH{30} $\\         
  $+ \V{4 }  \V{23} \sH{7 } \sH{30} \sV{31} \sV{44} $         
 &$+ \V{4 }  \V{23} \sH{15} \sH{30} \sV{44} \sV{47} $         
 &$+ \V{4 }  \V{33} \sH{6 } \sH{15} \sV{32} \sV{41} $\\
\end{tabular}
\nolabel
\eeqn

\no $W_6(\hid)$ continued:
\beqn
\begin{tabular}{lll}
  $+ \H{25}  \V{9 }  \H{26}  \V{37} \sV{41} \sV{50} $         
 &$+ \H{25}  \V{39}  \P{3 } \bpp{4} \sH{30} \sV{31} $         
 &$+ \H{25}  \V{39} \sH{7 } \sH{8 } \sH{30} \sV{31} $\\         
  $+ \H{25}  \V{39} \sH{8 } \sH{15} \sH{30} \sV{47} $         
 &$+ \H{25}  \V{39} \sH{9 } \sH{10} \sH{30} \sV{31} $         
 &$+ \H{28}  \H{28}  \H{23}  \H{23}  \p{23} \bp{56} $\\         
  $+ \H{28}  \V{9 }  \H{13}  \H{23} \sH{15} \sV{41} $         
 &$+ \H{28}  \V{9 }  \H{23}  \V{37} \sV{41} \sV{50} $         
 &$+ \H{28}  \V{9 }  \V{15}  \V{37} \sH{3 } \sV{48} $\\         
  $+ \H{28}  \V{20}  \P{1 } \bp{56} \sH{7 } \sV{44} $         
 &$+ \H{28}  \V{20} \sH{15} \sV{32} \sV{43} \sV{48} $         
 &$+ \H{28}  \V{30}  \H{11}  \H{13} \sH{15} \sV{32} $\\         
  $+ \H{28}  \V{30}  \H{11}  \V{37} \sV{32} \sV{50} $         
 &$+ \H{28}  \V{30}  \P{2 }  \p{4 } \sH{15} \sV{32} $         
 &$+ \H{28}  \V{30} \sH{3 } \sH{4 } \sH{15} \sV{32} $\\         
  $+ \H{28}  \V{30} \sH{5 } \sH{6 } \sH{15} \sV{32} $         
 &$+ \H{28}  \V{30} \sH{7 } \sV{31} \sV{32} \sV{48} $         
 &$+ \H{28}  \V{30} \sH{15} \sV{32} \sV{41} \sV{42} $\\         
  $+ \H{28}  \V{30} \sH{15} \sV{32} \sV{43} \sV{44} $         
 &$+ \H{28}  \V{30} \sH{15} \sV{32} \sV{47} \sV{48} $         
 &$+ \H{28}  \V{30} \sH{15} \sV{32} \sV{49} \sV{50} $\\         
  $+ \H{28}  \V{40}  \P{2 }  \P{2 } \sH{37} \sV{32} $         
 &$+ \H{28}  \V{40}  \p{13} \bp{13} \sH{37} \sV{32} $         
 &$+ \H{28}  \V{40}  \p{4 } \bp{4 } \sH{37} \sV{32} $\\
\end{tabular}
\nolabel
\eeqn

\no $W_6(\hid)$ continued:
\beqn
\begin{tabular}{lll}
  $+ \V{5 }  \V{17} \sH{15} \sH{30} \sV{47} \sV{48} $         
 &$+ \V{5 }  \V{17} \sH{15} \sH{30} \sV{49} \sV{50} $         
 &$+ \V{5 }  \V{25} \sH{9 } \sH{30} \sV{31} \sV{44} $\\         
  $+ \V{5 }  \V{27} \sH{7 } \sH{30} \sV{31} \sV{44} $         
 &$+ \V{5 }  \V{27} \sH{15} \sH{30} \sV{44} \sV{47} $         
 &$+ \V{5 }  \V{37} \sH{30} \sH{37} \sV{44} \sV{47} $\\         
  $+ \V{5 }  \V{37} \sH{31} \sH{38} \sV{41} \sV{50} $         
 &$+ \V{7 }  \V{15}  \P{3 } \bpp{4} \sH{17} \sH{30} $         
 &$+ \V{7 }  \V{15} \sH{7 } \sH{8 } \sH{17} \sH{30} $\\         
  $+ \V{7 }  \V{15} \sH{9 } \sH{10} \sH{17} \sH{30} $         
 &$+ \V{7 }  \V{37} \sH{6 } \sH{15} \sV{32} \sV{43} $         
 &$+ \V{7 }  \V{37} \sH{7 } \sH{30} \sV{11} \sV{50} $\\         
  $+ \V{15}  \V{35}  \P{3 }  \pp{4} \sH{21} \sH{30} $         
 &$+ \V{15}  \V{35} \sH{3 } \sH{4 } \sH{21} \sH{30} $         
 &$+ \V{15}  \V{35} \sH{5 } \sH{6 } \sH{21} \sH{30} $\\         
  $+ \V{15}  \V{35} \sH{21} \sH{30} \sV{41} \sV{42} $         
 &$+ \V{15}  \V{35} \sH{21} \sH{30} \sV{43} \sV{44} $         
 &$+ \V{15}  \V{35} \sH{21} \sH{30} \sV{47} \sV{48} $\\         
  $+ \V{15}  \V{35} \sH{21} \sH{30} \sV{49} \sV{50} $        
 &$+ \V{15}  \V{37}  \P{3 }  \P{3 } \sH{19} \sH{31} $         
 &$+ \V{15}  \V{37}  \p{12} \bp{12} \sH{19} \sH{31} $\\         
  $+ \V{15}  \V{37}  \pp{4} \bpp{4} \sH{19} \sH{31} $         
 &$+ \V{17}  \V{37} \sH{7 } \sH{30} \sV{1 } \sV{50} $         
 &$+ \V{25}  \V{27} \sH{29} \sH{30} \sV{31} \sV{32} $\\         
  $+ \V{25}  \V{35} \sH{21} \sH{30} \sV{41} \sV{50} $         
 &$+ \V{35}  \V{37} \sH{3 } \sH{30} \sV{21} \sV{50} $         
\end{tabular}
\label{w6hid}
\eeqn

}

\hfill\vfill\eject

\section{Maximally Orthogonal $D$--flat FNY Basis Set} 

\def\HS#1{$H^s_{#1}$}
\def\VS#1{$V^s_{#1}$}
\def\NC#1{$N^c_{#1}$} 

\noindent

\no Table C.I: Maximally orthogonal basis set of 
singlet VEVs directions that are $D$--flat for all 
non-anomalous $U(1)$ local symmetries and do not break $U(1)_Y$.
The first entry specifies the designation for a flat direction, while the second
entry gives the anomalous charge associated with it. The remaining entries
give the ratios of the norms of the component VEVs in each direction.  
A ``$1-1$'' entry for a $K_j$ flat direction component from field $\Phi_m$ implies that
$\mvev{\Phi_m} = \mvev{\overline{\Phi}_m}$.  
\hfill\vfill\eject

\noindent
\begin{tabular}{|c||c|cccccccccccccc|}
\hline
\hline
$D$--      &$\frac{Q^{(A)}}{112}$&
            \f{12}&\f{13}&\f{23}&\fb{56}&\fp{56}&\fpx{4}&\VS{31}&\VS{32}&\VS{21}&\VS{1}&\VS{2}&\VS{11}&\VS{22}&\VS{12}\\
flat      &&&&&&&&&&&&&&&\\
dir.      &&&&&&&&&&&&&&&\\
\hline
\hline
$L^{'}_{1}$&0&1&-1&-1& 0& 0& 0& 0& 0& 0& 0& 0& 0& 0& 0\\
$L^{'}_{2}$&0&0& 0&-1& 0& 0& 0& 1& 1& 0& 0& 0& 0& 0& 0\\
$L^{'}_{3}$&0&0& 1& 0& 0& 0& 0& 0& 0& 0& 0& 0& 1& 0& 1\\
$L^{'}_{4}$&0&0& 1& 1& 0& 0& 0& 0& 0& 1& 0& 0& 0& 1& 0\\
$L^{'}_{5}$&0&0& 2& 1& 0& 1& 0& 0& 0& 2& 0& 0& 0& 0& 2\\
\hline
\hline
\end{tabular}

\no Table C.II: Maximally orthogonal basis set of non-trivial singlet VEVs 
directions that are $D$--flat for all non-anomalous $U(1)$ local symmetries 
and that break neither $U(1)_Y$ nor $U(1)_{Z'}$. None of these directions
carry anomalous charge $Q^{(A)}$

\hfill\vfill\eject

\section{$F$--flat Directions} 

\def\f#1{$\Phi_{#1}$}
\def\fb#1{$\overline{\Phi}_{#1}$}
\def\fp#1{$\Phi^{'}_{#1}$}
\def\fbp#1{$\overline{\Phi}^{'}_{#1}$}
\def\fpb#1{$\overline{\Phi}^{'}_{#1}$}
\def\fpx#1{$\Phi^{(')}_{#1}$}
\def\fbpx#1{$\overline{\Phi}^{(')}_{#1}$}
\def\bfbpx#1{$\Phi^{(')}_{#1}, \overline{\Phi}^{(')}_{#1}$}
\def\HS#1{$H^s_{#1}$}
\def\VS#1{$V^s_{#1}$}
\def\NC#1{$N^c_{#1}$} 

\def\ify{$\infty$ }
\def\ifw{${\infty}^{\ast}$ }

\def\y{$\ast$}
\def\my{$\bar{\ast}$}
\def\ny{${(-)}\atop{\ast}$}

\begin{flushleft}
\begin{tabular}{|r||r|l|ccccccccccccc|}
\hline 
\hline
$c\#$ &$\#$&$\cal{O}(S)$  &$\{VEV_1\}$&\fb{56}&\fp{56}&\HS{19}&\VS{31}&\HS{20}&\NC{1}&\NC{3}&\HS{21}&\HS{17}&\VS{12}&\HS{18}&\HS{39}\\
      &VEVs& term         &           &       &       &      &      &      &      &      &      &      &      &      &      \\
\hline
 1    & 8+3  & $\infty$   &\y         &\y     &       &      &      &      &      &      &      &      &      &      &      \\ 
 2    & 9+3  & $\infty$   &\y         &\y     &\my    &      &      &      &      &      &      &      &      &      &      \\  
 3    & 9+3  & $\infty$   &\y         &\y     &       &\y    &      &      &      &      &      &      &      &      &      \\   
 4    & 9+3  & 12--1,2    &\y         &\y     &       &      &      &\y    &      &      &      &      &      &      &      \\
 5    &10+3  &    7--1    &\y         &\y     &       &      &\y    &      &\y    &      &      &      &      &      &      \\
 6    &11+3  &    7--1    &\y         &\y     &\my    &      &\y    &      &\y    &      &      &      &      &      &      \\
 7    &11+3  &    7--1    &\y         &\y     &       &\y    &\y    &      &\y    &      &      &      &      &      &      \\ 
 8    &11+3  &    7--1    &\y         &\y     &       &      &\y    &\y    &\y    &      &      &      &      &      &      \\
 9    &10+3  &    6--1    &\y         &\y     &       &      &\y    &      &      &\y    &      &      &      &      &      \\   
10    &11+3  &    6--1    &\y         &\y     &       &      &\y    &      &\y    &\y    &      &      &      &      &      \\   
11    &11+3  &    6--1    &\y         &\y     &       &      &\y    &\y    &      &\y    &      &      &      &      &      \\   
12    &12+3  &    6--1    &\y         &\y     &       &      &\y    &\y    &\y    &\y    &      &      &      &      &      \\   
13    &11+3  &    6--1    &\y         &\y     &\my    &      &\y    &      &      &\y    &      &      &      &      &      \\   
14    & 9+3  &    6--1    &\y         &       &       &      &\y    &      &      &\y    &      &      &      &      &      \\   
15    &11+3  &    6--1    &\y         &       &       &      &\y    &\y    &\y    &\y    &      &      &      &      &      \\   
16    &12+3  &    6--1,2  &\y         &       &       &      &\y    &      &\y    &\y    &      &      &\y    &      &\y    \\   
17    &10+3  &    6--1    &\y         &       &       &      &\y    &      &\y    &\y    &      &      &      &      &      \\   
18    &10+3  &    6--1    &\y         &       &       &      &\y    &      &      &\y    &      &\y    &      &      &      \\   
19    &10+3  &    6--1    &\y         &       &       &      &\y    &      &      &\y    &      &      &      &\y    &      \\   
20    &11+3  &    6--1    &\y         &       &\my    &      &\y    &      &\y    &\y    &      &      &      &      &      \\   
21    &12+3  &    6--1    &\y         &       &\my    &      &\y    &      &\y    &\y    &      &      &      &\y    &      \\   
22    &10+3  &    6--1    &\y         &       &\my    &      &\y    &      &      &\y    &      &      &      &      &      \\   
23    &11+3  &    6--1    &\y         &       &\my    &      &\y    &      &      &\y    &\y    &      &      &      &      \\   
24    &12+3  &    6--1,2  &\y         &       &\my    &      &\y    &      &      &\y    &      &      &\y    &      &\y    \\   
25    &10+4  &    6--1    &\y$^{'}$   &       &\my    &      &\y    &      &      &\y    &      &\y    &      &      &      \\   
26    &11+4  &    6--1    &\y$^{'}$   &       &\my    &      &\y    &      &\y    &\y    &      &\y    &      &      &      \\   
27    &11+3  &    6--1    &\y         &\my    &       &      &\y    &\y    &      &\y    &      &      &      &      &      \\   
28    &12+3  &    6--1,2  &\y         &\my    &       &      &\y    &      &      &\y    &      &      &\y    &      &\y    \\   
29    &12+3  &    6--1    &\y         &\my    &\my    &      &\y    &      &\y    &\y    &      &      &      &      &      \\   
30    &11+3  &    6--1    &\y         &\my    &\my    &      &\y    &      &      &\y    &      &      &      &      &      \\   
\hline
\hline
\end{tabular}
\end{flushleft}

\noindent Table D.I: Classes of $D$--flat directions producing MSSM massless
field content that are $F$--flat to at least $6^{th}$ order in the superpotential. 
These classes are defined by their respective set of non-Abelian singlet states that develop
VEVs and are identified by their class identification number appearing in column one.
Column two indicates the number of singlets that acquire VEVs in each direction, with the first entry
being the number of states that take on VEVs as a result of FI cancellation and the second entry
being the number of $\Phi_4$--related states for which we additionally require VEVs. 
Column three gives the order 
in the FNY superpotential at which flatness is broken and the designation of the superpotential term
in Table D.II responsible for the breaking. For example, a flat direction belonging to class 5 is broken by the
seventh order term designated 7--1. 

The remaining columns indicate which states receive VEVs.
These are indicated by the $*$'s in the respective columns. 
An $*$ implies the given state takes on a VEV, while a $\bar{\ast}$ implies the vector partner instead. 
$\{VEV_1\}$ denotes the set of states   
$\{$\f{12},\f{23},(\f{4},\fp{4},\fb{4},\fpb{4}),\HS{30}\HS{38},\HS{15},\HS{31},$\}$. 
An $*$ in the $\{VEV_1\}$ column implies that all of the states in $\{VEV_1\}$ receive VEVs.
Furthermore, $*^{'}$ in the $\{VEV_1\}$ column indicates that, while present, none of the \f{4} VEVs are required for FI--term
cancellation.  
\vskip 2.0truein

\begin{flushleft}
\begin{tabular}{|r|l|}
\hline 
\hline
designation& FNY Superpotential Term \\ 
\hline
 6--1      & \HS{15} \HS{30} \HS{31} \HS{36} \VS{31} \Nc{3} \\
 6--2      & \HS{15} \HS{30} \HS{31} \HS{39} \VS{31} \VS{22} \\
 7--1      & \fp{4}  \HS{15} \HS{20} \HS{30} \HS{31} \VS{31} \NC{1} \\ 
12--1      & \f{23}  \fb{56} \f{4}$^2$  \HS{15}$^2$\HS{20}$^2$\HS{31}$^2$\HS{38}$^2$\\  
12--2      & \f{23}  \fb{56} \fp{4}$^2$\HS{15}$^2$\HS{20}$^2$\HS{31}$^2$\HS{38}$^2$\\  
\hline
\hline
\end{tabular}
\end{flushleft}

\noindent Table D.II: FNY superpotential terms that break $F$--flatness of 
$D$--flat directions 
at sixth order or higher. The first column gives a designation for each term. 

\def\f#1{$\Phi_{#1}$}
\def\fb#1{$\overline{\Phi}_{#1}$}
\def\fp#1{$\Phi^{'}_{#1}$}
\def\fbp#1{$\overline{\Phi}^{'}_{#1}$}
\def\fpx#1{$\Phi^{(')}_{#1}$}
\def\fbpx#1{$\overline{\Phi}^{(')}_{#1}$}
\def\HS#1{$H^s_{#1}$}
\def\VS#1{$V^s_{#1}$}
\def\NC#1{$N^c_{#1}$} 

\def\ify{$\infty$ }
\def\ifw{${\infty}^{\ast}$ }

\def\x{$\phantom{1}$}
\def\m{$\phantom{-}$}

\begin{flushleft}
\begin{tabular}{|r||r||rrrrrrr|}
\hline 
\hline
$c\#$  
&$\frac{Q^{(A)}}{112}$&$\{$\f{12},\f{23},$($\f{4}$)$,\HS{30}\HS{38},\HS{15},\HS{31},$\}$                                       
                                  &\fb{56}&\fp{56}&\HS{19}&\VS{31}&\HS{20}&\Nc{1}\\
&                     &           &\Nc{3} &\HS{21}&\HS{17}&\VS{12}&\HS{18}&\HS{39}\\
\hline
\hline
 1  &  -2 &{\x}3, 1, 1, {\x}3, {\x}2, {\x}2,  1  &{\m}1  &{\m}0 &0  &0  &0  &0\\  
                                                                    &&&0 &0 &0 &0 &0 &0 \\  
 2  &  -6 &   10, 2, 2, {\x}8, {\x}6, {\x}4,  2  &{\m}3  &   -1 &0  &0  &0  &0\\  
                                                                    &&&0 &0 &0 &0 &0 &0 \\  
 3  &  -4 &{\x}4, 3, 2, {\x}8, {\x}2, {\x}6,  2  &{\m}1  &{\m}0 &2  &0  &0  &0\\  
                                                                    &&&0 &0 &0 &0 &0 &0 \\  
 4  &  -2 &{\x}3, 2, 3, {\x}3, {\x}4, {\x}4,  3  &{\m}2  &{\m}0 &0  &0  &2  &0\\  
                                                                    &&&0 &0 &0 &0 &0 &0 \\  
 5  &  -4 &{\x}6, 1, 2, {\x}8, {\x}2, {\x}4,  2  &{\m}1  &{\m}0 &0  &2  &0  &2\\  
                                                                    &&&0 &0 &0 &0 &0 &0 \\  
 6  &  -6 &   10, 1, 2,    10, {\x}4, {\x}4,  2  &{\m}2  &   -1 &0  &2  &0  &2\\  
                                                                    &&&0 &0 &0 &0 &0 &0 \\  
 7  &  -8 &   10, 3, 4,    18, {\x}2,    10,  4  &{\m}1  &{\m}0 &2  &4  &0  &4\\  
                                                                    &&&0 &0 &0 &0 &0 &0 \\  
 8  &  -4 &{\x}6, 1, 4,    10, {\x}2, {\x}6,  4  &{\m}1  &{\m}0 &0  &4  &2  &4\\  
                                                                    &&&0 &0 &0 &0 &0 &0 \\  
 9  &  -4 &{\x}5, 2, 2, {\x}8, {\x}3, {\x}5,  2  &{\m}1  &{\m}0 &0  &1  &0  &0\\  
                                                                    &&&1 &0 &0 &0 &0 &0 \\  
10  &  -8 &   10, 3, 4,    18, {\x}4,    10,  4  &{\m}1  &{\m}0 &0  &4  &0  &2\\  
                                                                    &&&2 &0 &0 &0 &0 &0 \\  
11  &  -4 &{\x}4, 3, 4,    10, {\x}4, {\x}8,  4  &{\m}1  &{\m}0 &0  &2  &2  &0\\  
                                                                    &&&2 &0 &0 &0 &0 &0 \\  
12  &  -4 &{\x}4, 3, 6,    12, {\x}4,    10,  6  &{\m}1  &{\m}0 &0  &4  &4  &2\\  
                                                                    &&&2 &0 &0 &0 &0 &0 \\  
13  &  -6 &{\x}8, 2, 2,    12, {\x}4, {\x}6,  2  &{\m}1  &   -1 &0  &2  &0  &0\\  
                                                                    &&&2 &0 &0 &0 &0 &0 \\  
14  &  -2 &{\x}2, 1, 1, {\x}5, {\x}1, {\x}3,  1  &{\m}0  &{\m}0 &0  &1  &0  &0\\  
                                                                    &&&1 &0 &0 &0 &0 &0 \\  
25  &  -4 &{\x}4, 2, 4,    12, {\x}2, {\x}8,  4  &{\m}0  &{\m}0 &0  &4  &2  &2\\  
                                                                    &&&2 &0 &0 &0 &0 &0 \\  
16  &  -7 &{\x}9, 1, 3,    18, {\x}2, {\x}9,  4  &{\m}0  &    0 &0  &6  &0  &3\\  
                                                                    &&&2 &0 &0 &1 &0 &1 \\  
17  &  -4 &{\x}5, 1, 2,    10, {\x}1, {\x}5,  2  &{\m}0  &{\m}0 &0  &3  &0  &2\\  
                                                                    &&&1 &0 &0 &0 &0 &0 \\  
18  &  -3 &{\x}2, 2, 1, {\x}9, {\x}1, {\x}5,  2  &{\m}0  &{\m}0 &0  &2  &0  &0\\  
                                                                    &&&2 &0 &1 &0 &0 &0 \\  
19  &  -6 &{\x}6, 3, 5,    16, {\x}2,    10,  2  &{\m}0  &{\m}0 &0  &2  &0  &0\\  
                                                                    &&&4 &0 &0 &0 &2 &0 \\  
20  &  -6 &{\x}8, 1, 2,    14, {\x}2, {\x}6,  2  &{\m}0  &   -1 &0  &4  &0  &2\\  
                                                                    &&&2 &0 &0 &0 &0 &0 \\  
\hline
\end{tabular}
\end{flushleft}

\begin{flushleft}
\begin{tabular}{|r||r||rrrrrrr|}
\hline 
\hline
$c\#$  
&$\frac{Q^{(A)}}{112}$&$\{$\f{12},\f{23},$($\f{4}$)$,\HS{30}\HS{38},\HS{15},\HS{31},$\}$
                                                 &\fb{56}&\fp{56}&\HS{19}&\VS{31}&\HS{20}&\Nc{1}\\
&                     &                          &\Nc{3} &\HS{21}&\HS{17}&\VS{12}&\HS{18}&\HS{39}\\
\hline
\hline
21  &  -8 &   10, 2, 5,    20, {\x}2,    10,  2  &{\m}0  &   -1 &0  &4  &0  &2\\  
                                                                    &&&4 &0 &0 &0 &2 &0 \\  
22  &  -4 &{\x}5, 1, 1, {\x}9, {\x}2, {\x}4,  1  &{\m}0  &   -1 &0  &2  &0  &0\\  
                                                                    &&&2 &0 &0 &0 &0 &0 \\  
23  &  -8 &{\x}8, 3, 4,    20, {\x}2,    10,  2  &{\m}0  &   -1 &0  &4  &0  &0\\  
                                                                    &&&4 &2 &0 &0 &0 &0 \\  
24  &  -8 &   10, 1, 2,    20, {\x}4,    10,  4  &{\m}0  &   -1 &0  &6  &0  &0\\  
                                                                    &&&4 &0 &0 &2 &0 &2 \\  
25  &  -3 &{\x}3, 1, 0, {\x}8, {\x}1, {\x}3,  1  &{\m}0  &   -1 &0  &2  &0  &0\\  
                                                                    &&&2 &0 &1 &0 &0 &0 \\  
26  &  -8 &   10, 1, 0,    20, {\x}2, {\x}6,  2  &{\m}0  &   -3 &0  &6  &0  &2\\  
                                                                    &&&4 &0 &2 &0 &0 &0 \\  
27  &  -4 &{\x}2, 3, 4,    14, {\x}2,    10,  4  &   -1  &{\m}0 &0  &4  &2  &0\\  
                                                                    &&&4 &0 &0 &0 &0 &0 \\  
28  &  -6 &{\x}6, 1, 2,    18, {\x}2,    10,  4  &   -1  &{\m}0 &0  &6  &0  &0\\  
                                                                    &&&4 &0 &0 &2 &0 &2 \\  
29  &  -8 &   10, 1, 2,    20, {\x}2, {\x}8,  2  &   -1  &   -2 &0  &6  &0  &2\\  
                                                                    &&&4 &0 &0 &0 &0 &0 \\  
30  &  -6 &{\x}6, 2, 2,    16, {\x}2, {\x}8,  2  &   -1  &   -1 &0  &4  &0  &0\\  
                                                                    &&&4 &0 &0 &0 &0 &0 \\  
\hline
\hline
\end{tabular}
\end{flushleft}

\noindent Table D.III: 
Examples of flat directions from the various classes presented in Table D.I.a.
Column one indicates the class of directions to which the example belongs.
Column two gives the anomalous charge $Q^{(A)}/112$ of the example flat direction.
The remaining column entries specify the ratios of the norms of the VEVs in the
flat direction. 
$\{$\f{12},\f{23},$($\f{4}$)$,\HS{30}\HS{38},\HS{15},\HS{31},$\}$$\equiv\{ VEV_1 \}$ 
is the set of the norms that are non-zero for all flat directions.
The third component VEV, involving all of the $\Phi_4$--related states, is the net value of
$\mvev{\Phi_4} + \mvev{\Phi^{'}_4} - \mvev{\bar{\Phi}_4} - \mvev{\bar{\Phi}^{'}_4}$. 
Thus, a ``0'' in the $\Phi_4$ column for classes 14 and 27 implies that   
$\mvev{\Phi_4} + \mvev{\Phi^{'}_4} - \mvev{\bar{\Phi}_4} - \mvev{\bar{\Phi}^{'}_4} = 0$, while
each of the four states still takes on a VEV.      
\hfill\vfill\eject

{                     
\def\L#1{L_{#1}}
\def\K#1{K_{#1}}

\noindent 
\begin{flushleft}
\begin{tabular}{|r||r|l|l|r|r|r|}
\hline 
\hline
c$\#$& $\#$ MOBD& MOBD expression \\ 
\hline
\hline
 1  & 5+2 &$ 2 \L{6} +  6 \L{12} +   \L{13} +  4 \L{15} +  3 \L{16} + \K{7} + \K{8}                       $\\
 2  & 5+2 &$ 3 \L{6} + 10 \L{12} +   \L{13} +  4 \L{15} +  4 \L{16} + \K{7} + \K{8}                       $\\
 3  & 6+2 &$   \L{6} +  4 \L{12} +   \L{13} +  6 \L{15} +  4 \L{16} +   \L{8} + \K{7} + \K{8}             $\\     
 4  & 6+2 &$ 3 \L{6} +  6 \L{12} + 2 \L{13} +  6 \L{15} +  3 \L{16} +   \L{20}+ \K{7} + \K{8}             $\\     
 5  & 6+2 &$   \L{6} +  6 \L{12} +   \L{13} +  4 \L{15} +  4 \L{16} +   \L{7} + \K{7} + \K{8}             $\\     
 6  & 6+2 &$ 2 \L{6} + 10 \L{12} +   \L{13} +  4 \L{15} +  5 \L{16} +   \L{7} + \K{7} + \K{8}             $\\     
 7  & 7+2 &$   \L{6} + 10 \L{12} + 2 \L{13} + 10 \L{15} +  9 \L{16} + 2 \L{7} +   \L{8}  + \K{7} + \K{8}  $\\    
 8  & 7+2 &$   \L{6} +  6 \L{12} + 2 \L{13} +  6 \L{15} +  5 \L{16} + 2 \L{7} +   \L{20} + \K{7} + \K{8}  $\\       
 9  & 6+2 &$ 3 \L{6} + 10 \L{12} + 2 \L{13} + 10 \L{15} +  8 \L{16} +   \L{7} + \K{7} + \K{8}             $\\     
10  & 6+2 &$ 2 \L{6} + 10 \L{12} + 2 \L{13} + 10 \L{15} +  9 \L{16} + 2 \L{7} + \K{7} + \K{8}             $\\  
11  & 7+2 &$ 2 \L{6} +  4 \L{12} + 2 \L{13} +  8 \L{15} +  5 \L{16} +   \L{7} +   \L{20} + \K{7} + \K{8}  $\\      
12  & 7+2 &$ 2 \L{6} +  4 \L{12} + 3 \L{13} + 10 \L{15} +  6 \L{16} + 2 \L{7} + 2 \L{20} + \K{7} + \K{8}  $\\   
13  & 6+2 &$ 2 \L{6} +  8 \L{12} +   \L{13} +  6 \L{15} +  6 \L{16} +   \L{7}            + \K{7} + \K{8}  $\\     
14  & 6+2 &$   \L{6} +  4 \L{12} +   \L{13} +  6 \L{15} +  5 \L{16} +   \L{7}            + \K{7} + \K{8}  $\\     
15  & 7+2 &$   \L{6} +  4 \L{12} + 2 \L{13} +  8 \L{15} +  6 \L{16} + 2 \L{7} +   \L{20} + \K{7} + \K{8}  $\\       
16  & 7+2 &$   \L{6} +  9 \L{12} + 2 \L{13} +  9 \L{15} +  9 \L{16} + 3 \L{7} +   \L{17} + \K{7} + \K{8}  $\\       
17  & 6+2 &$   \L{6} + 10 \L{12} + 2 \L{13} + 10 \L{15} + 10 \L{16} + 3 \L{7}            + \K{7} + \K{8}  $\\   
18  & 7+2 &$   \L{6} +  4 \L{12} + 2 \L{13} + 10 \L{15} +  9 \L{16} + 2 \L{7} +   \L{10} + \K{7} + \K{8}  $\\       
19  & 7+2 &$   \L{6} +  6 \L{12} +   \L{13} + 10 \L{15} +  8 \L{16} +   \L{7} +   \L{2}  + \K{7} + \K{8}  $\\      
20  & 6+2 &$   \L{6} +  8 \L{12} +   \L{13} +  6 \L{15} + \L{16}* 7 + 2 \L{7}            + \K{7} + \K{8}  $\\  
21  & 7+2 &$   \L{6} + 10 \L{12} +   \L{13} + 10 \L{15} + 10 \L{16} + 2 \L{7} +   \L{2}  + \K{7} + \K{8}  $\\      
22  & 6+2 &$ 2 \L{6} + 10 \L{12} +   \L{13} +  8 \L{15} +  9 \L{16} + 2 \L{7}            + \K{7} + \K{8}  $\\  
23  & 7+2 &$   \L{6} +  8 \L{12} +   \L{13} + 10 \L{15} + 10 \L{16} + 2 \L{7} +   \L{5}  + \K{7} + \K{8}  $\\      
24  & 7+2 &$ 2 \L{6} + 10 \L{12} + 2 \L{13} + 10 \L{15} + 10 \L{16} + 3 \L{7} + 2 \L{17} + \K{7} + \K{8}  $\\    
25  & 7+2 &$   \L{6} +  6 \L{12} +   \L{13} +  6 \L{15} +  8 \L{16} + 2 \L{7} +   \L{10} + \K{7} + \K{8}  $\\       
26  & 7+2 &$   \L{6} + 10 \L{12} +   \L{13} +  6 \L{15} + 10 \L{16} + 3 \L{7} +   \L{10} + \K{7} + \K{8}  $\\       
27  & 7+2 &$   \L{6} +  2 \L{12} + 2 \L{13} + 10 \L{15} +  7 \L{16} + 2 \L{7} +   \L{20} + \K{7} + \K{8}  $\\       
28  & 7+2 &$   \L{6} +  6 \L{12} + 2 \L{13} + 10 \L{15} +  9 \L{16} + 3 \L{7} + 2 \L{17} + \K{7} + \K{8}  $\\    
29  & 6+2 &$   \L{6} + 10 \L{12} +   \L{13} +  8 \L{15} + 10 \L{16} + 3 \L{7}            + \K{7} + \K{8}  $\\  
30  & 6+2 &$   \L{6} +  6 \L{12} +   \L{13} +  8 \L{15} +  8 \L{16} + 2 \L{7}            + \K{7} + \K{8}  $\\
\hline
\hline
\end{tabular}
\end{flushleft} 

Table D.IV: The flat direction in Table D.III expressed as linear combinations of the one--dimensional 
maximally orthogonal basis directions (MOBD) presented in Table D.I. 
Column one gives the class number of the flat direction.
Column two specifices the number of unique MOBD's
associated with each flat direction, the first entry being the number of $L$--class basis directions
and the second entry the number of $K$--class. Column three gives the linear combination. Since the
$K$--class terms contribution to each $D$--term is zero, their coefficients are not specified.}
\hfill\vfill\eject

\begin{flushleft}
\begin{tabular}{|r||r||r|rrrrrrr|}
\hline 
\hline
$c\#$  & $\#$
&$\frac{Q^{(A)}}{112}$&$\{$\f{12},\f{23},$($\f{4}$)$,\HS{30}\HS{38},\HS{15},\HS{31},$\}$
                                  &\fb{56}&\fp{56}&\HS{19}&\VS{31}&\HS{20}&\Nc{1}\\
&&                    &           & \Nc{3}&\HS{21} &\HS{17}&\VS{12}&\HS{18}&\HS{39}\\
\hline
\hline
$X$&  8&  -2& 3, 1, 1, {\x}3, 2, {\x}2,  1&   1&   0&  0&   0&   0&   0\\  
                                      &&&&    0&   0&  0&   0&   0&   0\\    
$Y$&  9&  -2& 2, 1, 1, {\x}5, 1, {\x}3,  1&   0&   0&  0&   1&   0&   0\\    
                                      &&&&    1&   0&  0&   0&   0&   0\\    
$A$&  7&   0& 0, 1, 2, {\x}0, 2, {\x}2,  2&   1&   0&  0&   0&   2&   0\\      
                                      &&&&    0&   0&  0&   0&   0&   0\\    
$B$&  7&  -2& 1, 2, 1, {\x}5, 0, {\x}4,  1&   0&   0&  2&   0&   0&   0\\
                                      &&&&    0&   0&  0&   0&   0&   0\\    
$C$&  7&  -2& 3, 0, 1, {\x}5, 0, {\x}2,  1&   0&   0&  0&   2&   0&   2\\
                                      &&&&    0&   0&  0&   0&   0&   0\\    
$D$&  7&  -2& 3, 0, 0, {\x}4, 1, {\x}1,  0&   0&  -1&  0&   1&   0&   0\\     
                                      &&&&    1&   0&  0&   0&   0&   0\\    
$E$&  9&  -2& 0, 3, 5, {\x}9, 3, {\x}9,  5&   0&   0&  0&   3&   4&   0\\
                                      &&&&    3&   0&  0&   0&   0&   0\\    
$F$&  9&  -4& 0, 3, 4,    18, 0,    12,  4&  -3&   0&  0&   6&   2&   0\\
                                      &&&&    6&   0&  0&   0&   0&   0\\    
$G$&  7&   0& 0, 0, 1, {\x}1, 0, {\x}1,  1&   0&   0&  0&   1&   1&   1\\
                                      &&&&    0&   0&  0&   0&   0&   0\\    
$H$&  5&  -2& 4, 0, 0, {\x}2, 2, {\x}0,  0&   1&  -1&  0&   0&   0&   0\\
                                       &&&&   0&   0&  0&   0&   0&   0\\    
$I$&  7&  -1& 0, 1, 0, {\x}4, 0, {\x}2,  1&   0&   0&  0&   1&   0&   0\\
                                       &&&&   1&   0&  1&   0&   0&   0\\
$J$&  7&  -2& 2, 1, 3, {\x}6, 0, {\x}4,  0&   0&   0&  0&   0&   0&   0\\
                                       &&&&   2&   0&  0&   0&   2&   0\\                             
$M$& 10&  -4& 5, 0, 1,    11, 2, {\x}6,  3&   0&   0&  0&   4&   0&   0\\
                                       &&&&   2&   0&  0&   2&   0&   2\\
$N$&  7&  -2& 2, 0, 0, {\x}6, 0, {\x}2,  0&  -1&  -1&  0&   2&   0&   0\\
                                       &&&&   2&   0&  0&   0&   0&   0\\              
$P$&  9&  -2& 1, 1, 1, {\x}7, 0, {\x}4,  1&  -1&   0&  0&   2&   0&   0\\
                                       &&&&   2&   0&  0&   0&   0&   0\\
$Q$&  9&  -4& 3, 2, 3,    11, 0, {\x}6,  1&   0&   0&  0&   2&   0&   0\\
                                       &&&&   2&   2&  0&   0&   0&   0\\
$R$&  9&  -3& 4, 0, 0, {\x}8, 0, {\x}2,  1&   0&  -1&  0&   3&   0&   2\\
                                       &&&&   1&   0&  1&   0&   0&   0\\
\hline
\hline
\end{tabular}
\end{flushleft}

\noindent Table D.V: 
Some one--dimenional physical directions that are $D$--flat for all 
non-anomalous $U(1)_i$ 
and from which all $D$-- and $F$--flat directions in Table D.IV may be formed.

\hfill\vfill\eject

\noindent 
\begin{flushleft}
\begin{tabular}{|r||l|r|r|r|}
\hline 
\hline
c$\#$& physical basis expression 
                         & $\#$ VEV fields  & Dim$_{FI}$  &  $\#$ U(1) broken \\ 
\hline
\hline
 1  &$ X                 $ &  8     &  0          &  7 + 1 \\   
 2  &$ 2 X + H           $ &  9     &  1          &  7 + 1 \\
 3  &$ X + B             $ &  9     &  1          &  7 + 1 \\
 4  &$ X + A             $ &  9     &  1          &  7 + 1 \\
 5  &$ X + C             $ & 10     &  1          &  8 + 1 \\
 6  &$ X + C + H         $ & 11     &  2          &  8 + 1 \\
 7  &$ X + 2 C + B       $ & 11     &  2          &  8 + 1 \\
 8  &$ X + C + 2 G       $ & 11     &  2          &  8 + 1 \\
 9  &$ Y + C             $ & 10     &  1          &  8 + 1 \\
10  &$ Y                 $ &  9     &  0          &  8 + 1 \\      
11  &$ Y + D             $ & 10     &  1          &  8 + 1 \\
12  &$ X + Y             $ & 10     &  1          &  8 + 1 \\
13  &$ Y + I             $ & 10     &  1          &  8 + 1 \\
14  &$ D + I             $ & 10     &  1          &  8 + 1 \\
15  &$ 2 Y + J           $ & 10     &  1          &  8 + 1 \\
16  &$ 2 Y + N           $ & 11     &  1          &  9 + 1 \\
17  &$ 3 Y + E + F       $ & 11     &  2          &  8 + 1 \\ 
18  &$ X + Y + D         $ & 11     &  2          &  8 + 1 \\ 
19  &$ 2 X + Y + E       $ & 11     &  2          &  8 + 1 \\
20  &$ X + Y + C         $ & 11     &  2          &  8 + 1 \\
21  &$ Y + D + Q         $ & 11     &  2          &  8 + 1 \\
22  &$ Y + D + C         $ & 11     &  2          &  8 + 1 \\
23  &$ Y + G             $ & 11     &  1          &  9 + 1 \\ 
24  &$ Y + C + D + N     $ & 12     &  3          &  8 + 1 \\
25  &$ M + P             $ & 12     &  1          & 10 + 1 \\
26  &$ 2 X + Y + E + 4 G $ & 12     &  3          &  8 + 1 \\ 
27  &$ 2 D + I + R       $ & 12     &  2          &  9 + 1 \\
28  &$ Y + D + M         $ & 12     &  2          &  9 + 1 \\
29  &$ Y + C + D + J     $ & 12     &  3          &  8 + 1 \\ 
30  &$ 2 Y + 3 C + 2 M   $ & 12     &  2          &  9 + 1 \\
\hline
\hline
\end{tabular}
\end{flushleft}

\noindent Table D.VI: 
Expressions for flat direction examples given in terms of physical 
dimension--one directions.
Column two expresses the example flat direction for the class, indicated by
the column one entry,
in terms of the one--dimensional $D$--flat directions of Table D.V.
Column three indicates the number of {\it independent} VEVs
(which excludes the $\Phi^{'}_4$, $\overline{\Phi}_4$, 
and $\overline{\Phi}^{'}_4$ VEVs). 
Column four specifies the dimension of each flat direction
(excluding the two degrees of freedom in the 
$\overline{\Phi}_4$, and $\overline{\Phi}^{'}_4$
VEVs) following cancellation of the FI term, which reduces a dimension by 1.
This dimension is one less
than the number of physical dimension--one $D$--flat--directions in column two 
since, prior to FI term cancellation, 
the degrees of freedom in a given class of flat directions are the 
coefficients $w_{j,k}$ of the physical flat directions, in (\ref{cfb1}), 
associated with a given class.
The number of $U(1)_i$ broken by a flat direction, including the anomalous
$U(1)_A$, is specified in column five. This equals the difference between 
the number of independent VEVs in column three and the
number of degrees of freedom in column four.  

Since the gauge charges of $\Phi^{'}_4$, 
$\overline{\Phi}_4$, and $\overline{\Phi}^{'}_4$  
are not independent from those of $\Phi_4$, the related terms $K_7$ and $K_8$, that
contain $\Phi^{'}_4$, $\overline{\Phi}_4$, and 
$\overline{\Phi}^{'}_4$ VEVs do not offer 
independent degrees of freedom that result in
additional breaking of non--anomalous $U(1)_i$. Therefore, these terms
are excluded in column two, as are their two degrees of freedom in column four.

               
\vfill\eject

\bigskip
\medskip

\bibliographystyle{unsrt}

\hfill\vfill\eject
\end{document}